\documentclass{aa}  
\usepackage{natbib}
\usepackage{lscape}
\usepackage{color}
\usepackage{graphicx}
\usepackage[varg]{txfonts}
\DeclareRobustCommand{\APC}[1]{#1}

\DeclareRobustCommand{\etg}{ETG}
\DeclareRobustCommand{\etgs}{ETGs}

\begin{document} 

\title{VEGAS: A VST Early-type GAlaxy Survey}
   \subtitle{II. Photometric study of giant ellipticals and their stellar halos}

   \author{Marilena Spavone 
          \inst{1}
          \and
          Massimo Capaccioli \inst{1,2}
\and
Nicola R. Napolitano \inst{1}
\and
          Enrichetta Iodice \inst{1}
\and
          Aniello Grado \inst{1}
\and
          Luca Limatola \inst{1}
\and
           Andrew P. Cooper\inst{4}
\and
          Michele Cantiello \inst{1,3}
\and
          Duncan A. Forbes \inst{5}
\and
Maurizio Paolillo \inst{2}
\and
Pietro Schipani \inst{1}
          }
          
 \institute{INAF-Astronomical Observatory of Capodimonte, Salita Moiariello 16, I80131, Naples, Italy\\
              \email{spavone@na.astro.it}
         \and
             University of Naples Federico II, C.U. Monte Sant'Angelo,
             Via Cinthia, 80126, Naples, Italy
\and
INAF-Astronomical Observatory of Teramo, Via Maggini,
             64100, Teramo, Italy
\and
Institute for Computational Cosmology, Durham, UK
\and
Centre for Astrophysics \& Supercomputing, Swinburne University, Hawthorn, VIC 3122, Australia}

\date{Received ....; accepted ...}

\abstract{Observations of diffuse starlight in the outskirts of galaxies
are thought to be a fundamental source of constraints on the
cosmological context of galaxy assembly in the $\Lambda$CDM model.
Such observations are not trivial because of the extreme faintness
of such regions.  In this work, we investigate the photometric properties of
six massive early type galaxies (ETGs) in the VEGAS sample (NGC 1399, NGC 3923,
NGC 4365, NGC 4472, NGC 5044, and NGC 5846) out to extremely low surface
brightness levels, with the goal of characterizing the global structure of
their light profiles for comparison to state-of-the-art galaxy formation
models.  We carry out deep and detailed photometric mapping of our
ETG sample taking advantage of deep imaging with VST/OmegaCAM in
the g and i bands.  By fitting the light profiles, {  and comparing
  the results to simulations of elliptical galaxy assembly, we identify} signatures
of a transition between ``relaxed'' and ``unrelaxed'' accreted components and
can constrain the balance between in situ and accreted stars. The very
good agreement of our results with predictions from theoretical
simulations demonstrates that the full VEGAS sample of $\sim 100$ ETGs will
allow us to use the distribution of diffuse light as a robust statistical
probe of the hierarchical assembly of massive galaxies.}


\authorrunning{Spavone et al.}
\titlerunning{Stellar halos of giant ETGs}

\maketitle 


\section{Introduction}\label{intro}  

LCDM galaxy formation theories predict that galaxies grow through a combination
of in situ star formation and accretion of stars from other galaxies
\citep{White91}. The ratio of stellar mass contributed by these two modes of
growth is expected to change systematically over the lifetime of a galaxy as
its dark matter halo and star formation efficiency evolve
\citep[e.g.][]{Guo08}. Accreted stars are expected to dominate in the outer
parts of galaxies because they have much lower binding energies in the remnant
system than stars formed by dissipative collapse. Since dynamical timescales
are long in these outer regions, phase-space substructures related to
accretion, such as streams and caustics, can persist over many gigayears.

The structural properties of the outer parts of galaxies and their correlations
with stellar mass and other observables might therefore provide ways of testing
theoretical predictions of growth by accretion. For example, they could
constrain the overall ratio of stars accreted to those formed in situ, the
number of significant progenitors and the dynamics of the most important merger
events \citep{Johnston08,Cooper10,Deason13,Amorisco15}.  In support of this
idea, cosmological simulations of galaxy formation predict trends in these
quantities with mass that are responsible for systematic variations in the
average shape, amplitude and extent of azimuthally averaged surface brightness
profiles \citep{Font11,Cooper13,Pillepich14}.  Unfortunately, the accreted
debris is also expected to have extremely low surface brightness ($\mu_{V}
\sim$ 29 mag/arcsec$^{2}$) on average and therefore to be very hard to
separate from the background sky in conventional photometric observations.

In this paper we use extremely deep images of six massive early type galaxies
(\etgs{}) from the VEGAS survey (described below) to constrain the properties
of their accreted stellar components.  The brightest \etgs{} are thought to be
predominantly the central galaxies of the most massive DM halos at the present
day \citep[e.g.][]{Frenk85, Kauffmann93, DeLucia06}. In order to explain the
exponential high-mass cut-off of the galactic stellar mass function,
theoretical models have invoked processes whereby in situ star formation in
such halos is strongly suppressed at late times (e.g.  Bower et al.  2006). As
a consequence, massive ETGs in these models accumulate the bulk of their
stellar mass by accretion, predominantly through $\sim10$ mergers (in the case
of massive BCGs; \citealt{DeLucia07, Cooper15}) of low stellar mass ratio.
Their spheroidal morphology is thought to arise from violent relaxation of the
central potential during the most massive of these mergers, which can erase
coherent structures established by earlier dissipative star formation
\citep{Villumsen83,Cole00,Hilz12}. The same models predict a rather
different relationship between mass growth and structure for late type galaxies
(LTGs) in less massive dark matter halos, like the Milky Way. In LTGs, the vast
majority of stars are formed in situ in rotating discs that survive to the
present day, while the bulk of the accreted stars are contributed by one or two
objects of much lower mass and are mostly deposited on weakly-bound orbits
extending far from the center of the potential.  These models therefore imply a
natural but complex continuum between the properties of the dominant stellar
components of \etgs{} and those of the diffuse spheroidal `stellar halos'
around LTGs like the Milky Way, all of which originate from the same 
hierarchical accretion process \citep{Purcell07,Cooper13}. 

From an observational perspective, it is straightforward to separate the disc
and stellar halo components of LTGs because their mass, kinematics, spatial
distribution and stellar populations are very different. Moreover, with
relatively little ambiguity, the empirically-defined stellar halo can be
identified with the bulk of the accreted stellar mass in those galaxies and the
disc with the bulk of in situ stellar mass.  In \etgs{}, however, the
connections between different mechanisms of mass growth and the `structural
components' inferred from images are much less straightforward. If the bulk of
the stars are really accreted, then the `stellar halo' (or `spheroid' or
`classical bulge') should be identified with at least the structural component
that dominates the observed stellar mass. However, other empirical `components' might
also be accreted.  In situ stars in \etgs{} will be extremely difficult to
distinguish if they are also spheroidal, dispersion-supported and have old,
metal-rich stellar populations resembling those of the dominant accreted
component(s), with which they have been thoroughly mixed by violent
relaxation\footnote{In the extreme case of a merger between two identical
elliptical galaxies in identical DM halos, it would clearly be impossible to
differentiate between the ‘in situ’ component and the ‘accreted’ component in
the surface brightness profile of the remnant.}.  

Cosmological dynamical simulations can help by suggesting plausible
interpretations for features in the surface brightness profiles of ETGs in the
context of specific galaxy formation theories.  In particular, simulated
galaxies show evidence for substructure in the form of inflections (`breaks'),
at which the surface brightness profile either becomes steeper or shallower
\citep[e.g.][]{Cooper13, Rodriguez15}. As noted above, models predict that the
accreted components of \etgs{} comprise several equal-size chunks of stellar
mass from different contributors, but also that these contributors have a range
of relative total (virial) mass ratios, such that their debris will relax to
distributions with a range of characteristic radii \citep{Amorisco15}. As shown
for example in figure 10 of \citet{Cooper15}, inflections in the profile may
therefore correspond not only to the transition between in situ and
accretion dominance (which is responsible for the most obvious profile
inflection for most LTGs, although this occurs at relatively low SB;
\citealt{Abadi06, Font11, Pillepich14, Rodriguez15}) but more generally to
variations in the ratio between individual accreted components as a function of
radius \citep{Cooper10, Deason13, Amorisco15}.  Indeed, for the reasons given
above, most simulations\footnote{The exception being simulations that do not
strongly suppress in situ star formation in very massive galaxies, which are
generally in disagreement with observed luminosity functions.} predict that the
transition between in situ and accretion-dominated regions should be almost
imperceptible in the azimuthally averaged profiles of typical \etgs{}, even
though it occurs at relatively high SB.  This transition may still be
detectable as a change in shape or stellar population.

Observational work has clearly identified features in ETG surface brightness profiles and
kinematic distributions indicative of substructure in the accreted component
\citep{Bender15, Longobardi15a, Longobardi15b, Foster16}.  Some of these
feature occur at distances of many effective radii and hence at very low SB,
which could be interpreted as evidence for multiple accreted components in the
context of the models mentioned above.   Using different techniques with
observations of different depths, several authors have concluded that the 
profiles of massive ETGs are not well described by a single S{\'e}rsic
$r^{1/n}$ law component, once thought to be near-universal for spheroidal
galaxies. For example, \citet{Gonzalez05} found that the surface brightness profiles of BCGs
in their sample were well fitted by a double $r^{1/4}$ de Vaucouleurs law,
while other works have argued that a combination of a S{\'e}rsic and an
exponential ($n=1$) component provide the best description of particular \etg{}
light profiles \citep{Seigar07, Donzelli11, Bender15}. \citet{Huang13} have
suggested three or more S{\'e}rsic components may be required for an accurate
description of typical \etgs{}.

A recent revival of this topic, historically hampered by the faintness of
galaxian halos \citep{deV79, Donofrio94}, has been fostered by the coming into
operation of a new generation of very-wide-field imaging
instruments \citep{Ferrarese12, Mihos13, Duc15, Cap15, Iodice16}. We are taking
advantage of the wide field of view and high spatial resolution of the VLT
Survey Telescope (VST; \citealt{Capaccioli11}) at the ESO Cerro Paranal
Observatory (Chile) to carry out a multiband imaging survey of nearby
ETGs, named VEGAS (VST Early-type GAlaxies Survey) (\citealt{Cap15};
hereafter Paper I). The primary aim of the survey is to obtain
maps of the surface brightness of galaxies with Heliocentric redshifts
$V_{rad}<4000$ km/s and $M_{B_{tot}}<-19.2$ mag sufficiently deep to
detect the signatures of diffuse stellar components (the expected depths are
$\mu_{g} \sim 27.3$, $\mu_{r} \sim 26.8$, and $\mu_{i} \sim 26$
mag/arcsec$^{2}$).  The large, homogeneous and deep sample from VEGAS will
enable a more systematic analysis of \etg{} light profiles and other
photometric trends with radius than has previously been possible. We
expect to obtain deep images for a sample of $\sim 100$ ETGs across a range of
environments. To date, we have collected data for about thirty galaxies.
The present status of VEGAS observations is regularly updated at
the link \url{http://www.na.astro.it/vegas/VEGAS/VEGAS_Targets.html}.

The subsample of VEGAS galaxies that we study here are listed in
Tab. \ref{basic} together with their basic parameters. These galaxies
have been chosen because of their large angular dimensions, and because
they are the brightest galaxies in their local environment. Here we
present and analyse the {\it g} and {\it i}-band photometric properties of the
four galaxies NGC 4365, NGC 3923, NGC 5044 and NGC 5846, which are in
small groups and clusters.  We combine these results with those
presented in Paper I for NGC 4472 in the Virgo cluster, and in \citet{Iodice16}
for NGC 1399 in the Fornax cluster. Since both NGC 4472 and NGC 1399 are
targets of the VEGAS survey and have been observed with VST, together with the
other objects they represent a homogeneous sample of bright ETGs.

The paper is organized as follows. In Sec. \ref{sample} we describe the
galaxies in our sample and their main properties. The observing strategy
and the data reduction procedure are described in Sec. \ref{vegas}, while the
data analysis is presented in Sec. \ref{phot} and \ref{sb}. In Sec. \ref{teor}
we compare our data with theoretical predictions, and in Sec. \ref{disc} we
discuss our results and draw conclusions. 

\section{The sample}\label{sample}

We briefly describe here the main properties of the four galaxies analyzed in
this work, NGC 3923, NGC 4365, NGC 5044 and NGC 5846, and those of NGC
4472 and NGC 1399 (published in Paper I and by \citet{Iodice16},
respectively).

\begin{table}
\setlength{\tabcolsep}{2.1pt}
\caption{\label{basic}Basic properties of the galaxies studied in this paper.} \centering
\begin{tabular}{lccc}
\hline\hline
Parameter & Value & Ref.\\
\hline \vspace{-7pt}\\
 & {\it NGC 1399} & \\
\hline \vspace{-7pt}\\
Morphological type & E1 & RC3 \\
R.A. (J2000)& 03h38m29.0s& NED \\
Dec. (J2000) & -35d27m02s & NED\\
Helio. radial velocity & 1425 km/s&NED\\
Distance& 20.9 Mpc& \citet{Blak09} \\
Absolute magnitude $M_{g}$& -23.05\tablefootmark{b}&
\citet{Iodice16}\\
\hline \vspace{-7pt}\\
 & {\it NGC 3923} & \\
\hline \vspace{-7pt}\\
Morphological type & E4 & RC3 \\
R.A. (J2000)& 11h51m01.7s& NED \\
Dec. (J2000) & -28d48m22s & NED\\
Helio. radial velocity & 1739 km/s&NED\\
Distance& 22.9 Mpc& \citet{Tonry01} \\
Absolute magnitude $M_{g}$& -22.66\tablefootmark{b}& This work\\
\hline \vspace{-7pt}\\
 & {\it NGC 4365} & \\
\hline \vspace{-7pt}\\
Morphological type & E3 & RC3 \\
R.A. (J2000)& 12h24m28.3s & NED \\
Dec. (J2000) & +07d19m04s& NED\\
Helio. radial velocity & 1243 km/s&NED\\
Distance& 23.1 Mpc& \citet{Blak09} \\
Absolute magnitude $M_{g}$& -22.41\tablefootmark{a}& This work\\
\hline \vspace{-7pt}\\
 & {\it NGC 4472} & \\
\hline \vspace{-7pt}\\
Morphological type & E2 & RC3 \\
R.A. (J2000)& 12h29m46.7s& NED \\
Dec. (J2000) & +08d00m02s & NED\\
Helio. radial velocity & 981 km/s&NED\\
Distance& 16.9 Mpc& \citet{Mei07} \\
Absolute magnitude $M_{g}$& -22.77\tablefootmark{a}& This work\\
\hline \vspace{-7pt}\\
 & {\it NGC 5044} & \\
\hline \vspace{-7pt}\\
Morphological type & E0 & RC3 \\
R.A. (J2000)& 13h15m24.0s & NED \\
Dec. (J2000) & -16d23m08s & NED\\
Helio. radial velocity & 2782 km/s&NED\\
Distance& 31.19 Mpc& \citet{Tonry01} \\
Absolute magnitude $M_{g}$& -22.59\tablefootmark{b}& This work\\
\hline \vspace{-7pt}\\
 & {\it NGC 5846} & \\
\hline \vspace{-7pt}\\
Morphological type & E0 & RC3 \\
R.A. (J2000)& 15h06m29.3s& NED \\
Dec. (J2000) & +01d36m20s & NED\\
Helio. radial velocity & 1712 km/s&NED\\
Distance& 24.89 Mpc& \citet{Tonry01} \\
Absolute magnitude $M_{g}$& -22.77\tablefootmark{b}& This work\\
\hline
\end{tabular}
\tablefoot{ \tablefoottext{a}{Corrected for interstellar extinction as in
 \citet{Battaia12}.} \tablefoottext{b}{Corrected for interstellar extinction as in \citet{Burstein82}.}}
\end{table}

\subsection{NGC 1399}

NGC 1399 (FCC 213) is the central and brightest elliptical galaxy in the Fornax
cluster. It has been extensively studied in a wide range of wavelengths. This
galaxy has a vast and diffuse stellar envelope surrounding it \citep{Iodice16},
as well as an extended and asymmetric gaseous halo in X-rays
\citep{Paolillo02}.  NGC 1399 is the subject of a paper by \citet{Iodice16}, in
which its main properties are discussed at length.

\subsection{NGC 3923}

NGC 3923, first studied by \citet{Malin80}, is one of the best examples of
a ``shell galaxy'', with a vast system of peculiar, concentric shells.
\citet{Sandage94} found at least seven dwarf ellipticals in the field centered
on NGC 3923, which is the brightest member of this small group.
\citet{Carter98} found that the galaxy has minor axis rotation. This
could be either the result of a kinematically decoupled core, or evidence
for a triaxial geometry. \citet{Buote98}, by studying the X-ray isophotes of
NGC 3923, found evidence for a dark matter distribution more flattened and
extended than the optical luminosity distribution.

\subsection{NGC 4365}

NGC 4365 (VCC 0731) is a bright and giant elliptical galaxy located in
a group slightly behind the ``Virgo subcluster B'' centered on NGC 4472. It
shows the presence of a small scale disc in the center, bluer than the
surrounding galaxy \citep{Ferrarese06}. 

\citet{Peng06} detected a significant population of globular clusters (GCs)
around this object, with a specific frequency higher than in other
galaxies of comparable luminosity, probably due to a larger number of
red GCs.  Moreover, \citet{Blom12a, Blom12b} found the GC system to consist of
three subpopulations.

\citet{Surma93} found evidence for two kinematically distinct stellar components:
a bulge-like component and a disc-like subsystem. The most interesting kinematic features of NGC 4365 are its
kinematically decoupled core \citep{Davies01} and that the galaxy rolls rather
than rotates \citep{Arnold14}.  \citet{Bogdan12} found an extended faint
tidal tail between NGC 4365 and NGC 4342 in deep B-band images ($\mu_{B} \sim\
29.5$ mag/arcsec$^{2}$), which \citet{Blom14} claimed to be a sign of ongoing
interaction between the two galaxies.

\subsection{NGC 4472}

NGC 4472 (M49), a giant elliptical galaxy, is the brightest member of the Virgo
cluster. It lies in the center of the southern grouping of the cluster, also
known as ``Virgo subcluster B''. This galaxy has been extensively studied
by many authors (see e.g. \citealt{Kim00, Ferrarese06, Kormendy09,
Janowiecki10, Mihos13}) and, together with its companion galaxies, it was
the subject of the first VEGAS paper \citep{Cap15}. NGC 4472 presents
extended shells and fans of diffuse material, and \citet{Cap15} also
detected an Intra Cluster Light (ICL) contribution to the surface
brightness profile. \citet{Irwin96} found that the X-ray contours of NGC 4472
are elongated in the northeast-southwest direction, probably as a result of
motion toward the center of the Virgo cluster.

\subsection{NGC 5044}

NGC 5044 is the central and brightest galaxy of a rich and X-ray bright group
\citep{Mendel08}. Dust patches near the center of the galaxy have been detected
by \citet{Colbert01}. X-ray observations of NGC 5044 show two
cavities and cooler filamentary X-ray emission, thought to have been
caused by a dynamical perturbation induced by the accretion of a low mass
satellite \citep{Buote03, Gastaldello09}. 

\subsection{NGC 5846}

NGC 5846 is the optically most luminous galaxy in the center of its group. HST
images show the presence of a ``parabolic'' dust lane around the nucleus of the
galaxy. X-ray observations \citep{Trinchieri02} show a disturbed
hot gas morphology on arcsecond scales. \citet{Haynes91} found that
the three dimensional distribution of galaxies around NGC 5846 is a
prolate cloud, pointing toward the Virgo cluster.

\section{The Vegas Survey: observation and data reduction}\label{vegas}

The VEGAS survey and the data reduction procedure adopted in this work
are described by \citet{Cap15}. The only significant change we
make here concerns the sky background subtraction for objects with large
angular extent (i.e. those filling the VST field of view). In Paper
I we used a direct polynomial interpolation of regions surrounding
our target galaxies that were identified as `background'
after masking of bright sources.  This procedure, according
to our experience, works very well to isolate a statistically significant
area for the background interpolation when the target has a angular extent
much smaller than the VST field. However, experience gained in reducing
images of NGC 4472 highlighted the need for a different
approach, which we develop here. This modified approach involves
both observations and data reduction. 

We decided to adopt a {\it step-dither} observing strategy for galaxies
with large angular extent, consisting of a cycle of short exposures centered
on the target and on offset fields ($\Delta = \pm 1$ degree). With such
a technique the background can be estimated from exposures taken as
close as possible, in space and time, to the scientific ones. This ensures
better accuracy, reducing the uncertainties at very faint surface brightness
levels, as found by \citet{Ferrarese12}. This observing strategy
allowed us to build an average sky background of the night which was subtracted
from each science image.

The main advantage of the {\it step-dither} technique is the possibility
of obtaining a larger field around galaxies with larger angular
sizes. This reduces uncertainties in distinguishing  pixels
belonging to the target galaxy from the background, particularly in the
outskirts of the target. On the other hand, with this technique
galaxy and background exposures are not simultaneous, and more telescope
time is consumed with respect to the standard dithering strategy.

In the subsample of VEGAS analyzed in this work, the images of NGC 4472
and NGC 4365 were acquired with standard dithering strategy, while for NGC
1399, NGC 3923, NGC 5044 and NGC5846 we adopted the {\it step-dither}
strategy.

The data used in this paper consist of exposures in {\it g} and {\it
  i} SDSS bands (Table \ref{data}) obtained with VST + OmegaCAM, both in service and in visitor mode with the following constraints: 
\begin{itemize}
\item S/N $\geq$ 3 per arcsec$^{2}$;
\item dark time;
\item seeing $\sim$ 1'';
\item airmass $\leq$ 1.2.
\end{itemize}

\begin{table}
\caption{\label{data}VST exposures used in this work.} \centering
\begin{tabular}{lccccc}
\hline\hline
Object & Band & $T_{exp}$ & FWHM\tablefootmark{a}\\ 
    & & [sec] & [arcsec] & \\
\hline \vspace{-7pt}\\
NGC 3923 & g & 9600 & 1.08\\ 
        & i & 12450 & 0.70\\ 
NGC 4365 & g & 5475 & 0.86\\ 
                 & i & 6125 & 0.73\\ 
NGC 5044 & g& 14100 & 1.13\\ 
                 & i& 4800 & 1.06\\ 
NGC 5846 & g & 8550 & 0.96\\ 
                & i & 1650 & 0.90\\ 
\hline
\end{tabular}\label{tab:tab_data}
\tablefoot{ \tablefoottext{a}{Median value of the FWHM.}}
\end{table}
By assuming the above values for the total integration time, the ETC
  for OmegaCam (version 6.2.0) ensures that we should reach a limiting
  magnitude in the B band of $\mu$ = 29 mag/arcsec$^{2}$ ($\mu_{g}
  \sim$ 28.6 mag/arcsec$^{2}$ in the g band) with a S/N $\sim$ 3 per arcsec$^{2}$. The fainter magnitude levels of the azimuthally averaged surface brightness profiles are due to the increasing collecting area with the galactocentric radius.

All the data were processed with the VST-Tube pipeline
\citep{Grado12}.
Details about the various steps of the data reduction are described in
Appendix A of \citet{Cap15}. The sky background
subtraction procedure for data taken with the standard strategy is
illustrated in \citet{Cap15}. 

For images taken with the {\it step-dither} technique, the only difference
in our analysis concerns the estimation of the background. In these
cases, an average sky image, relative to each night, was derived from
the exposures of the offset fields, in which all bright sources
were masked. The VST mosaics of galaxies in this work, showing
  the masking of bright sources in the field, are in Appendix \ref{mask}. This mean sky frame was then scaled
to take into account transparency variations, and subtracted from the
science frame (see \citealt{Iodice16} for details). {  In this
  paper, all magnitudes are quoted in the AB system.}
  
  \begin{figure*}
\centering
\hspace{-0.cm}
 \includegraphics[width=14cm]{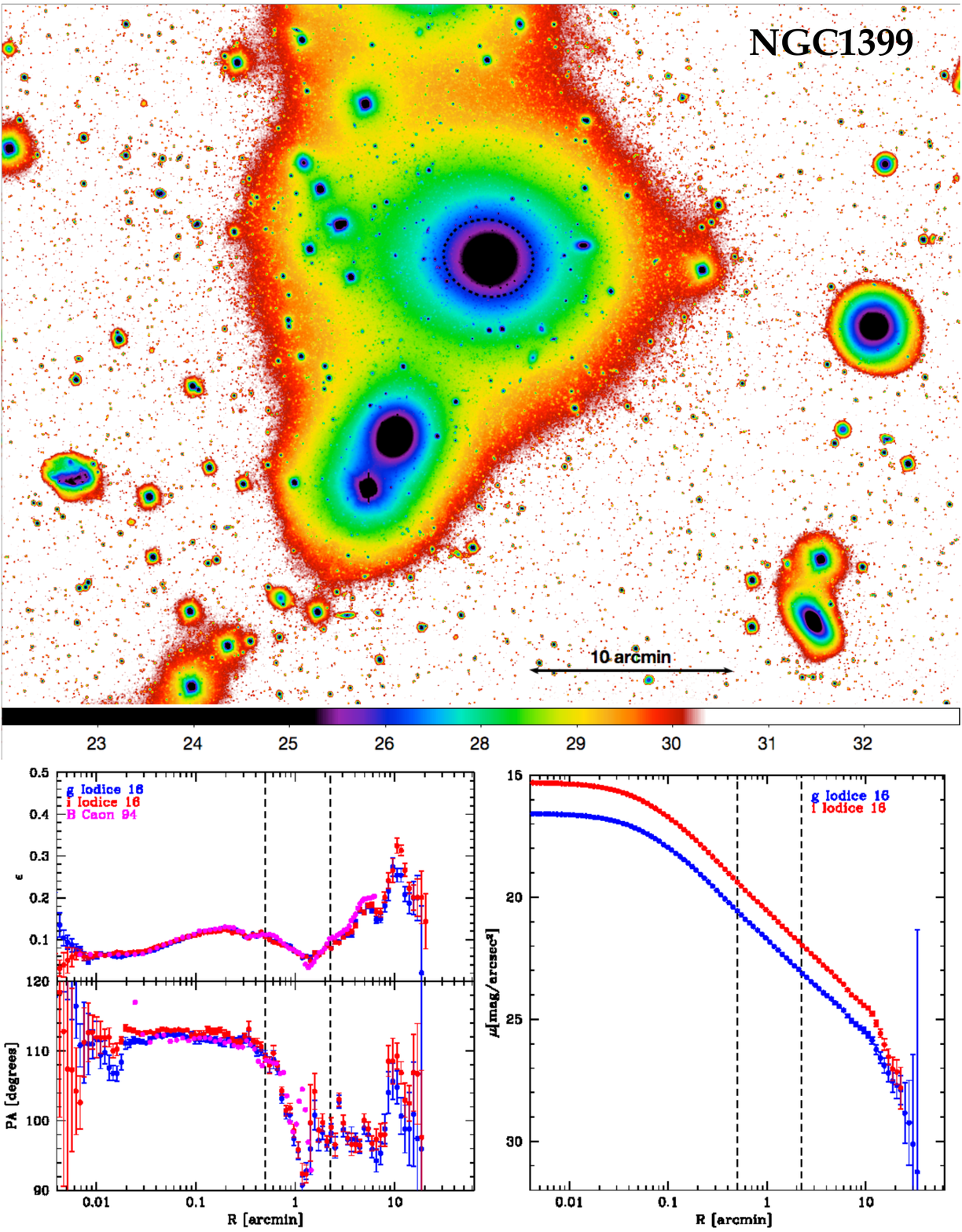}
\caption{{\it Top}: VST {\it g} band sky-subtracted image of the field around NGC 1399. The color scale represents
surface brightness in mag/arcsec$^{2}$. The dashed black contour marks the
transition radius in the surface brightness profile, defined in Sec. \ref{sb}.
North is up, East is to the left. {\it Bottom left}: ellipticity
($\epsilon$) and position angle (P.A.) profiles for NGC 1399, in the {\it g}
band (blue points) and {\it i} band (red points). {\it Bottom right}: Azimuthally averaged
surface brightness profiles in the {\it g} (blue) and {\it i} (red)
bands. {  The dashed lines mark the
transition radii in the surface brightness profile, listed in Tab. \ref{tabfit3comp}, which represents
the location of the transition between two fit conponents, as described in Sec. \ref{sb}.}}\label{mag1399}
\end{figure*}

  \begin{figure*}
\centering
\hspace{-0.cm}
 \includegraphics[width=15cm]{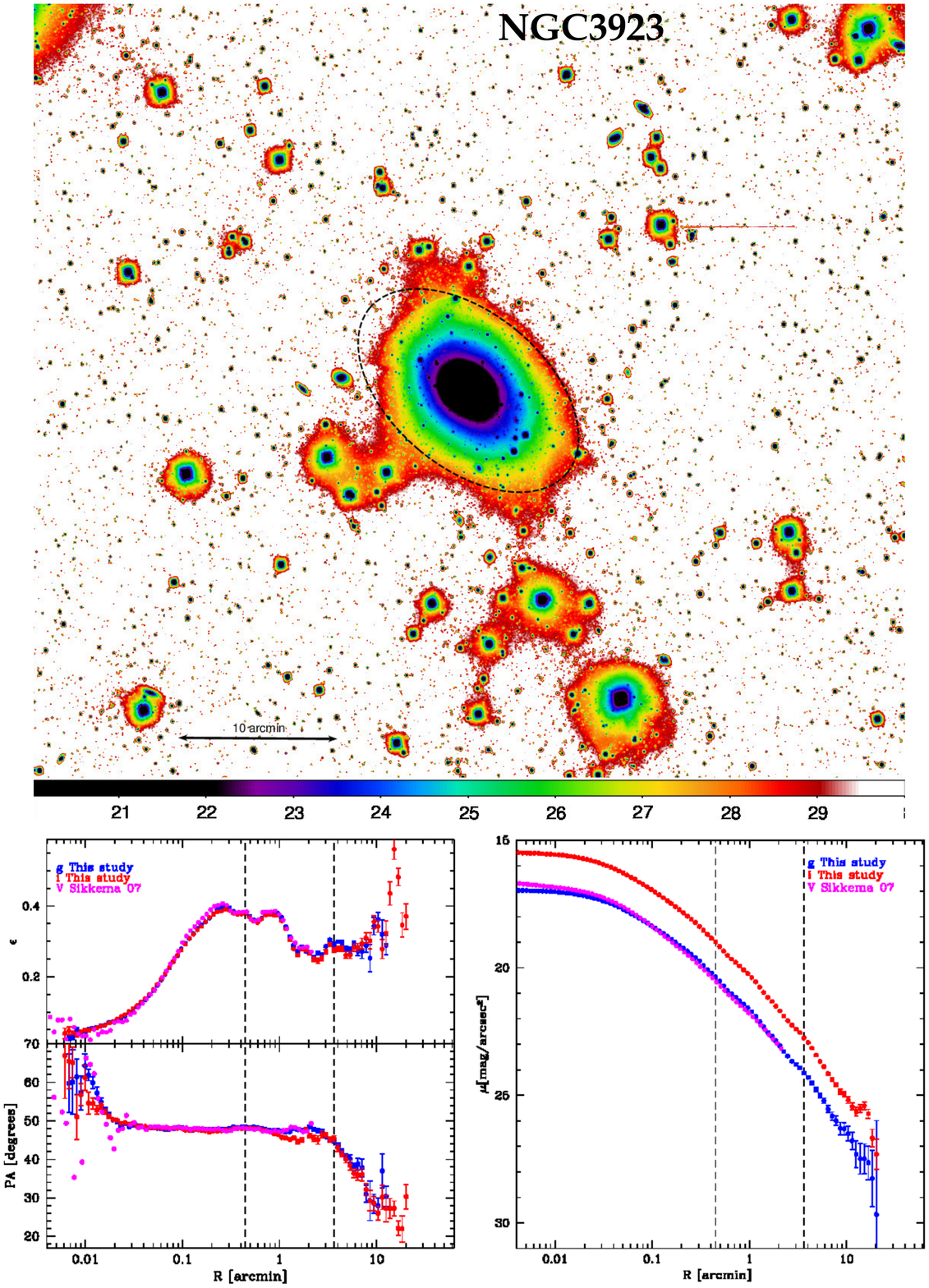}
\caption{The same as Fig. \ref{mag1399} for NGC 3923. The V band measurements of \citet{Sikkema07}
(magenta) shifted by +0.7 mag to match the amplitude of the VEGAS
profile.}\label{mag3923}

\end{figure*}

\begin{figure*}
\centering
\hspace{-0.cm}
 \includegraphics[width=15cm]{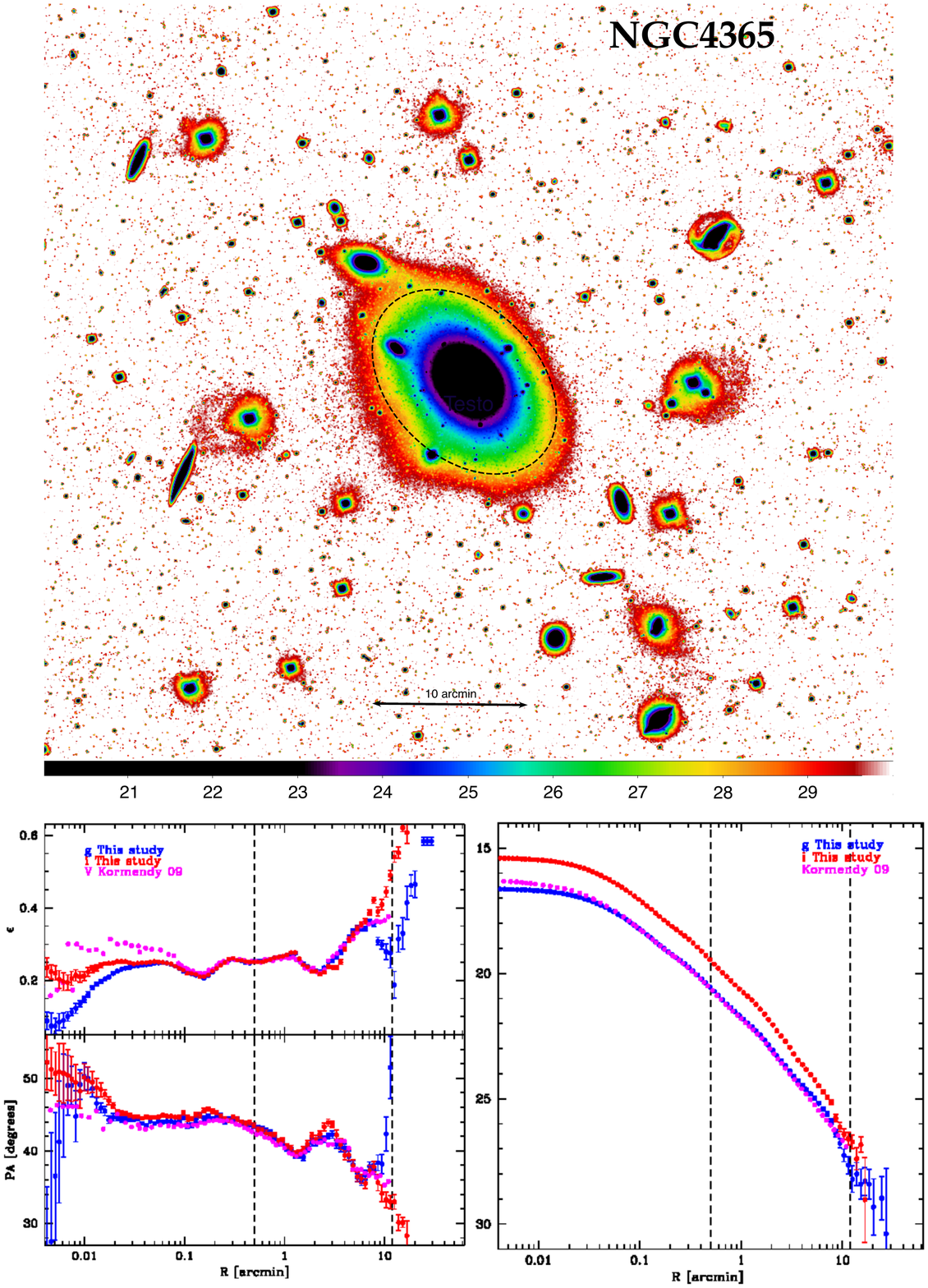}
\caption{The same as Fig. \ref{mag3923} for NGC 4365. The V band
  surface brightness profile by \citet{Kormendy09} (magenta points in
  the bottom panel) has been shifted by +0.2 mag to match the amplitude of
  the VEGAS profile.}\label{mag4365}
\end{figure*}

\begin{figure*}
\centering
\hspace{-0.cm}
 \includegraphics[width=15cm]{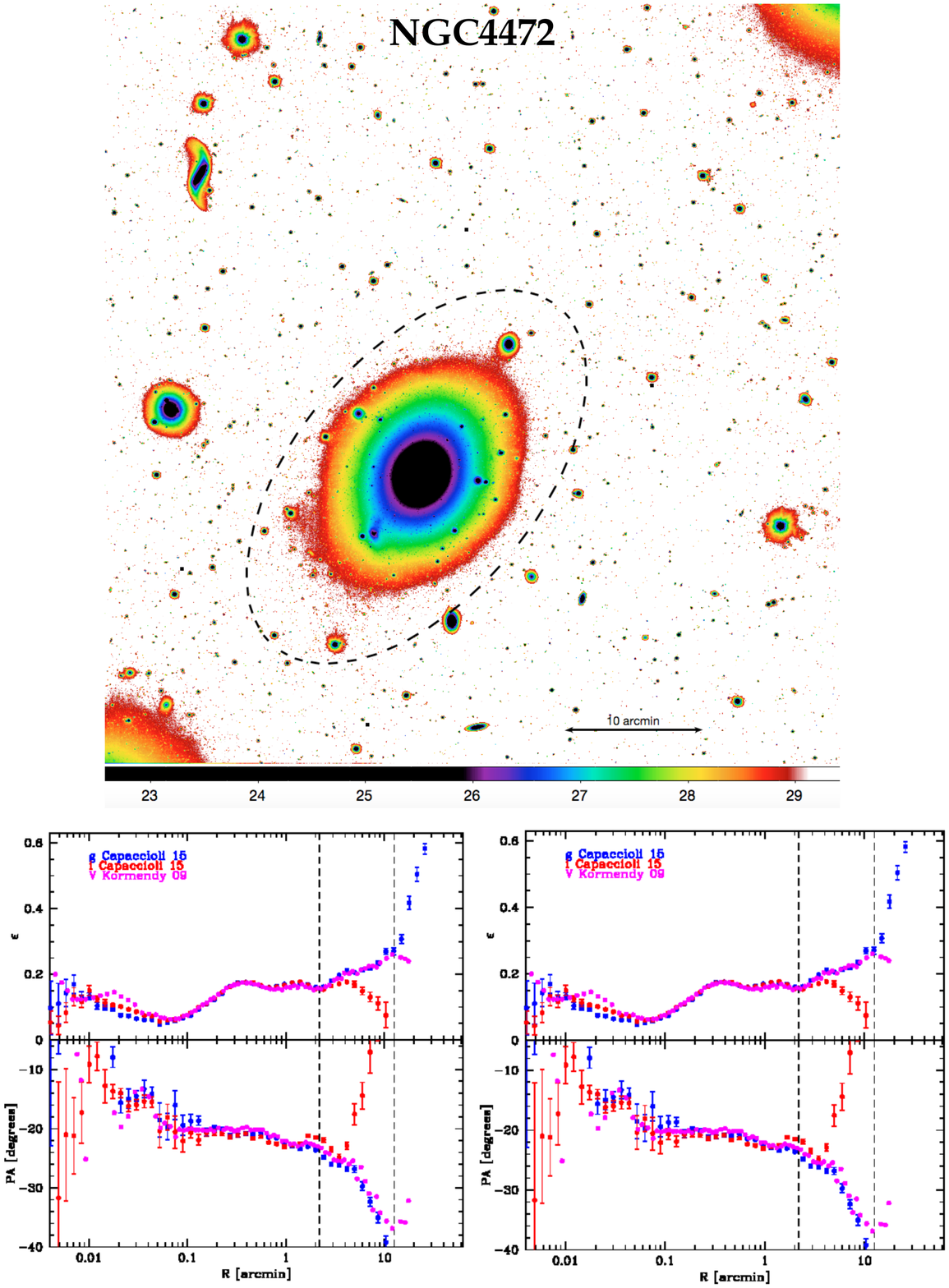}
\caption{The same as Fig. \ref{mag3923} for NGC 4472. The V band
  surface brightness profile by \citet{Kormendy09} (magenta points in
  the bottom panel) has been shifted by +0.33 mag to match the amplitude of
  the VEGAS profile.}\label{mag4472}
\end{figure*}

\begin{figure*}
\centering
\hspace{-0.cm}
 \includegraphics[width=15cm]{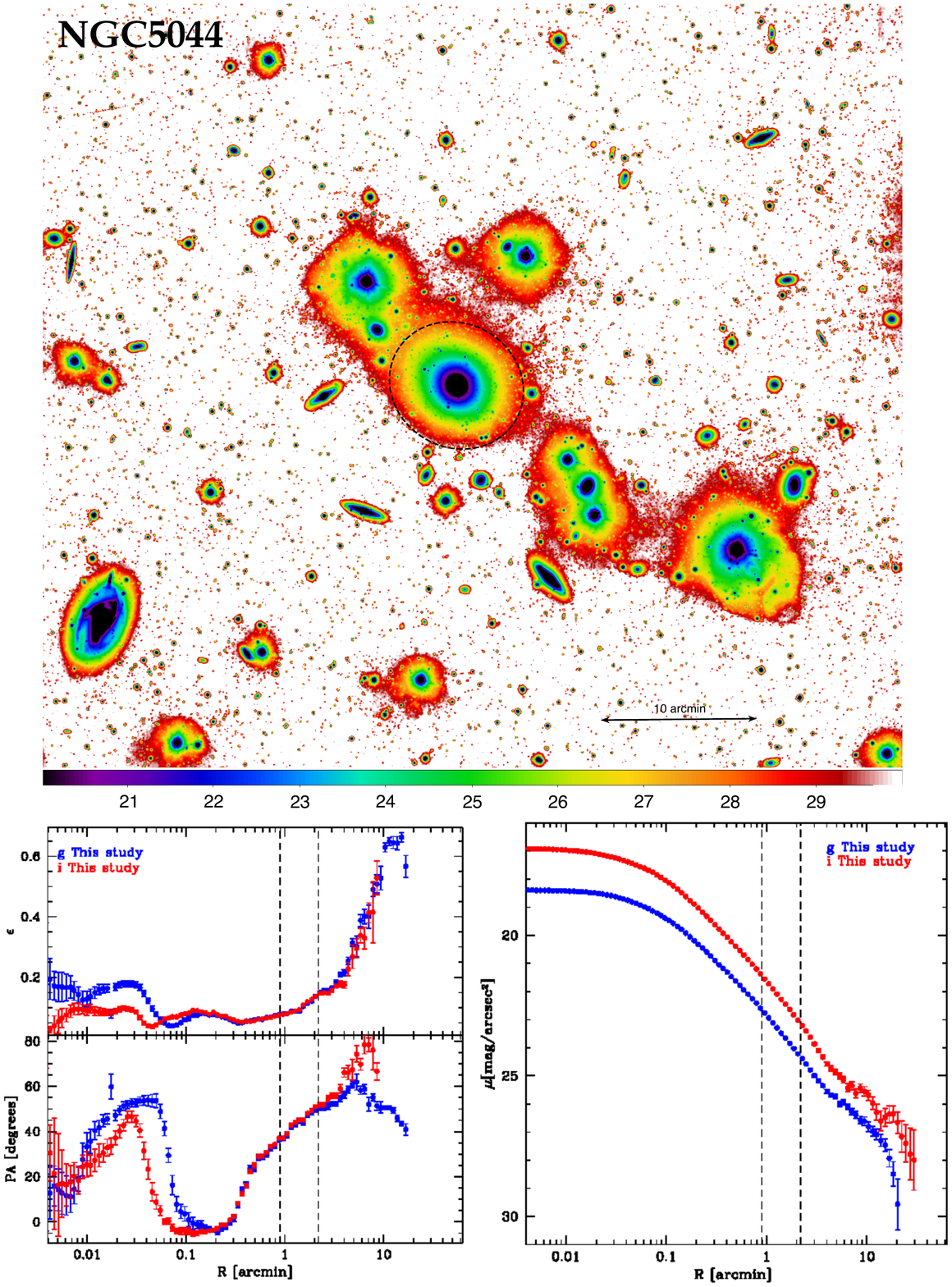}
\caption{The same as Fig. \ref{mag3923} for NGC 5044.}\label{mag5044}
\end{figure*}

\begin{figure*}
\centering
\hspace{-0.cm}
 \includegraphics[width=15cm]{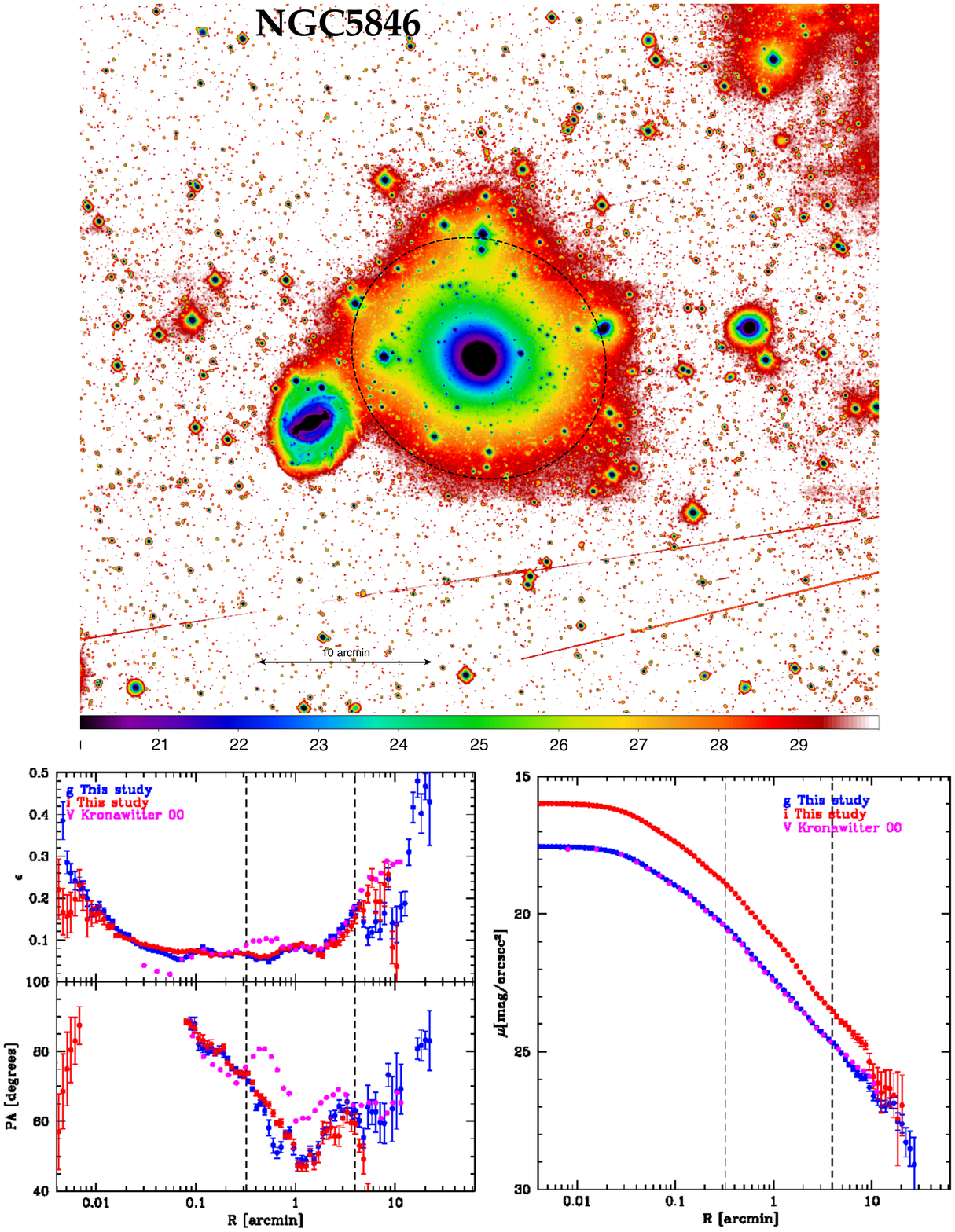}
\caption{The same as Fig. \ref{mag3923} for NGC 5846.  The V band
  surface brightness profile by \citet{Kronawitter00} (magenta points in
  the bottom panel) has been shifted by +0.23 mag to match with
  the VEGAS result.}\label{mag5846}
\end{figure*}

\section{Data analysis}\label{phot}

The most critical operation in deep photometric analysis is the
estimation and the subtraction of the sky background. In Paper I we discuss
pros and cons of a background subtraction made with the direct polynomial
interpolation and by using a  {\it step-dither}  technique. After the
background subtraction, described in Sec. \ref{vegas}, we applied the
methodology described by \citet{Pohlen06} to quantify  the sky variations and
subtract any possible residuals. Following this procedure, on the
sky-subtracted mosaics in both the {\it g} and {\it i} bands, we
masked all the bright sources and used the IRAF\footnote{IRAF (Image
Reduction and Analysis Facility) is distributed by the National Optical
Astronomy Observatories, which is operated by the Associated Universities for
Research in Astronomy, Inc.  under cooperative agreement with the National
Science Foundation.} task {\small ELLIPSE} to extract azimuthally
averaged intensity profiles in elliptical annuli out to the edges of the
frame by fixing both the ellipticity ($\epsilon$) and the position angle (P.A.)
to the values best describing the isophotes at the
semi-major axis $R\sim 1$ arcminute. By
extrapolating the trend in the outer regions of these intensity profiles
for each galaxy in our sample, we found that the error on the average value of
the residual background is $0.1\leq\sigma_{g}\leq 0.2$ counts in the {\it g}
band, and $0.2\leq\sigma_{i}\leq 0.5$ counts in the {\it i} band. These
fluctuations of the sky background have been taken into account in the error
estimates we quote on our surface brightness measurements. VST {\it
g} band images of NGC 1399, NGC 3923, NGC 4365, NGC 4472, NGC 5044 and NGC 5846, colored by
surface brightness, are shown in the top panels of figures
\ref{mag1399}, \ref{mag3923},
\ref{mag4365}, \ref{mag4472}, \ref{mag5044}, and \ref{mag5846} respectively. 

In the bottom left panels of the same figures we show the ellipticity
($\epsilon$) and position angle (P.A.) profiles resulting from our isophotal
analysis, performed by using the {\small ELLIPSE} task, leaving
ellipticity and position angle free and sampling the semi-major axis
with a variable bin (as described in Paper I). {  ELLIPSE computes
  the intensity, $I(a,\theta)$, azimuthally sampled along an
  elliptical path described by an initial guess for the isophote
  center, (X,Y), ellipticity, $\epsilon$, and semi-major axis position
  angle, $\theta$, at different semi-major axis lengths, $a$. At a given $a_0$, $I(a_0, \theta)$ is expanded into a Fourier series as

\begin{equation}
I(a_{0}, \theta)=I_0+\sum _k (a_k \sin(k\theta)+b_k \cos (k\theta))
\end{equation}
according to \citet{iraf}. The best-fit parameters are those minimizing the 
residuals between the actual and the model isophotes; $a_k$ and $b_k$
are the coefficients measuring the deviations from a pure ellipse,
including the signature of boxiness and/or diskiness
\citep{Bender89}. At each iteration of the isophote fitting, an image
sampling is performed. Three methods are available for sampling an
image along the elliptical path: bi-linear interpolation, and either
mean or median over elliptical annulus sectors. Integration over
elliptical annulii is preferred, in order to extract all the
information from an image and to avoid that some pixels in the image
will never be sampled. For this reason, the SB profiles in this work have
been extracted by performing a median over elliptical annuli, and by
applying a k-sigma clipping algorithm for cleaning deviant sample
points at each annulus. The combined use of the median and the
  sigma clipping algorithm improves the isophotal fitting, ensuring
  the rejection of not entirely masked features such as stars,
  defects, and outskirts of neighbouring galaxies.}

For all the galaxies we find an
increase of the ellipticity in the outer regions, indicating the presence of an
more flattened outer component, with a significant gradient in the PA. We
shall return to this point in Sec. \ref{sb}.

In the bottom right panels we show {\it g} and {\it i} band VST surface
brightness profiles, including comparisons with other measurements
in the literature. No correction for seeing blurring is applied to the inner
regions of the profiles. The offsets relative to the VST {\it g} band
providing the best match to our photometry are +0.7 mag for the V band profile
of NGC 3923 by \citet{Sikkema07}, +0.33 mag for the V band profile of NGC 4472
by \citet{Kormendy09}, +0.2 mag for the V band profile of NGC 4365
by \citet{Kormendy09}, and +0.23 mag for the V band profile of NGC 5846 by
\citet{Kronawitter00}. Outside the seeing-blurred cores, despite the different
color bands, the agreement with the literature is good. The reliability of the surface brightness measurements in
  the faint outskirts of the galaxies, has been tested by extracting the surface
  brightness profiles in azimuthal sectors (see App. \ref{sect} for details). The errors associated
with our surface brightness measurements were computed as in Paper I,
and take into account both the uncertainties on the photometric
calibration and on sky subtraction.  All the profiles are very extended and
deep ($29 \leq\ \mu_{g}\ \leq 30.5$ mag/arcsec$^{2}$), and the azimuthally
averaged surface brightness of the four objects spans almost one dex.

As in Paper I, we use the luminosity profiles of the galaxies, together
with their ellipticity profiles, to compute the growth curves, which we
extrapolate to the derive total magnitudes $m_{T}$, effective radii
$R_{e}$ and corresponding effective magnitude $\mu_{e}$ (Tab.
\ref{par}). 

\begin{figure*}
\centering
\hspace{-0.cm}
 \includegraphics[width=9cm, angle=-0]{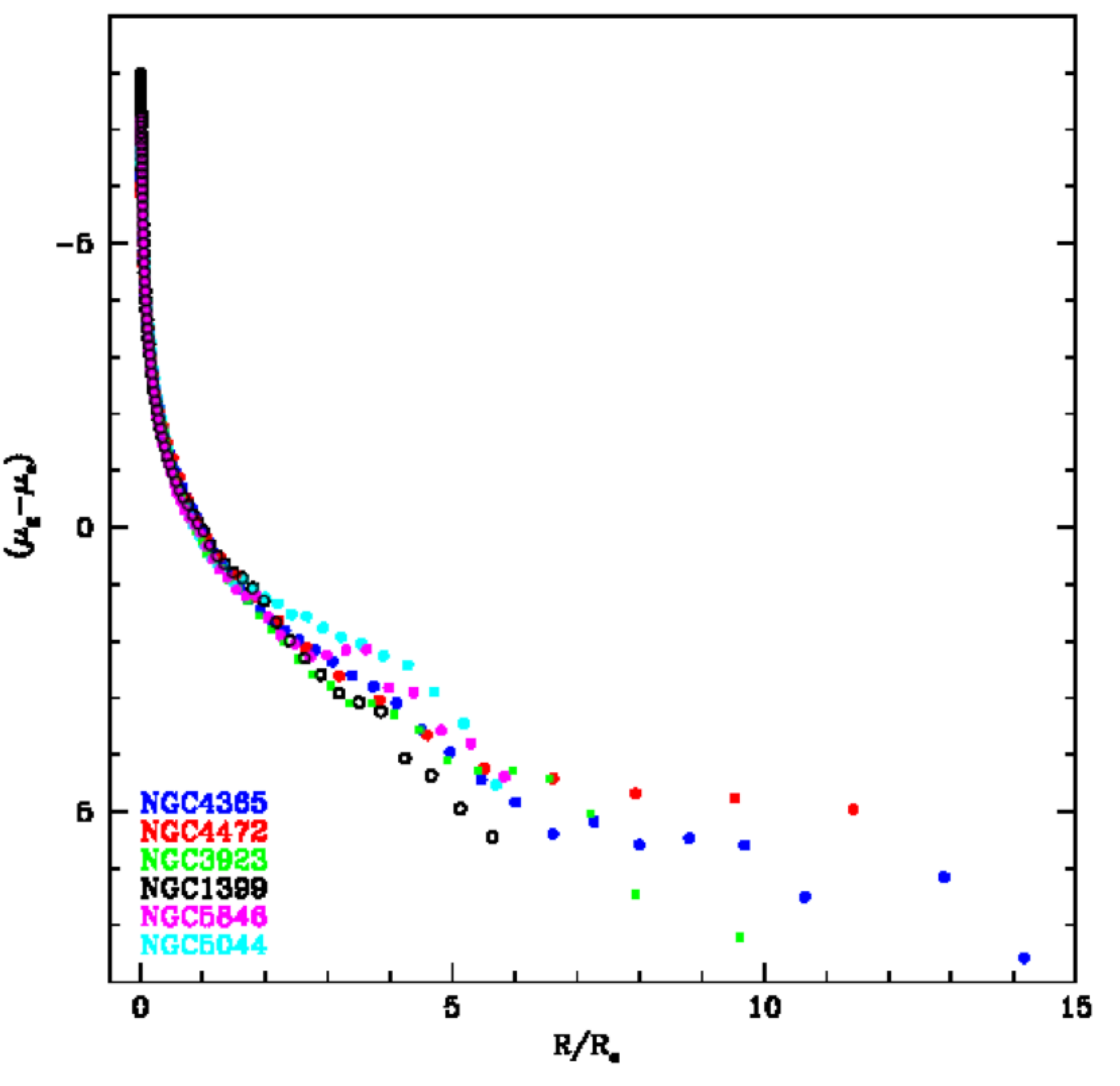}
 \includegraphics[width=9cm, angle=-0]{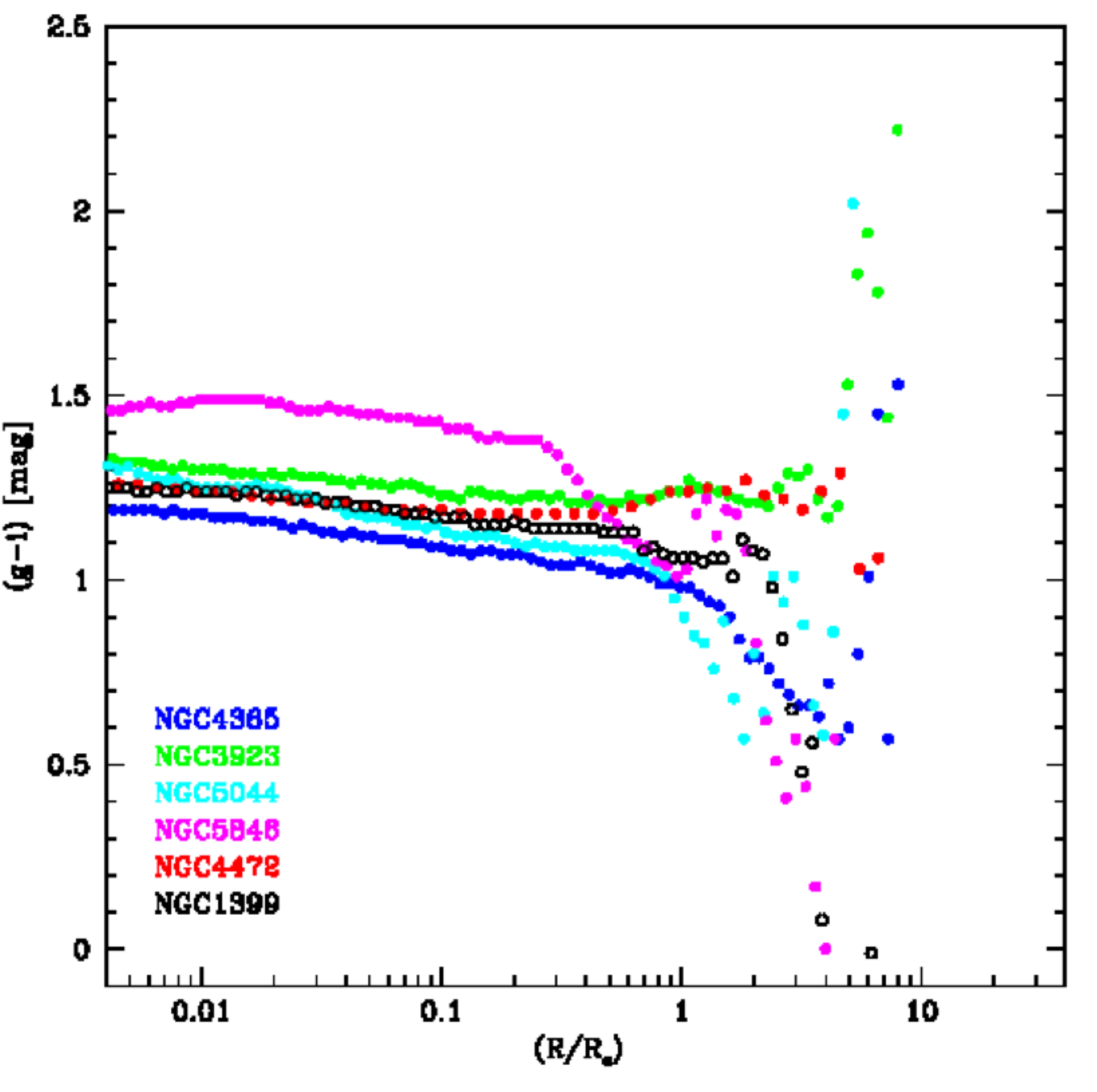}
\caption{Left: Azimuthally averaged surface brightness profiles in the {\it
g}-band {  scaled to their effective magnitude}, as a function of galactocentric radius normalized to the
effective radius, $R/R_{e}$. The innermost region is not corrected for seeing
convolution (typical FWHM=0.8 arcsec). Right: ({\it g-i}) color profiles of the
galaxies in our study.}\label{halo}
 \end{figure*}

In Fig.
\ref{halo}, we plot all the profiles as a function of
galactocentric radius normalized to the effective radius $R/R_{e}$.  This
comparison shows that, in our sample, we have two kinds of
profile: those having an `excess' of light in the outer regions,
which tend to flatten in the outer regions, and those for which the
surface brightness falls off more rapidly. The
differences in the shape of the light profiles may be signatures of different
dynamical states for the stellar halos surrounding these galaxies; we
will discuss this point in Sec.  \ref{disc}.

\begin{table*}
\setlength{\tabcolsep}{2.1pt}
\small
\begin{center}
\caption{Distances and photometric parameters for the
  sample galaxies.} \label{par}
\vspace{10pt}
\begin{tabular}{lcccccccccccccccc}
\hline\hline
Object & $D$ &$A_{g}$&$A_{i}$& $\mu_{e,g}$ &
$r_{e,g}$& $(R_{e,g})_{maj}$&$m_{T,g}$& $M_{T,g}$& $\mu_{e,i}$ &
$r_{e,i}$& $(R_{e,i})_{maj}$&$m_{T,i}$& $M_{T,i}$&$\langle\epsilon_{in}\rangle$&$\langle\epsilon_{out}\rangle$\\
    &[Mpc] &[mag]&[mag] & [mag/arcsec$^{2}$] & [arcmin] & [kpc]&[mag] &[mag] & [mag/arcsec$^{2}$] & [arcmin] & [kpc]&[mag] &[mag]&\\
&(a)&(b)&(b)& (c)&&&(c)&(c)&(c) &&&(c)&(c)&(d)&(d)\\
\hline \vspace{-7pt}\\
NGC 1399    &    20.9    &   0.05   & 0.04&   24.55    &   5.87   &    35.65&       8.55  &    -23.05   &    -   &5.05&  30.70  &7.51  &     -24.09&0.095 &0.203\\
 NGC 3923    &    22.9   &   0.31  &   0.17&   23.22  &     2.55  &     16.97   &        9.14 &     -22.66   &    22.30   &   3.15  &    20.98  &     7.78 &    -24.02&0.356 &0.292\\
 NGC 4365     &   23.1  &    0.08  &   0.04&   22.82  &     1.90   &    12.79  &        9.41  &    -22.41   &    21.15   &    1.40   &   9.41  &    8.52  &    -23.30&0.238 &0.282\\
 NGC 4472    &    16.9   &   0.08  &    0.04&  22.68  &     2.70  &     13.29    &       8.37   &   -22.77   &    22.23  &     4.22   &    20.73    &  6.92  &  -24.22&0.137 &0.205\\
NGC 5044   &     31.2  &    0.26   &   0.14&  25.04   &    3.55   &    32.20     &      9.88   &   -22.59    &   24.20  &    3.65    &  33.13   &   8.76   & -23.71&0.068 &0.340\\
 NGC 5846   &     24.9  &    0.21   & 0.11&   24.72  &     4.62  &     33.41    &       9.21   &   -22.77   &    23.32   &   3.70   &   26.80   &   7.99   &  -24.00& 0.067&0.167\\
\hline
\end{tabular}
\tablefoot{\tablefoottext{a}{Distances of NGC 3923, NGC 5044 and NGC 5846 are
  from \citet{Tonry01}; of NGC 4365 and NGC 1399 are from
  \citet{Blak09}; of NGC 4472 is from \citet{Mei07}.}
\tablefoottext{b}{Sources for the extinction correction in the {\it g}
  and {\it i}-band are {  \citet{Schlegel98}} for all objects but
  NGC 4472 for which we adopt the value used in 
Paper I.}
\tablefoottext{c}{Derived by integrating the growth curves and corrected for interstellar extinction.}
\tablefoottext{d}{{\it g} band mean ellipticity in the ranges
  $-1.5<log(R/R_{e})_{maj}<-0.5$ ($\epsilon_{in}$) and $0<log(R/R_{e})_{maj}<0.5$ ($\epsilon_{out}$).}}
\end{center}
\end{table*}

 The mean {\it (g-i)} color profiles for galaxies in our sample
  are shown in right panel of Fig. \ref{halo}. The {\it (g-i)} color
  profile for NGC 1399 is from \citet{Iodice16}, while that for NGC
  4472 is from \citet{Cap15}.
The colors
of all galaxies are redder than those those typically found for ETGs.
As shown by \citet{Ann2010}, galaxies that are beyond the red edge of the red
sequence can have different origins; many of them have dust lanes or
are merging galaxies, and the dust obscuration is the main responsible
for their red colors. Consistently with this picture, all the galaxies in our sample show
the presence of dust lanes, shells and other signs of ongoing interactions.
All the color profiles display in the central regions ($0.01 < R/R_{e} <
0.1$), where on average the
galaxies are redder, small but significant negative gradients typical
of early types \citep{LaBarbera2010, Tortora10}.
At a different radius for each galaxy, the colors
develop a steeper gradient, becoming bluer (for NGC 1399, NGC 4365, NGC 5044 and
NGC 5846) or redder (NGC3923 and NGC 4472). {  In Sec. \ref{sb} we
  will show that the discontinuities observed in the color profiles correspond to the transition radii observed in each
surface brightness profile, which are defined as the locations of the
transition between two different fit components.}

\subsection{Correction for the PSF}\label{psf}

It is well known \citep{Capaccioli83} that the faint outskirts of galaxies
can be contaminated by light scattered from the bright core by
the telescope and by the atmosphere. In order to account for the artificial
halos created by scattered light, the point spread function (PSF) must be
mapped out to a radial distance at least comparable to the extent of the
galaxian halo.  An extended PSF for VST has been derived in Appendix B of Paper
I.  There we also show how to estimate the amount of scattered light by
deconvolving a noiseless model of each galaxy with the IRAF task {\small
LUCY}, which is based on the Lucy-Richardson algorithm \citep{Lucy74,
Richardson72}.

As discussed in Paper I, the effect of the scattered light is quite different
at the same surface brightness level between galaxies with large and small
angular extent. We found that
the surface brightness profiles of smaller sources in the field of NGC
4472 were marginally affected ($\sim$ 0.2 mag) by the extended PSF at
$\mu_{g} \sim\ 29$ mag/arcsec$^{2}$. Galaxies with a large angular
extent, however, were not significantly affected by the PSF. 

We also performed a plain numerical convolution for a set of
azimuthally averaged light profiles for artificial galaxy models with different
sizes. The difference between the model and its convolution provides an
estimate of the effect of the extended PSF, at different surface brightness
levels, on galaxies with different sizes. Details and results of these
numerical experiments will be presented in a forthcoming paper (Spavone et al.
in preparation).  We conclude that no significant contribution of
scattered light to the  azimuthally averaged light profile is present in
the faintest, outermost regions of the six giant galaxies we study in this
paper.

\section{Fitting ellipticity and light distribution}\label{sb}

\subsection{Deprojected flattening}\label{ell}

It is not easy to make a meaningful comparison between the apparent ellipticity
curves of different ETGs when their intrinsic shapes and inclinations,
$\theta$, are unknown. As a first-order approximation to the average trend
in our sample we have deprojected the observed flattening profiles by
adopting the classical prescription \citep{Binney98}:

\begin{equation}
\xi^{2} sin^{2} \theta + cos^{2} \theta = \bigg \{ \begin{array} {rl} 
q^{2}  & \mbox{(oblate)} \\ 
1/q^{2} & \mbox{(prolate)}\end{array}
\end{equation} where $q$ and $\xi$ are the apparent and
true axial ratios respectively. We have assumed that the dominant spheroidal components of
all our six galaxies are oblate and have $\langle \xi \rangle= 0.62$
\citep{Binney81}.

\begin{figure}
\centering
 \includegraphics[width=9cm, angle=-0]{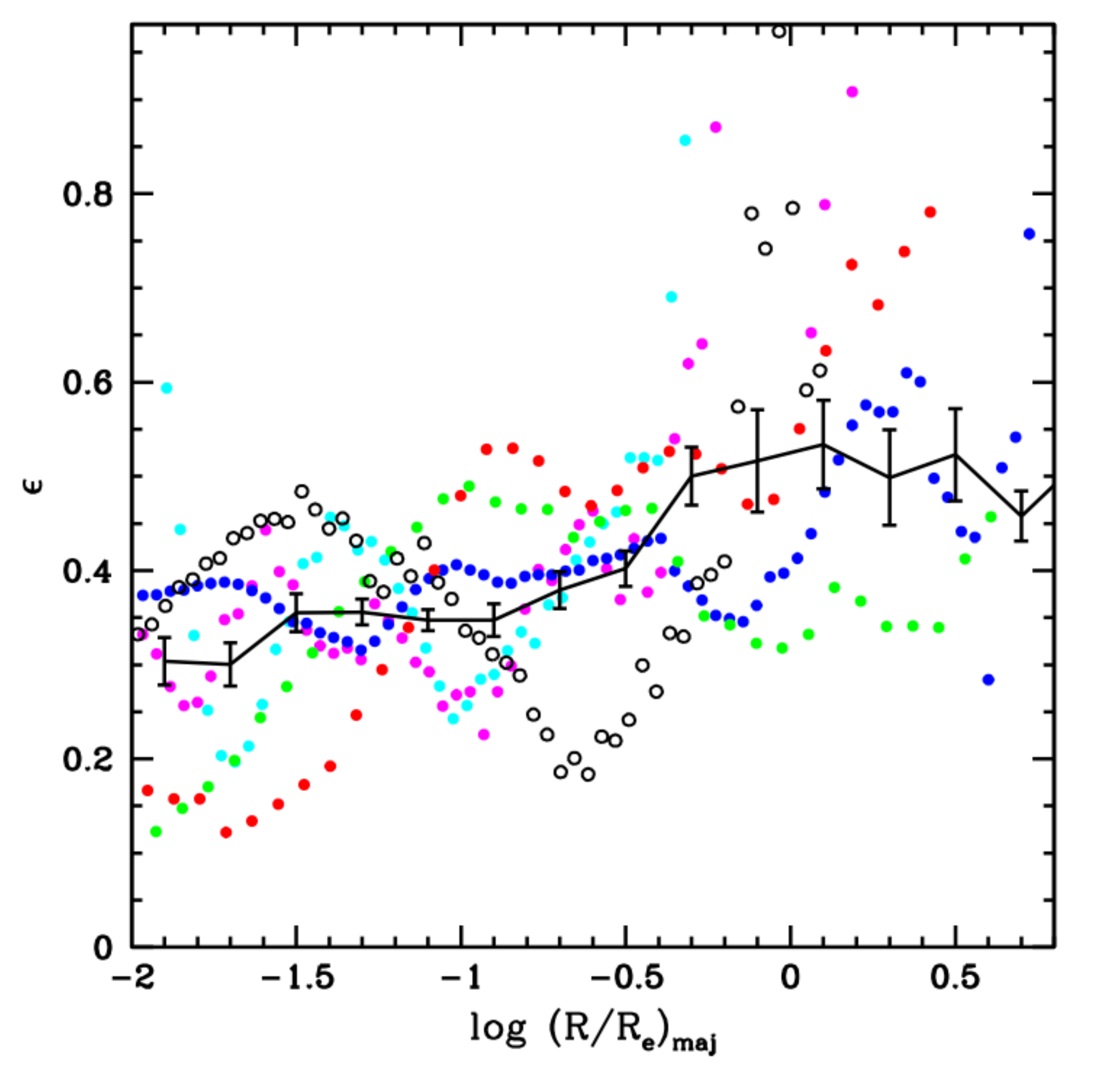}
 \caption{Deprojected ellipticity ($\epsilon$) profiles in the {\it g}-band, as
 a function of $log\ (R/R_{e})_{maj}$. The solid black line is the mean trend
 of the whole sample. The errorbars trace the dispersion of the points. The
 color code is the same as that in Fig. \ref{halo}.}\label{dep}
\end{figure}

Results are shown in Figure \ref{dep}, including ellipticity profiles for NGC
1399 and NGC 4472. The solid line represents a running mean of the complete
dataset. Individual profiles show a wide dispersion, but an overall
increase of flattening from 0.38 to 0.5 is apparent over a range $0.01 <
R/R_{\mathrm{e}} < 6$ along the major axis. This is broadly consistent
with previous findings. For example \citet{Gonzalez05} found that a sample of
BCGs to be well described by a two-component model, in which the outer
component was more elongated than the inner component which they associated
with the BCG.

\subsection{Fitting the light distribution with two-components models}\label{2compfit}

The remainder of our analysis focusses on the projected one-dimensional
(ellipsoidally-averaged) surface brightness profiles of our sample
galaxies.
As noted in the introduction, there is considerable evidence in the literature
that the light profiles of many of the most massive ETGs are not well fitted
by a single S{\'e}rsic law, and at least one additional component is
needed \citep{Seigar07,Donzelli11,Arnaboldi12,Iodice16}. This second
component is sometimes (but not universally) interpreted as evidence for a
stellar halo or `intracluster light'. 

{  Theoretical models suggest that massive ETGs accumulate the bulk
  of their stellar mass by accretion. For this reason, the `stellar
  halo' in these galaxies should be identified with the component
  dominating the stellar mass. From an observational point of view, it
is not straightforward to separate the in-situ and the accreted
component in ETGs, since they have similar physical properties and are
well mixed together. The overall profile is made up by different
contributions, and for this reason theory suggests that the surface
brightness profile of ETGs should be described by the superposition of
different components.}

Since we are interested in the study of the stellar distribution in
the outer envelopes of our sample galaxies, we do not use the
$\chi^{2}$ statistics in our fitting procedure. The data points
corresponding to the central regions of the galaxies in fact, with their
small uncertainties have considerable weight in determining the best
fit solution obtained by minimizing the $\chi^{2}$, while the outer
regions with bigger errors have no weight. This approach could produce
erroneous results when the functional form of the model is not known a priori,
and an empirical model has to be assumed to reproduce
the light distribution.
To avoid this problem, we adopt the same approach described by \citet{Seigar07}, and
performed least-square fits, by using a Levenberg–Marquardt algorithm, in which the function to be minimized is the
rms scatter, defined as $\Delta=\sqrt{ \frac{\sum_{i=1}^{m}
    \delta_{i}^{2}}{m}}$, where $m$ is the number of data points
and $\delta_{i}$ is the {\it i}th residual. In all the fit presented
below, the innermost seeing-dominated regions ($\sim 1.5\times FWHM$), marked with
dashed lines, have been excluded.

{  It's worth to notice that, although we average azimuthally in
  elliptical annulli, this does not mean that the components fit to
  the resulting radial profile should be mecessarily expected to be
  elliptical structures, or even azimuthally smooth. This is simply an
empirical approach to quantify the amount of stellar material as a
function of galaxy radius.}

\begin{figure*}
\centering
\hspace{-0.cm}
\includegraphics[width=7.5cm,]{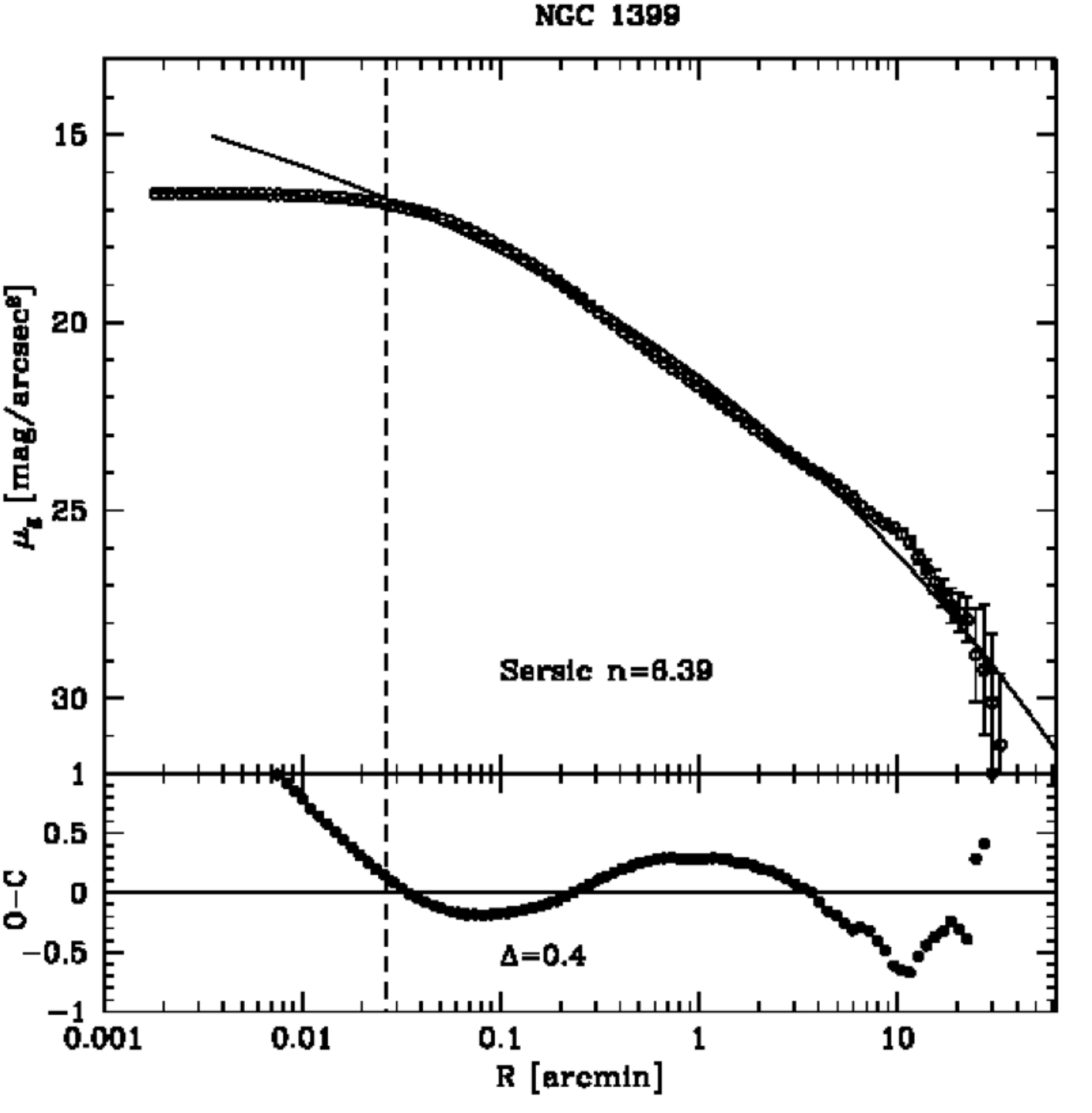}
 \includegraphics[width=7.5cm,]{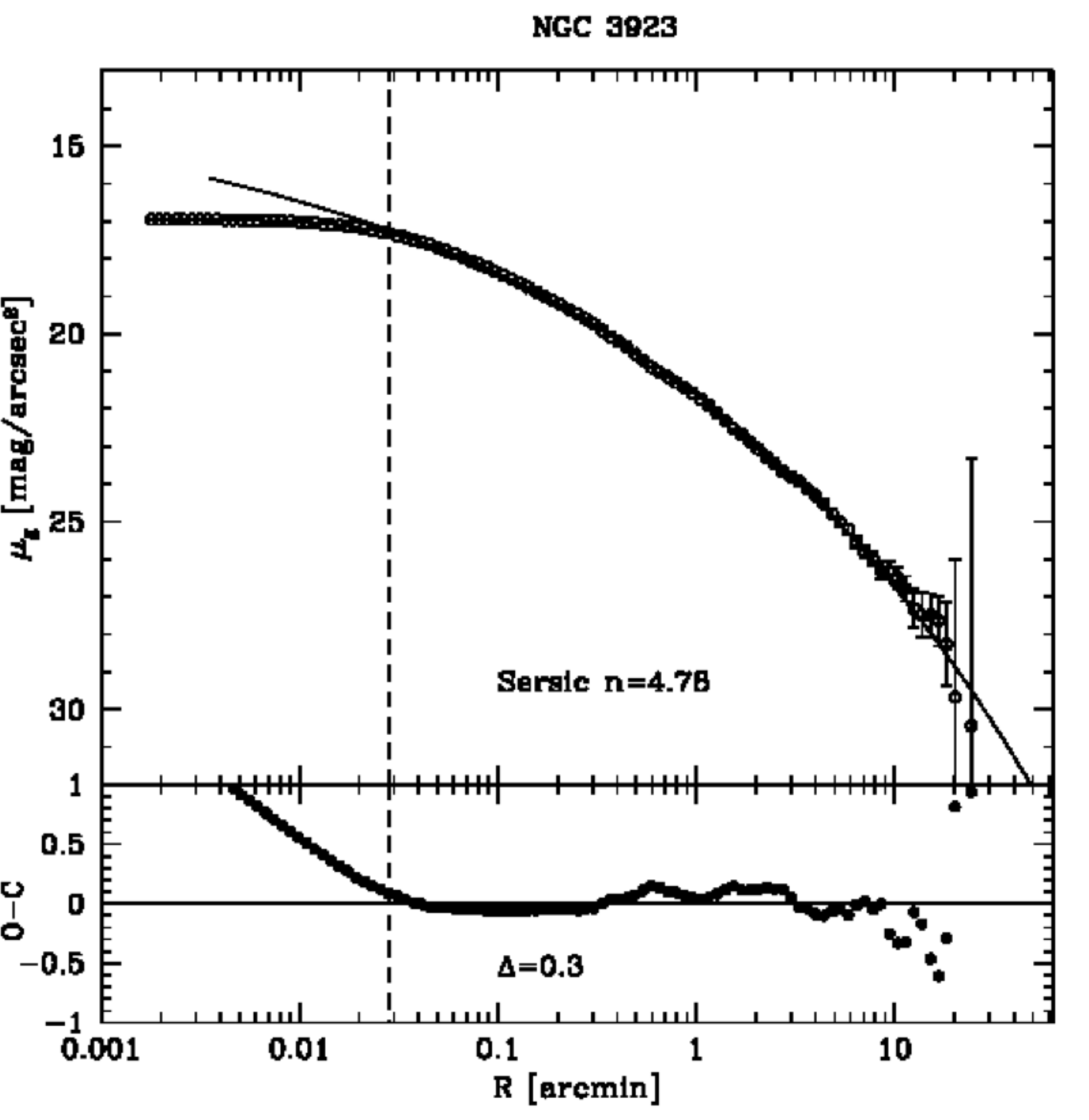}
 \includegraphics[width=7.5cm,]{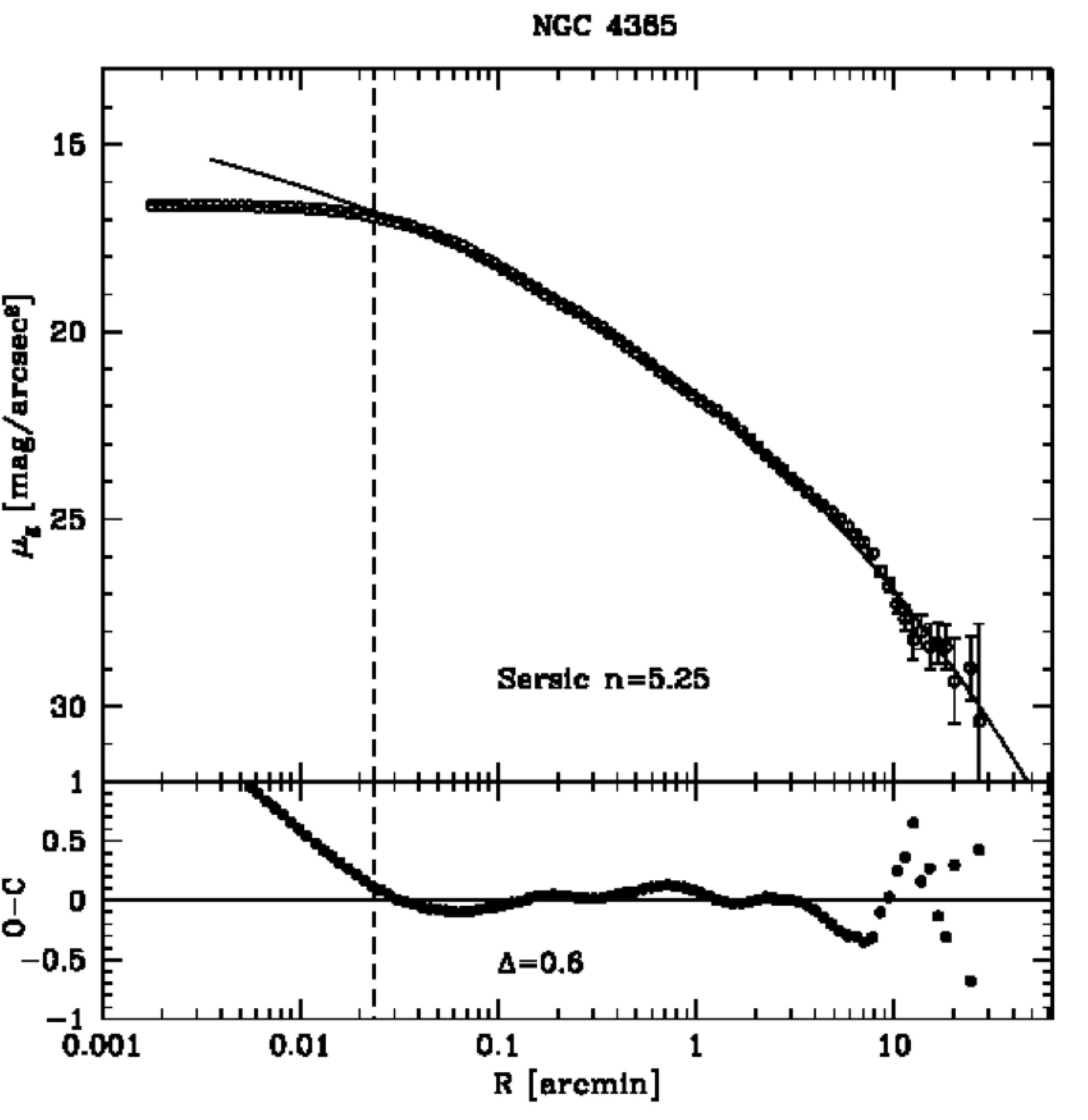}
\includegraphics[width=7.5cm,]{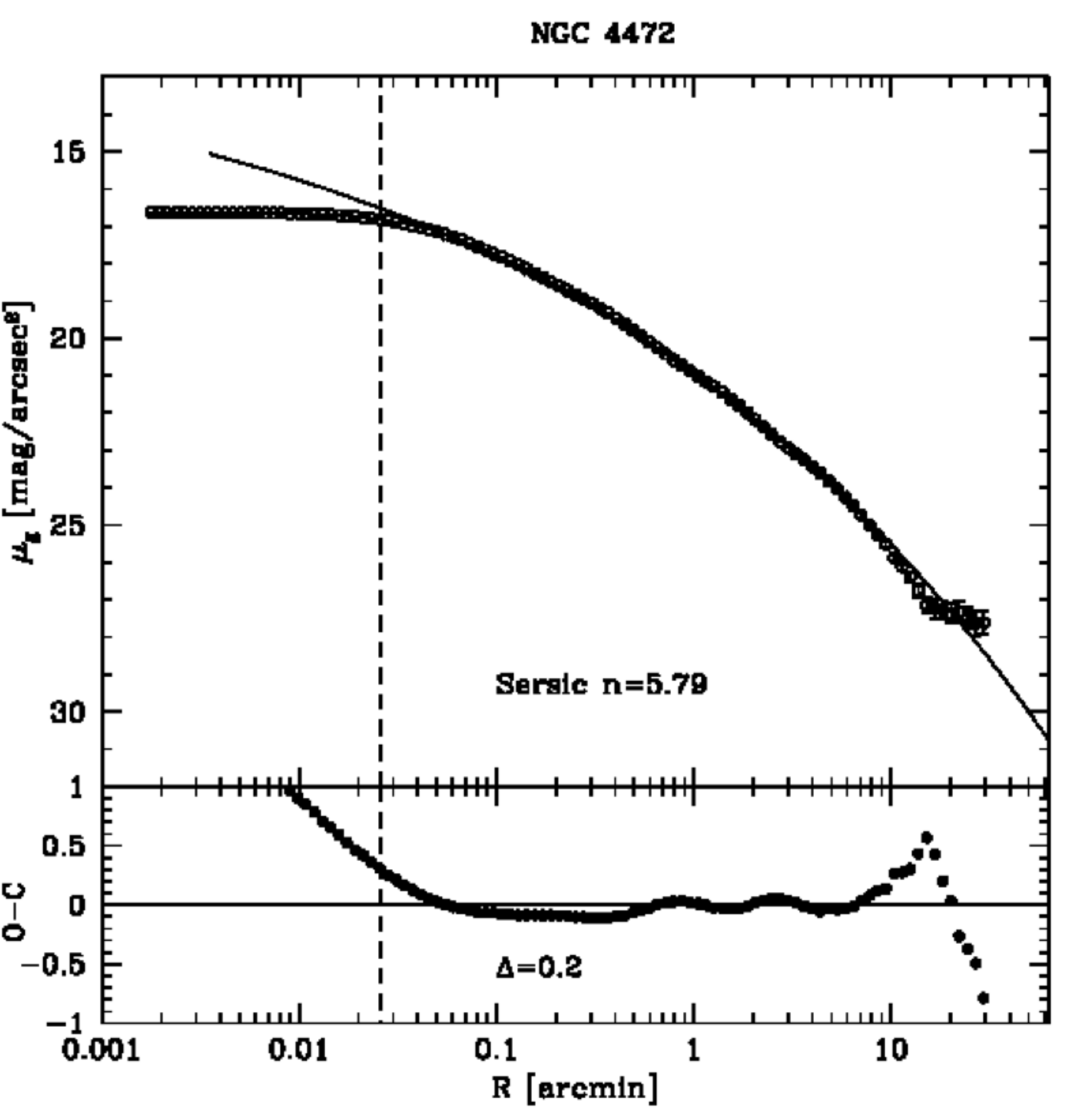}
\includegraphics[width=7.5cm,]{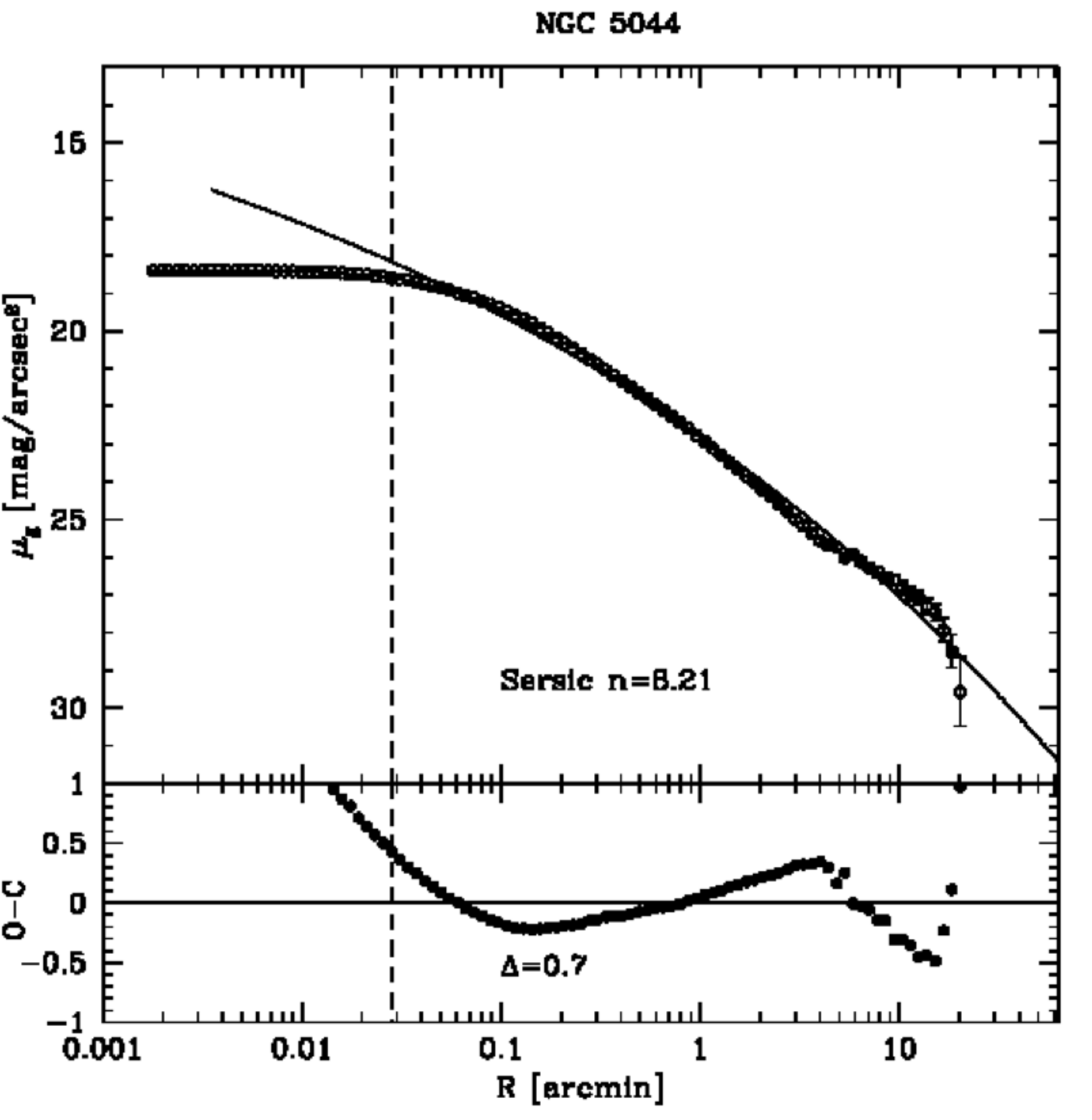}
\includegraphics[width=7.5cm,]{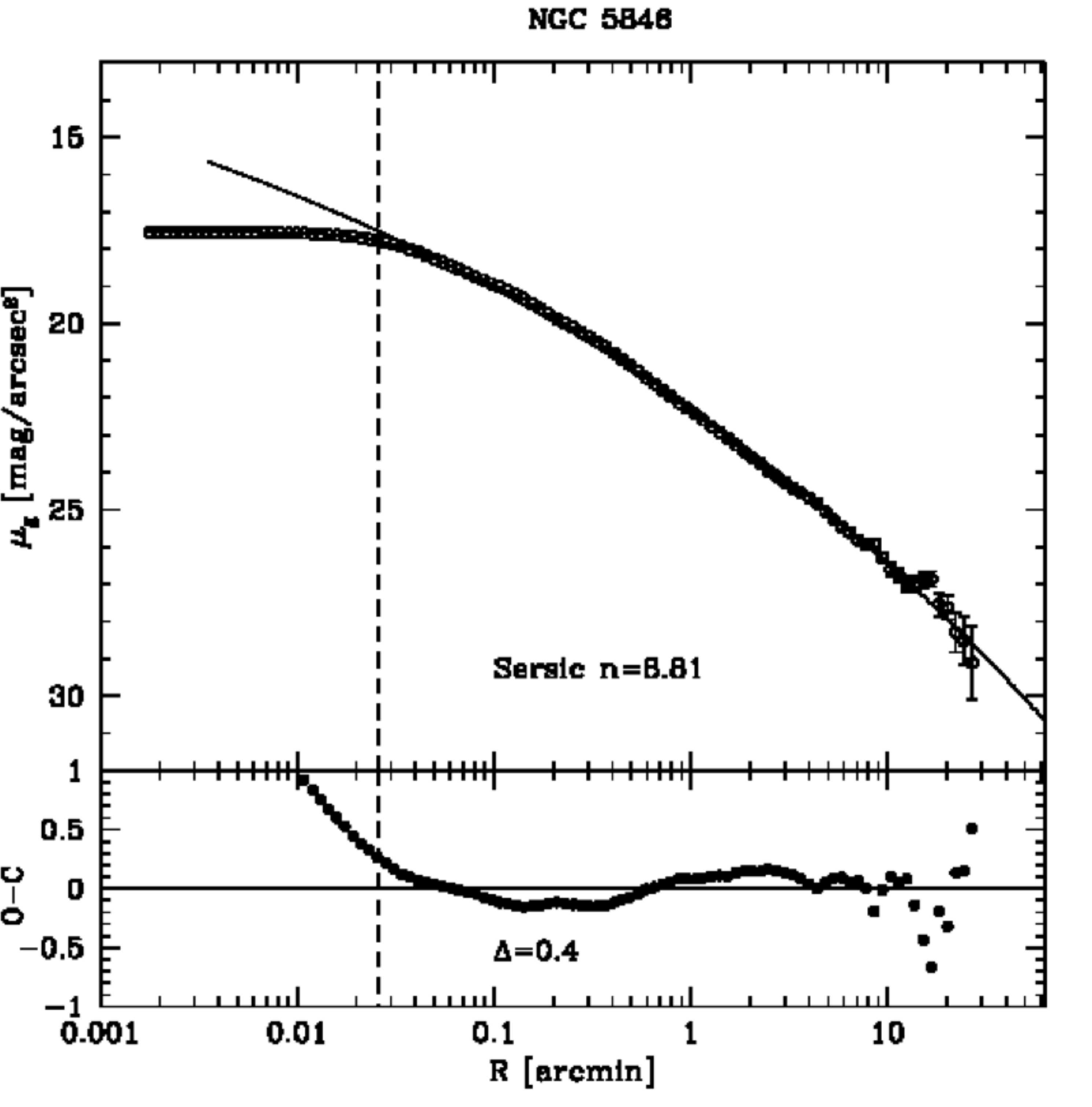}
\caption{VST {\it g} band profiles of NGC 1399, NGC 3923, NGC 4365, NGC 4472,
NGC 5044, and NGC 5846 plotted on a logarithmic scale. The black line is
a fit to the surface brightness profiles with a single S{\'e}rsic law. The dashed line marks the core of the galaxy
($\sim 1.5\times FWHM$), which
has been excluded in the fit.}
\label{fit1c}

\end{figure*}

As shown in Sec. \ref{phot}, we find
two apparent types of surface brightness profile in our sample: those
with a steeper slope in the outer regions and those with a shallower
slope.
In either case, the single S{\'e}rsic $R^{1/n}$ model provides a poor
description of the shape of the profile. This can be clearly seen from the
residuals with respect to a single S{\'e}rsic fit shown in 
Fig.  \ref{fit1c}.

For this reason, we first present models of the surface brightness profiles of galaxies in our
sample with a double S{\'e}rsic law \citep{Sersic63, Caon93},
\begin{equation}
\mu(R) = \mu_{e} + k(n) \left[ \left(\frac{R}{r_{e}}\right)^{1/n} -1 \right]
\end{equation} where $k(n)=2.17n - 0.355$, $R$ is the galactocentric radius and $r_{e}$
and $\mu_{e}$ are the effective radius and surface brightness.
  This asymptotic expansion provides an accurate fit for $n\geq 0.36$
  \citep{Ciotti99, Mac03}.
We found that this model converges to a best-fit solution, with
  physically meaningful value, for only two galaxies,
NGC 5044 and NGC 5846. In these cases the fit yields very small S{\'e}rsic indices
($n_{2} \sim 0.7$ and $n_{2} \sim 0.4$, respectively) for the outer components.



For the cases in which our double- S{\'e}rsic fit did not converge,
we imposed an exponential profile $(n=1)$ on the outer
component, given by the equation
\begin{equation} \mu(R) = \mu_{0} + 1.086 \times R/r_{h} \end{equation} 
where $\mu_{0}$ and $r_{h}$ are the central surface brightness and exponential
scale length respectively. In these fits a S{\'e}rsic component without
restrictions was retained for the inner regions.  

\begin{figure*}
\centering
\hspace{-0.cm}
\includegraphics[width=7.5cm,]{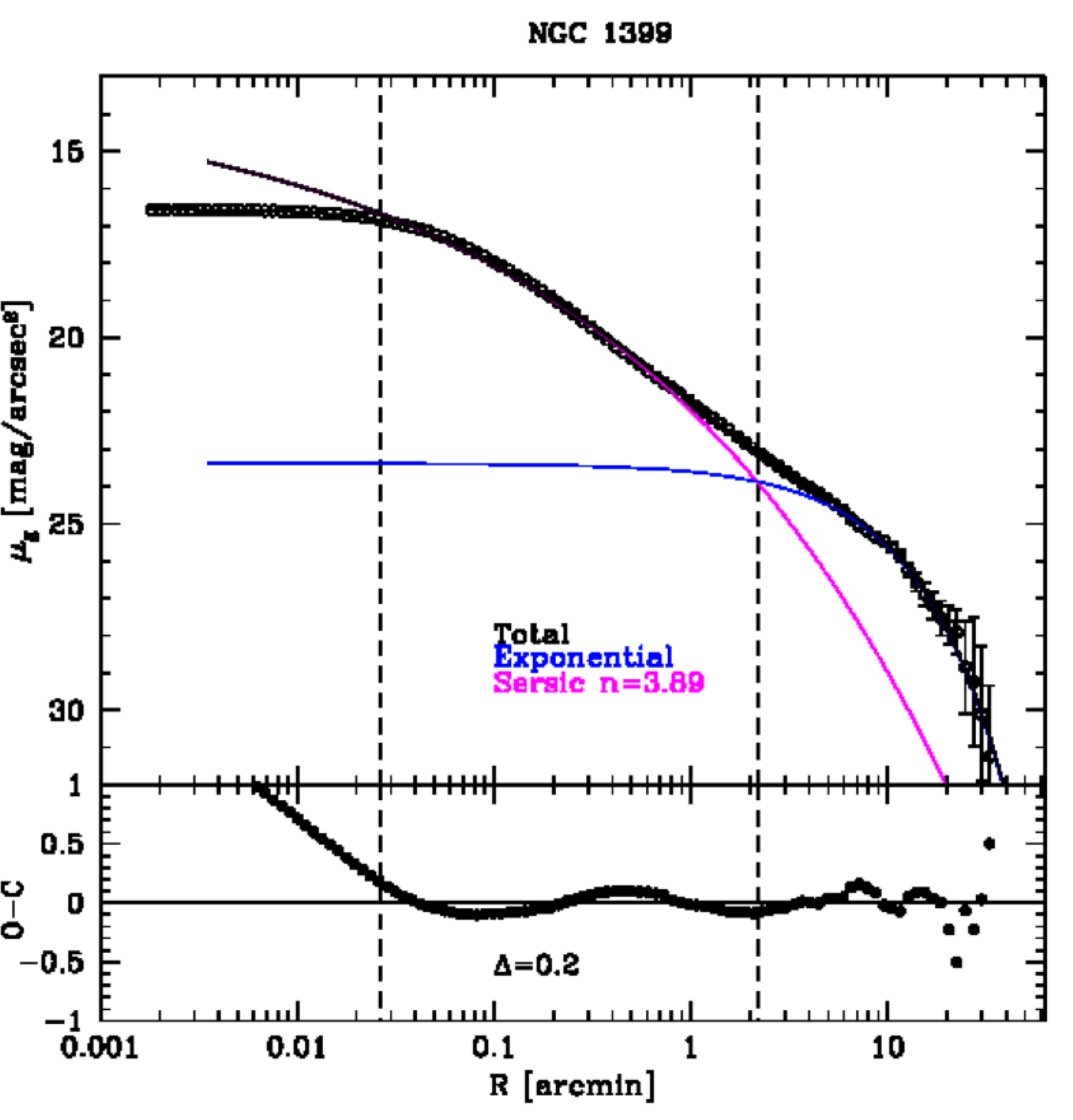}
 \includegraphics[width=7.5cm,]{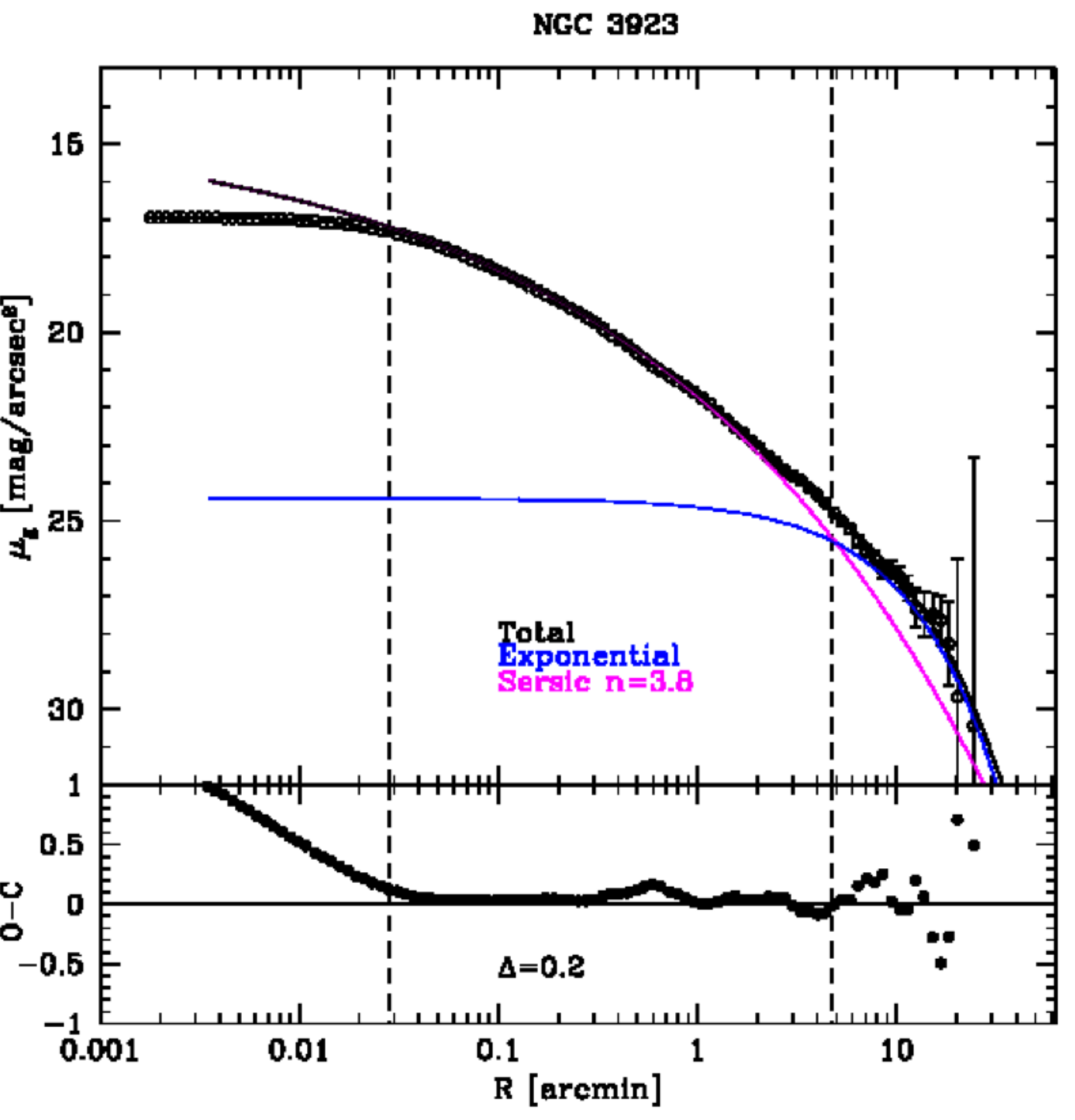}
 \includegraphics[width=7.5cm,]{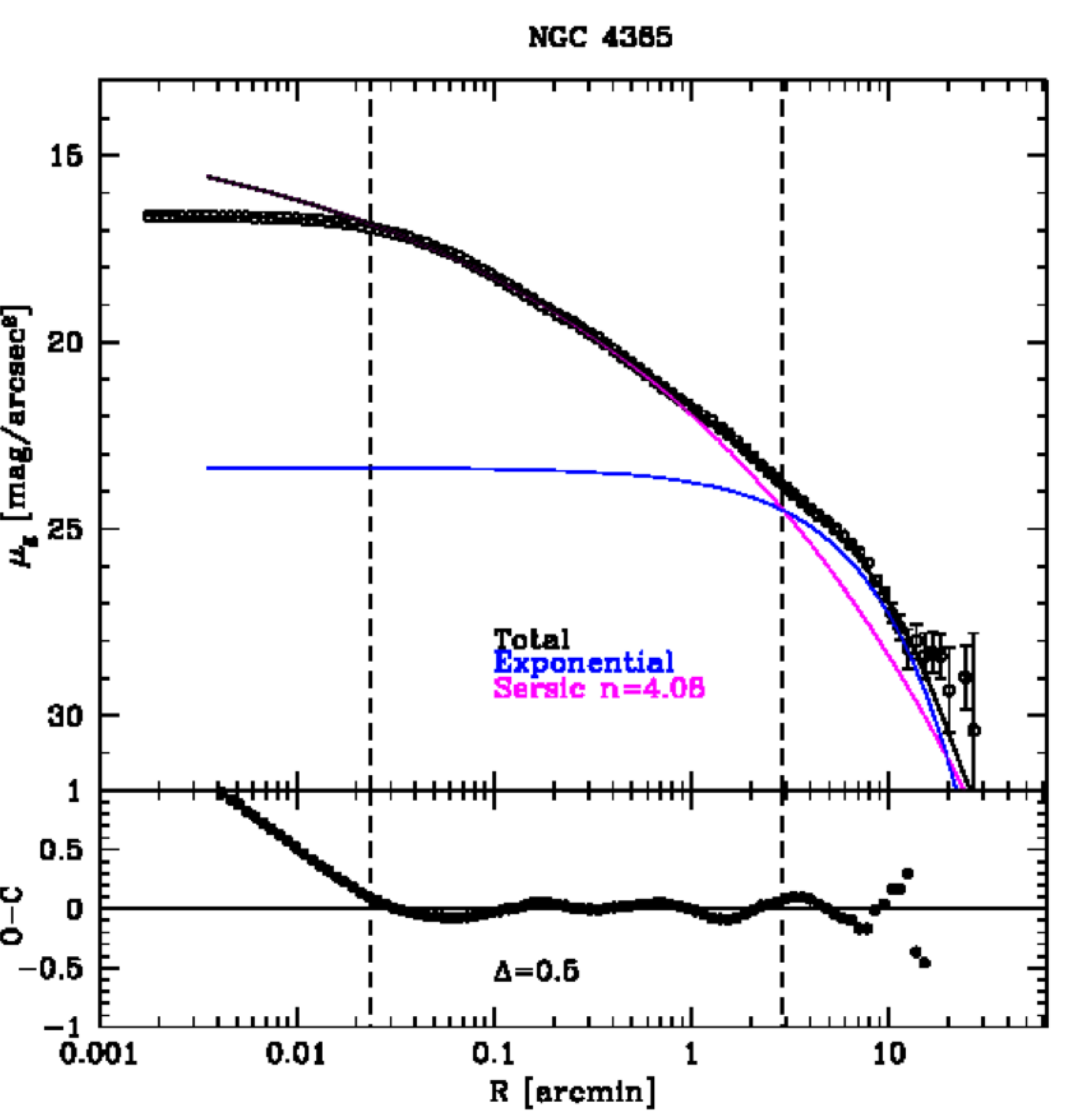}
\includegraphics[width=7.5cm,]{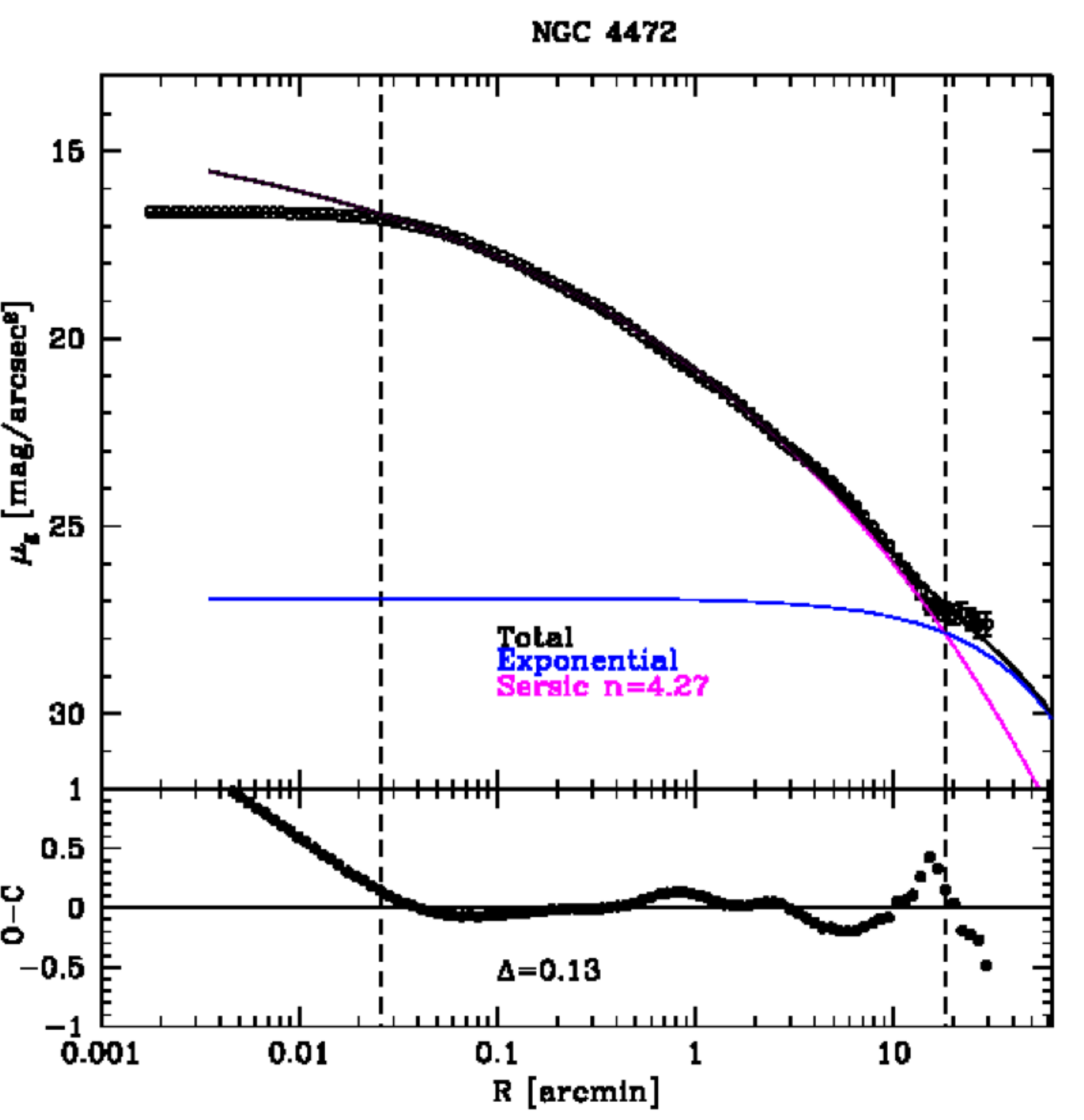}
\includegraphics[width=7.5cm,]{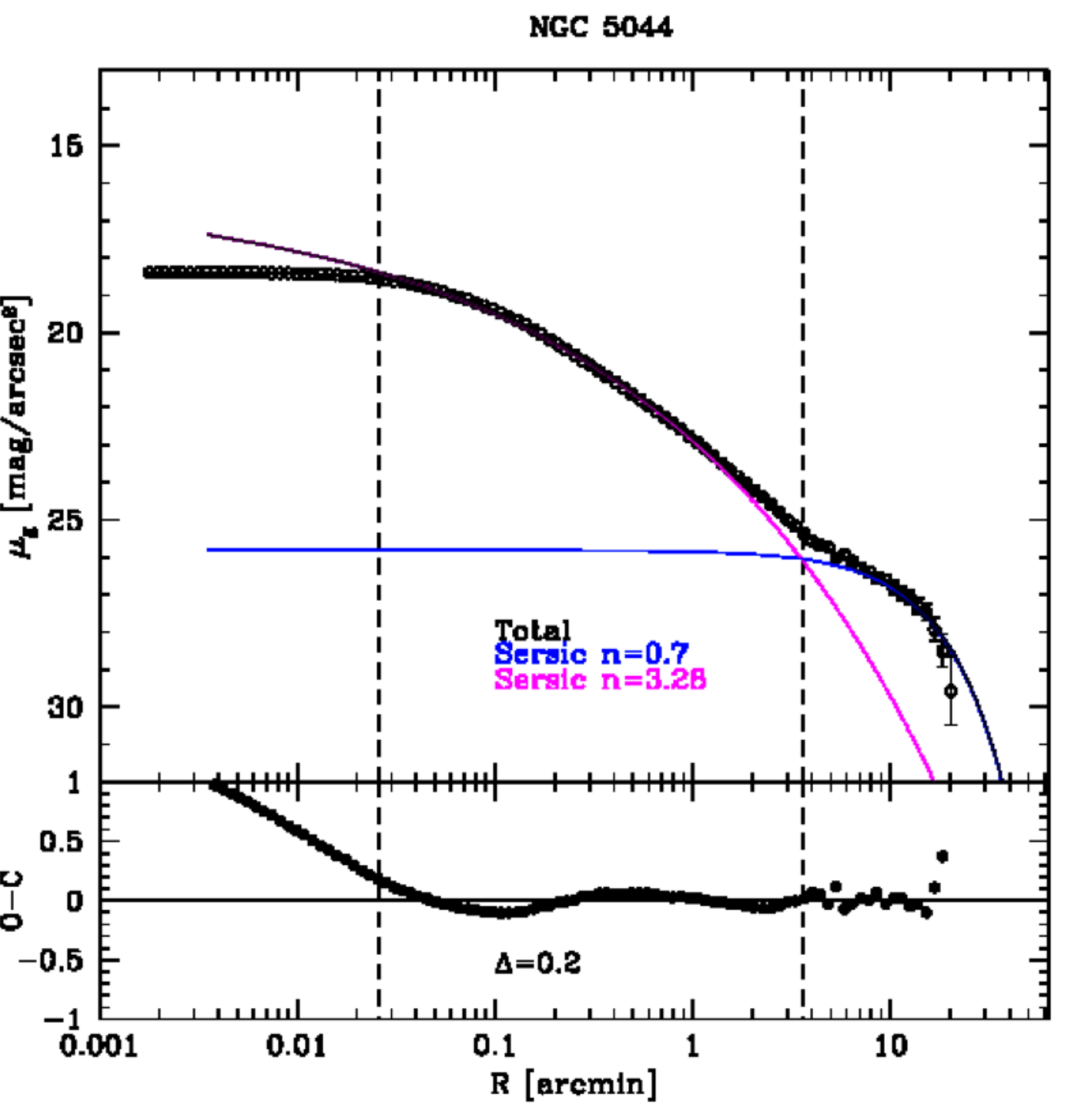}
\includegraphics[width=7.5cm,]{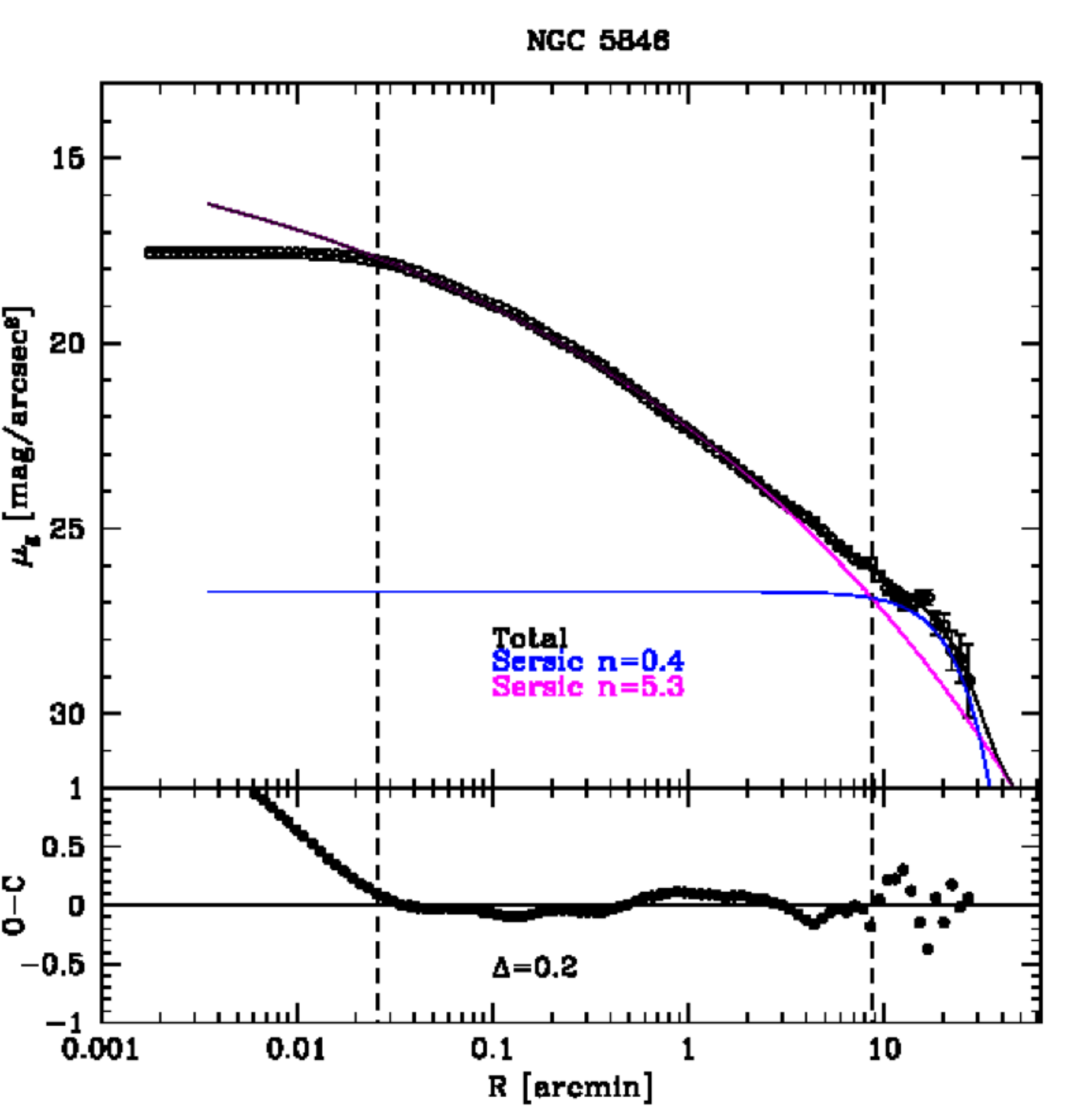}
\caption{VST {\it g} band profiles of NGC 1399, NGC 3923, NGC 4365, NGC 4472,
NGC 5044, and NGC 5846 plotted on a logarithmic scale. The blue line is
a fit to the outer regions with an exponential component, for NGC
1399, NGC 3923, NGC 4365 and NGC 4472, and with a S{\'e}rsic component
for NGC 5044 and NGC 5846. The magenta line is a fit to the inner
regions with a S{\'e}rsic profile, and the black line is the sum of the
components in each fit. The dashed lines mark the core of the galaxy
($1.5\times FWHM$), which
has been excluded in the fit, and the
transition point between the two components, respectively.}
\label{fit}
\end{figure*}

\begin{figure*}
\centering
\hspace{-0.cm}
\includegraphics[width=7.5cm]{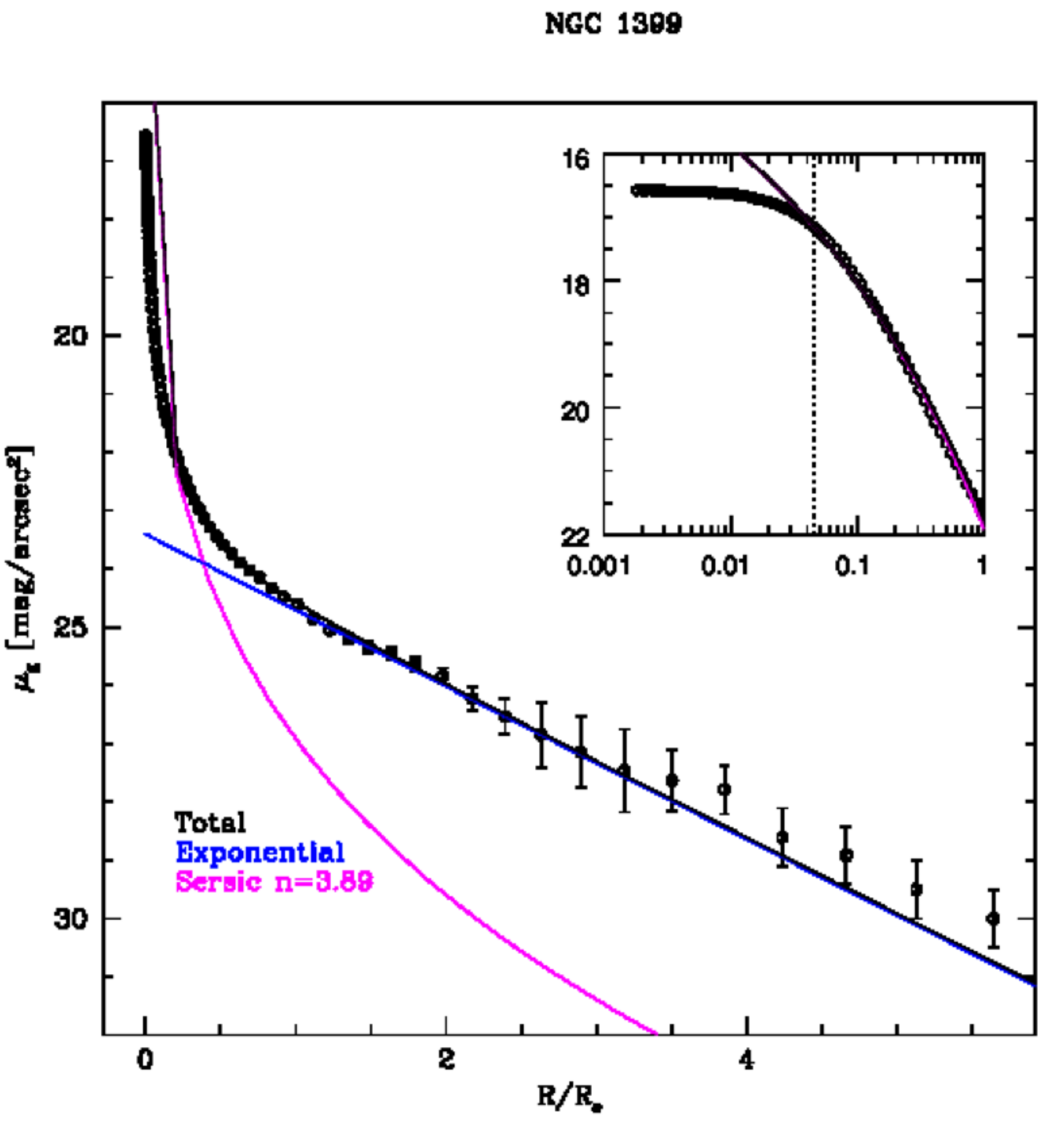}
 \includegraphics[width=7.5cm]{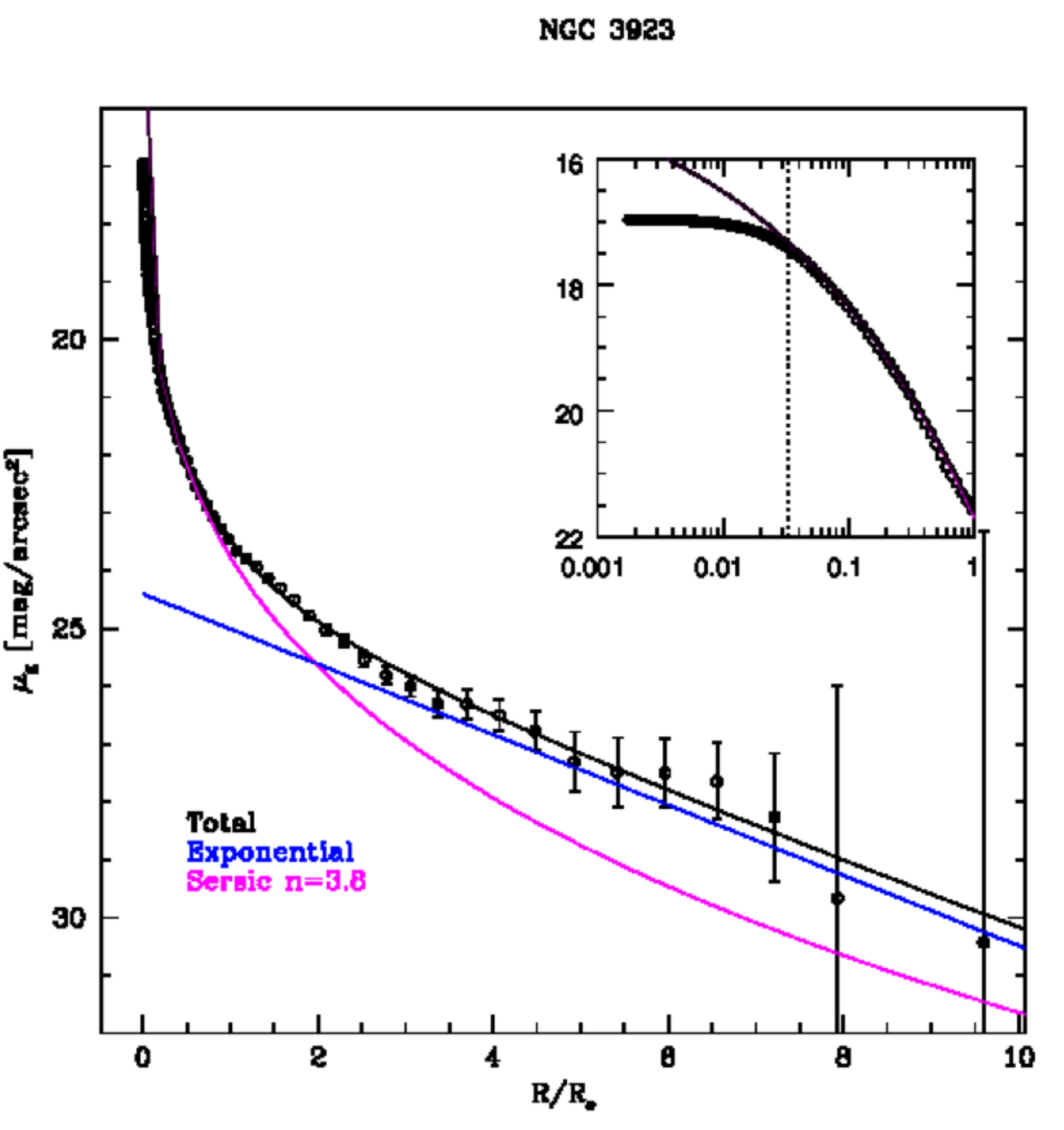}
 \includegraphics[width=7.5cm]{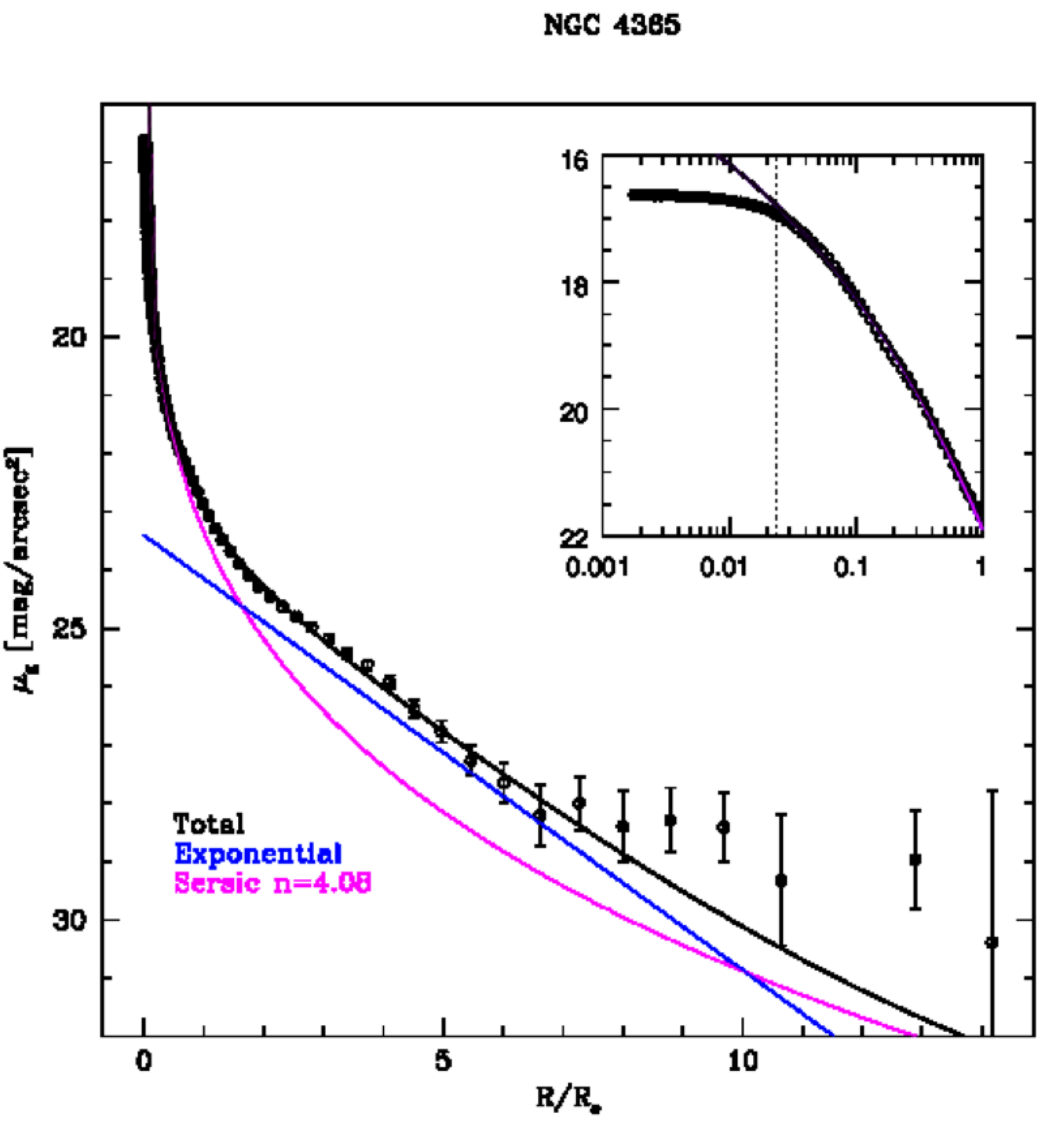}
\includegraphics[width=7.5cm]{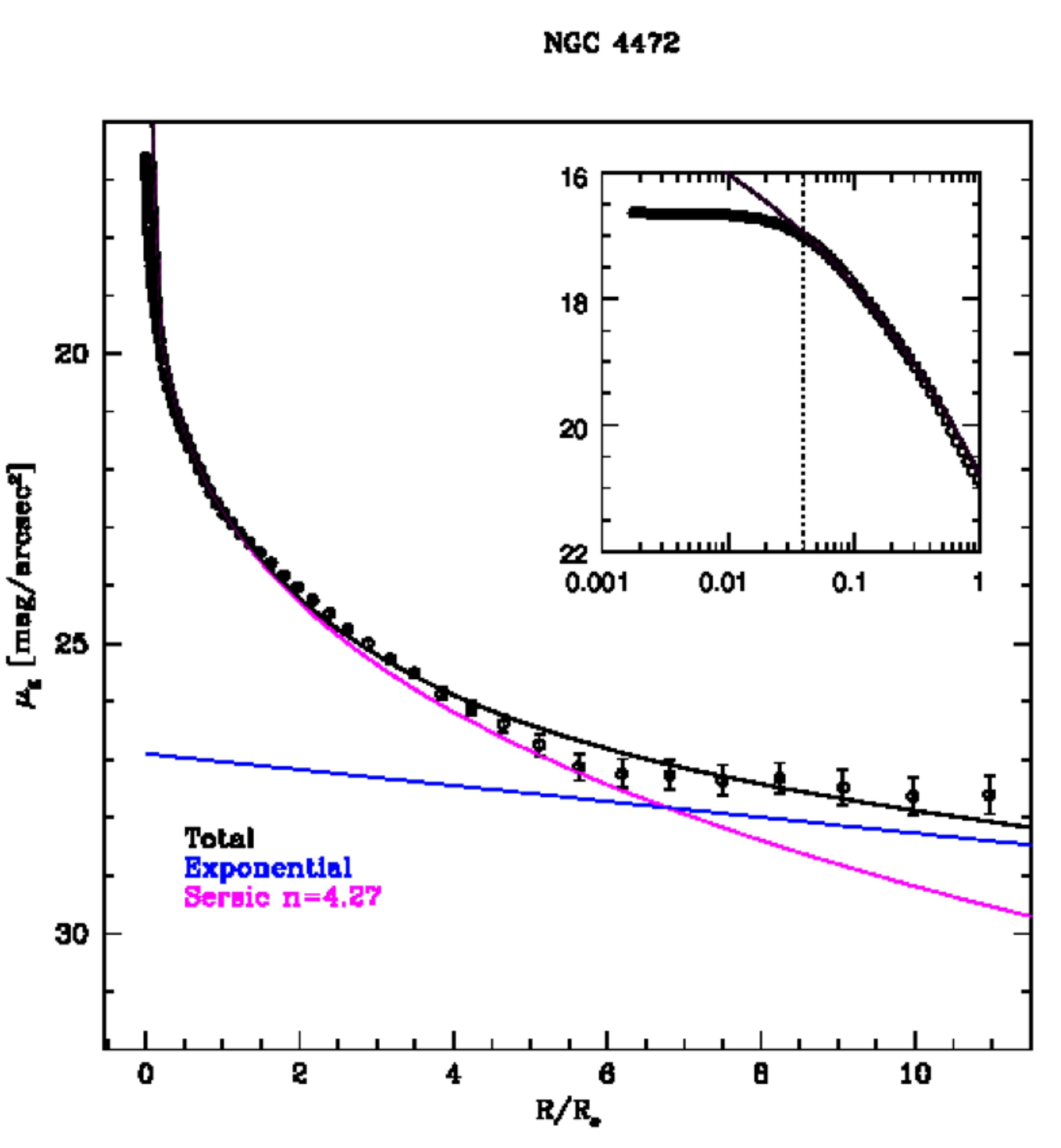}
\includegraphics[width=7.5cm]{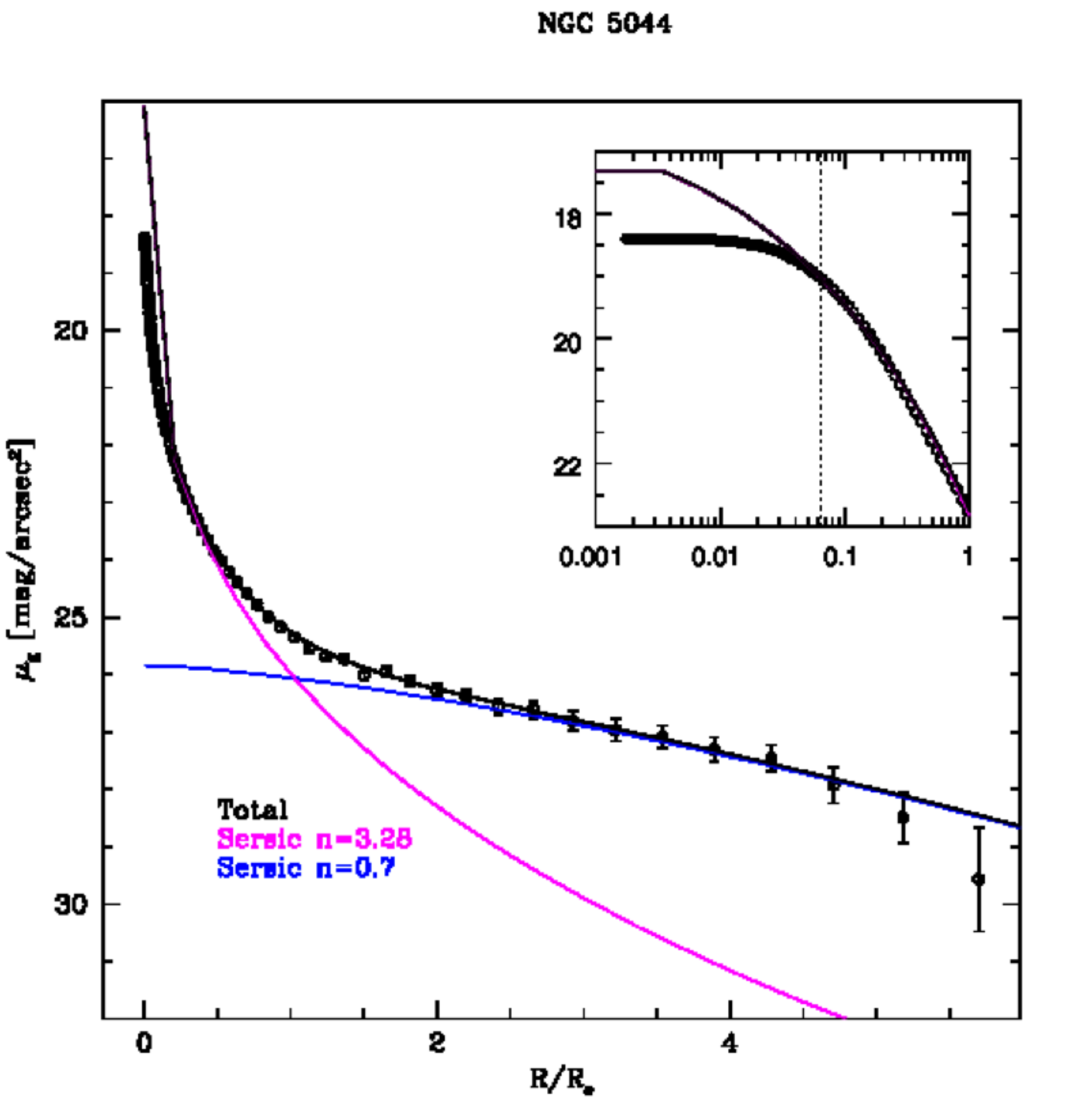}
\includegraphics[width=7.5cm]{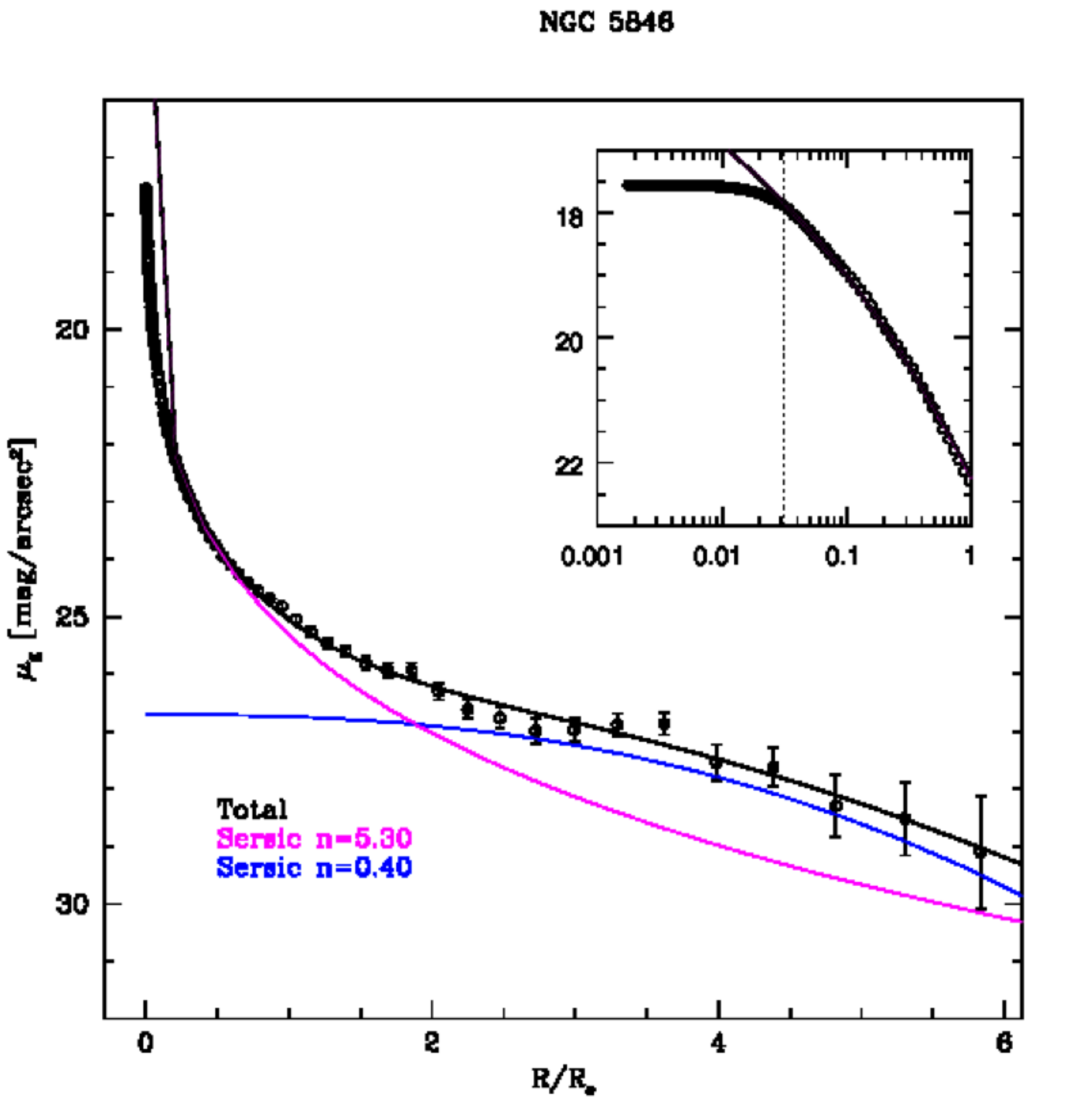}
\caption{The same as Fig. \ref{fit}, in linear scale as a function of
  $R/R_{e}$. {  In the small boxes there are the enlarged portions
    of the inner regions in a log radius scale, and the dashed lines mark the core of the galaxies ($\sim$ 1.5 $\times$ FWHM), which have been
    excluded from the fit.}}\label{fit_lin}
\end{figure*}

This double S{\'e}rsic / S{\'e}rsic plus exponential fits were performed on the
{\it g} band azimuthally averaged light profile, excluding the core of the
galaxies ($R\sim 2$ arcsec). The result of these fits and their residuals are shown in Fig.
\ref{fit}, in which the core of each galaxies is marked by a dashed
line. In Tab.
\ref{fit2comp} we report the structural parameters of the best fit for each
galaxy, with the relative 95\% confidence intervals. In Fig. \ref{fit_lin} we plot the results shown in Fig. \ref{fit}, but
on a linear scale, as a function of $R/R_{e}$. In this figure the distinction
between the two types of surface brightness profile are more clear.
Specifically, we can distinguish the profiles of NGC 1399, NGC 5044 and NGC
5846 which show a downward inflection, from those of NGC 3923, NGC 4365 and NGC
4472, which instead have an upward inflection. {  According to our
  multi-component modelling, all the upward inflections are the result
  of a transition from a high-$n$ inner component to a larger,
  lower-$n$, outer component. Downward inflection are, instead, the
  natural cut-off of a low-$n$ component. This implies that a downward
  inflection should be accompanied by an upward inflection at some
  smaller radius.}

We found that the inner components of each fit have effective radii
$r_{e}
\sim\ 5 - 25$ kpc ($45 - 202$ arcsec),  with an average value of $r_{e} \sim
12$ kpc, and S{\'e}rsic indices $n \sim 3 - 6$, with an average value of $n
\sim 4.3$. These values are consistent with those reported by
\citet{Gonzalez03} and \citet{Donzelli11}, who for their samples of BCGs found
$r_{e} \sim 5 - 15$ kpc and $n\sim 4.4$.

We used these fits to derive the total magnitude of the inner S{\'e}rsic
component, $m_{T,b}$, and outer (exponential or S{\'e}rsic) component,
$m_{T,h}$, as well as the relative contribution of the outer halo with respect
to the total galaxy light, $f_{h}$, reported in Tab. \ref{fit2comp}. 

From this analysis we have identified a radius for each galaxy in our sample
that marks the transition between the inner and outer components of our
two-component fit. We label this empirically-defined `transition
radius' $R_{tr}$. This transition occurs at very faint levels of surface
brightness ($\mu_{tr}$) for all our galaxies. As mentioned in Sec.  \ref{phot},
a discontinuity in the ellipticity, P.A., and ({\it g-i}) color profiles of all
six galaxies occurs around $R_{tr}$.

For NGC 1399 and NGC 5846 the outer halo component becomes even brighter than
the inner component at $R \geq R_{tr}$, where the relative contribution of the
outer halo with respect to the total galaxy light ($f_{h}$) reaches 60\% and
55\% respectively. For all the other galaxies the inner S{\'e}rsic component
dominates the light distribution, but the outer halo component is
never negligible, with relative fractions between 28\% and 39\%. Since
there is no clear reason to believe that in massive elliptical galaxies the
outer component in a fit such as this accounts for most of the accreted mass,
the halo mass fractions reported in Tab. \ref{fit2comp} should be considered as
a lower limit for the total accreted mass. We return to this point in the
following section.

\begin{figure}
\centering
\hspace{-0.cm}
 \includegraphics[width=9cm, angle=-0]{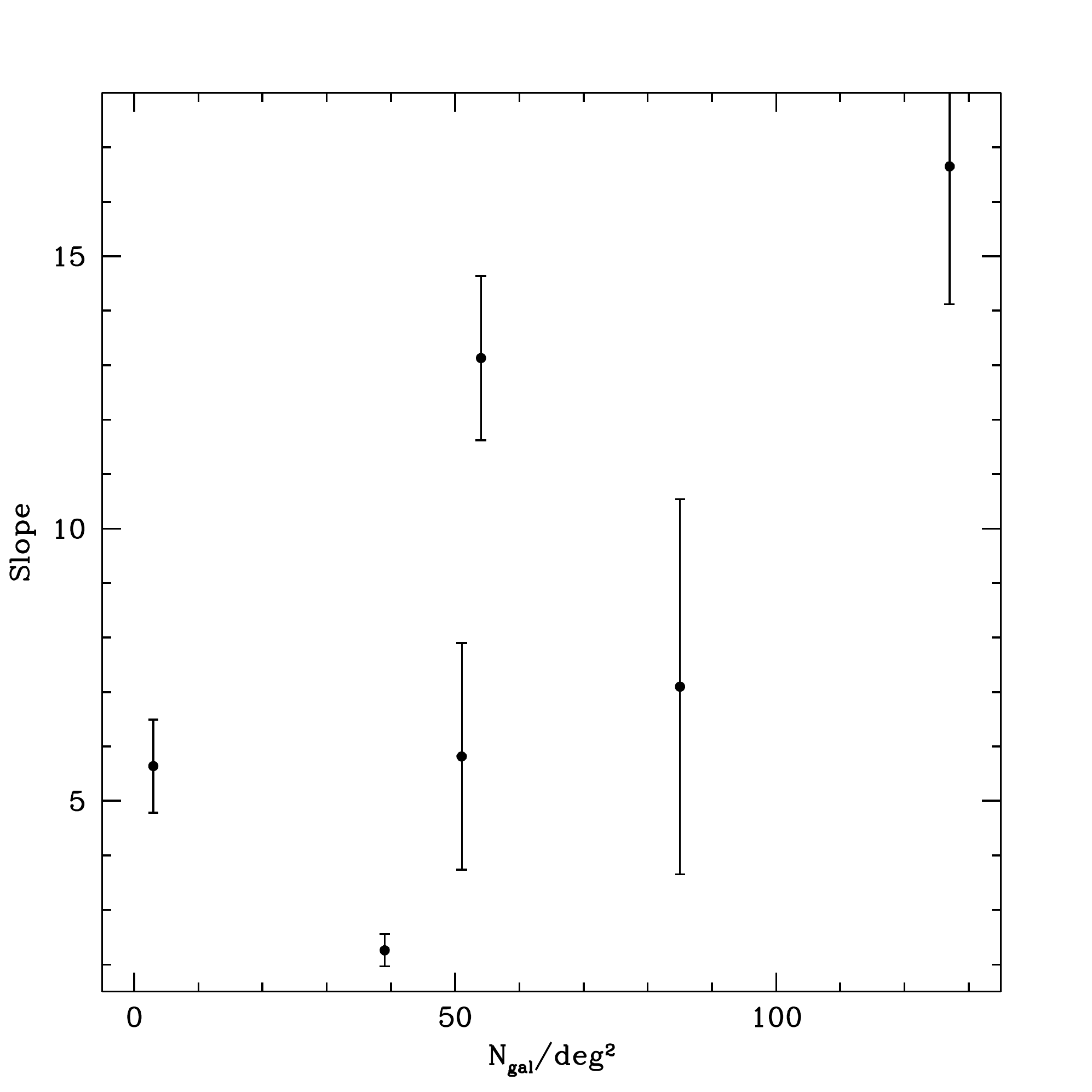}
\caption{Power-law slope of the outer halo surface brightness profiles (between $R_{tr}$ and the last
measured point), as a function of
the number of physically related galaxies {  (the spread of velocities for the individual galaxies is about 150 km/s)} in a field of 1 square
degree around each galaxy in our sample.}\label{density} 
\end{figure}

In order to investigate the possibility of a correlation between
environment and the shape of surface brightness profiles, in Fig.
\ref{density} we plot the slope of the outer halo surface brightness profile
(Fig.\ref{fit}) against the number of physically related galaxies
{  (the spread of velocities for the individual galaxies is about 150 km/s)} in a field of
1 square degree around each galaxy. The profile slope was obtained by
performing a power-law fit in the region of the outer envelope derived from
the two component model, that is between $R_{tr}$ and the last
measured point {  (see Tab. \ref{dens})}. This figure hints at a trend such that galaxies in
more dense environments have halo light profiles that are steeper than those in
less dense fields. In order to evaluate the robustness of the
correlation between the slopes and the density of the environment in
Fig. \ref{density}, we have estimated the Spearman's rank correlation
coefficient (SRCC, see Eq. 14.6.1 in Numerical Recipes), which ranges
in the [-1, 1] interval and measures the degree of correlation between
two variables. According to this test, two variables are
positively/negatively correlated if the SRCC is significantly
larger/smaller than zero for the two variables. As seen in
Fig. \ref{density}, the main uncertainty on the correlation is given
by the large error bars, hence we decided to use a Montecarlo approach
to the Spearman's test in order to establish the significance of the
correlation. We have ran 1000 experiments where we have perturbed the
data assuming a Normal distribution around each data point with the
standard deviation given by the data error and computed the SRCC at
every extraction. The mean SRCC we have obtained after the 1000 random
experiments is {  0.67$\pm0.20 (1\sigma)$. Even if we assume the
  more conservative standard error on SRCC, 
  i.e. $0.6325/\sqrt{N-1} = 0.3$ for N=6, this test suggests that the slope and
the density of the environment are significantly correlated at least
at 2$\sigma$ level. We
believe that this is a new and interesting result to check on larger samples.}

\begin{table*}
\setlength{\tabcolsep}{1.0pt}
\tiny
\caption{Best fitting structural parameters for a two component fit.} \label{fit2comp}
\vspace{10pt}
\begin{tabular}{lcccccccccccccccc}
\hline\hline
Object & $\mu_{e1}$ &$r_{e1}$&$n_{1}$& $\mu_{e2}$ &$r_{e2}$&$n_{2}$&$\mu_{0}$
&$r_{h}$&$m_{T,b}$&$m_{T,h}$&$f_{h}$&$R_{tr}$&$\mu_{tr}$\\
    & [mag/arcsec$^{2}$] &[arcsec]& & [mag/arcsec$^{2}$] & [arcsec]&& [mag/arcsec$^{2}$] & [arcsec]&[mag]&[mag]&&[arcmin]& [mag/arcsec$^{2}$]\\
\hline \vspace{-7pt}\\
NGC 1399  & 21.3$\pm$0.2   &   45$\pm$4  & 3.9$\pm$0.3&- &-&-&23.4$\pm$0.1 &292$\pm$4 &9.67&9.08&64\%&2.2$\pm$0.3&24.0\\
 NGC 3923    &    22.2$\pm$0.3   &   77$\pm$11  &   3.8$\pm$0.3&-
 &-&-& 24.4$\pm$0.2 &273$\pm$13 &9.43&10.21&32\%&4.8$\pm$2.2&25.3\\
 NGC 4365     &21.9$\pm$0.2  &60$\pm$7 &4.1$\pm$0.2&- &-&-&
 23.4$\pm$0.1 &166$\pm$4 &9.66&10.26&36\%&2.9$\pm$0.8&24.8\\
 NGC 4472    & 22.5$\pm$0.2  &148$\pm$14 &4.3$\pm$0.2&- &-&-&
 26.9$\pm$0.6 &1286$\pm$530 &8.29&9.36&27\%&18.2$\pm$8.9&27.8\\
NGC 5044   &22.9$\pm$0.1  &62$\pm$ 4 &3.3$\pm$0.1&27.00 $\pm$ 0.09&685
$\pm$ 33& 0.7 $\pm$ 0.1&- &-& 9.43&10.60&40\%&3.6$\pm$0.3&26.2\\
 NGC 5846   &24.2$\pm$0.3  &167$\pm$25 &5.3$\pm$0.3&27.22 $\pm$
 0.12&820 $\pm$ 17& 0.40 $\pm$ 0.07&- &-& 9.72&9.26&60\%&8.7$\pm$2.0&27.0\\
\hline
\end{tabular}

\tablefoot{Columns 2, 3 and 4 report effective magnitude, effective
radius and S{\'e}rsic index for the inner component of each fit. Columns
5, 6 and 7 list the same parameters for galaxies we fit with an
outer S{\'e}rsic component, whereas columns 8 and 9 list the
central surface brightness and scale length for galaxies we fit
with an outer exponential component. Columns 10 and 11 report the
total magnitude of the inner S{\'e}rsic ($m_{T ,b}$ ) and outer exponential or
S{\'e}rsic component ($m_{T ,h}$). Columns 12, 13 and 14
respectively give the relative fraction of the outer
component with respect to the luminosity of the galaxy; 
the transition radius; and the surface brightness at that radius.}
\end{table*}

\begin{table}
\setlength{\tabcolsep}{3.5pt}
\begin{center}
\caption{Correlation between outer profiles and environment.} \label{dens}
\begin{tabular}{lcccccccccccccccc}
\hline\hline
Object & $N_{gal}$ & slope \\
\hline \vspace{-7pt}\\
NGC 1399&54&13.13 $\pm$ 1.51\\
NGC 3923&3&5.64 $\pm$ 0.85\\
NGC 4365&51&5.82 $\pm$ 2.08\\
NGC 4472&39&2.26 $\pm$ 0.30\\
NGC 5044&127&16.65 $\pm$ 2.53\\
NGC 5846&85&7.10 $\pm$ 3.45\\
\hline
\end{tabular}

\tablefoot{Column 2 report the number of physically related galaxies (the spread of velocities for the individual galaxies is about 150 km/s) in a field of 1 square
degree around each galaxy in our sample. Column 3 is the slope of the
outer halo surface brightness profiles, obtained by performing a power-law fit in the region of the outer envelope derived from
the two component model, that is between $R_{tr}$ and the last
measured point.}
\end{center}
\end{table}

\subsection{Fitting the light distribution with three-components models}\label{3compfit}
\label{sec:threecomp}

Numerical simulations predict that stars accreted by BCGs account for most of
the total galaxy stellar mass ($\sim 90$\% on average), while in situ stars
significantly contribute significantly to the surface brightness profile only
out to $R \sim 10$~kpc \citep{Cooper13,Cooper15,Rodriguez15}. The overall
accreted profile is built up by contributions from several significant
progenitors, which can be divided in two broad types. Stars from massive
progenitors, which sink rapidly by dynamical friction,  are centrally
concentrated and have a shape resembling that of the overall galaxy profile,
while less significant progenitors have larger effective radii and lower
S{\'e}rsic indices \citep[a detailed discussion of the dynamical principles is
given by][]{Amorisco15}. 

For this reason, theory suggests that the surface brightness profile of an ETG
should be well described by the superposition of an inner S{\'e}rsic profile
representing the (sub-dominant) in situ component in the central regions,
another S{\'e}rsic profile representing the (dominant) superposition of the
``relaxed'', phase-mixed accreted components, and an outer diffuse component
representing ``unrelaxed'' accreted material (`streams' and other coherent
concentrations of debris) which does not contribute any significant surface
density to the brighter regions of the galaxy.  \citep{Cooper15} found an $n
\sim 1$ fitting component to be a reasonable empirical description of this
unrelaxed debris in the azimuthal average.  The stellar population of the in
situ component is expected to be quite similar to the dominant ``relaxed''
accreted component. As they are well mixed together, the sum of these two
components is expected to have a smooth distribution with only faint features
to suggest they are distinct. 

Following these theoretical predictions, we have described the surface
brightness profiles of our six galaxies with a three component model: a
S{\'e}rsic profile for the centrally concentrated in situ stars, a second
S{\'e}rsic for the ``relaxed'' accreted component, and an exponential or
S{\'e}rsic component for the diffuse and ``unrelaxed'' outer envelope.

A three-component fit may suffer from substantial degeneracy between
parameters. However, the numerical simulations cited above also predict that,
for early type galaxies in the mass range we are looking at, the variation of
$n$ with halo mass is weak, especially for the in situ component. There is
somewhat more variation with $n$ for the accreted stars, particularly with halo
mass rather than stellar mass. Therefore, in order to mitigate the degeneracy
in parameters and provide estimates of accreted components that are closely
comparable to the results of numerical simulations, we fixed $n \sim
2$ for the
in situ component of our three-component fits. We allowed small variations of $\pm 0.5$ around the mean
  value of $n=2$, since this would bracket the range of $n$ in the
  simulations, and allows us to obtain a better fit. The S{\'e}rsic parameters for
the accreted component were left free. We adopt this typical value of $n$ for
the in situ component on the basis of the results of \citet{Cooper13} for
massive galaxies (see their Fig. 7 in ). 

The reason for the weak variation of  $n$ with stellar mass and halo mass is
because in situ stars are relatively deeply embedded in the oldest and most
stable part of their host potential and initially have quasi-exponential
density profiles (i.e. are late-type galaxies\footnote{This is by construction
in the \citeauthor{Cooper13} models but obtained self-consistently in
hydrodynamical simulations such as those of \citet{Rodriguez15}.}). The
high-energy (large apocenter) tails of these distributions then diffuse outward
slightly in radius following relaxation in the near-equal mass ratio mergers
that are responsible for making these massive galaxies early types by $z=0$
\citep[e.g.][]{Hilz12}. Only very massive mergers (in terms of total mass)
affect the central regions of the potential in this way (hence the preservation
of stable discs with exponential profiles, characteristic of ongoing
dissipative star formation, as the dominant structural component of LTGs). Even
in very massive clusters, most of the $\sim$ 10 mergers that contribute
significantly to the observed light profile are associated with dark halos of
relatively low mass compared to the host (cluster) potential. Consequently,
even though each progenitors has roughly similar stellar mass, most of these
these mergers is (by the common convention) `minor', and the inner part of the
potential with the in situ stars is rarely disturbed. 

\begin{figure*}
\centering
\hspace{-0.cm}
\includegraphics[width=7.5cm,]{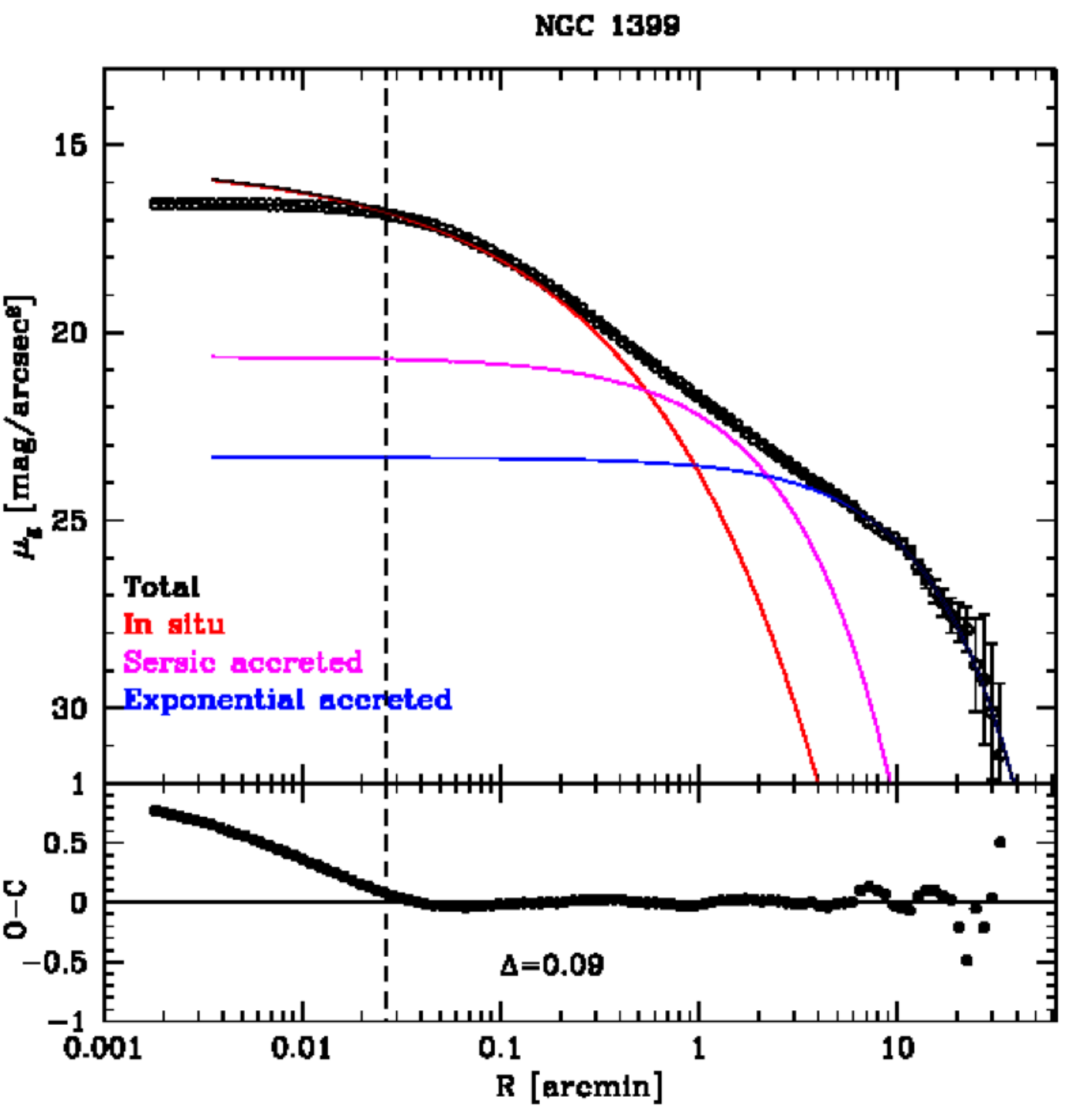}
 \includegraphics[width=7.5cm,]{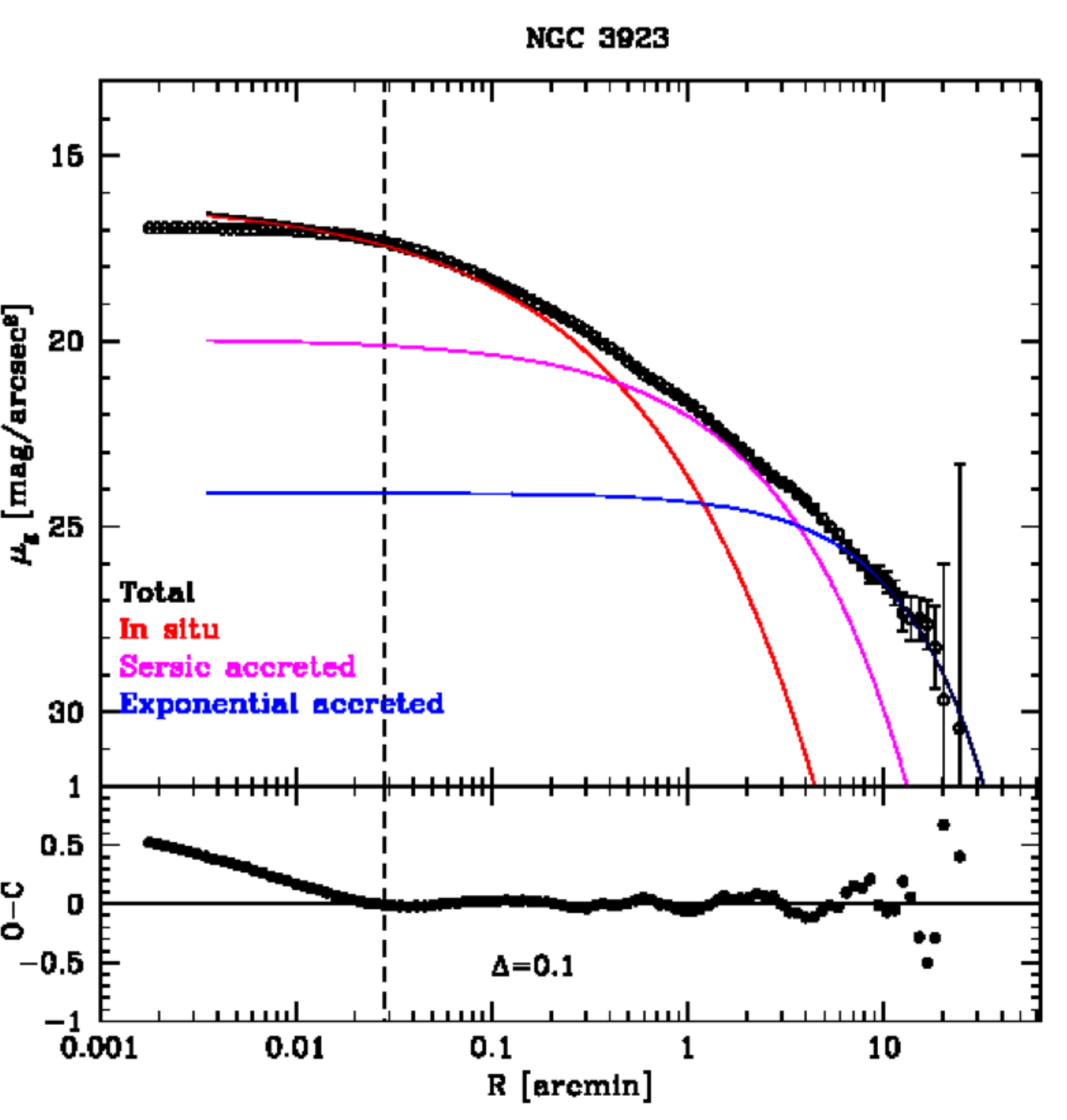}
 \includegraphics[width=7.5cm,]{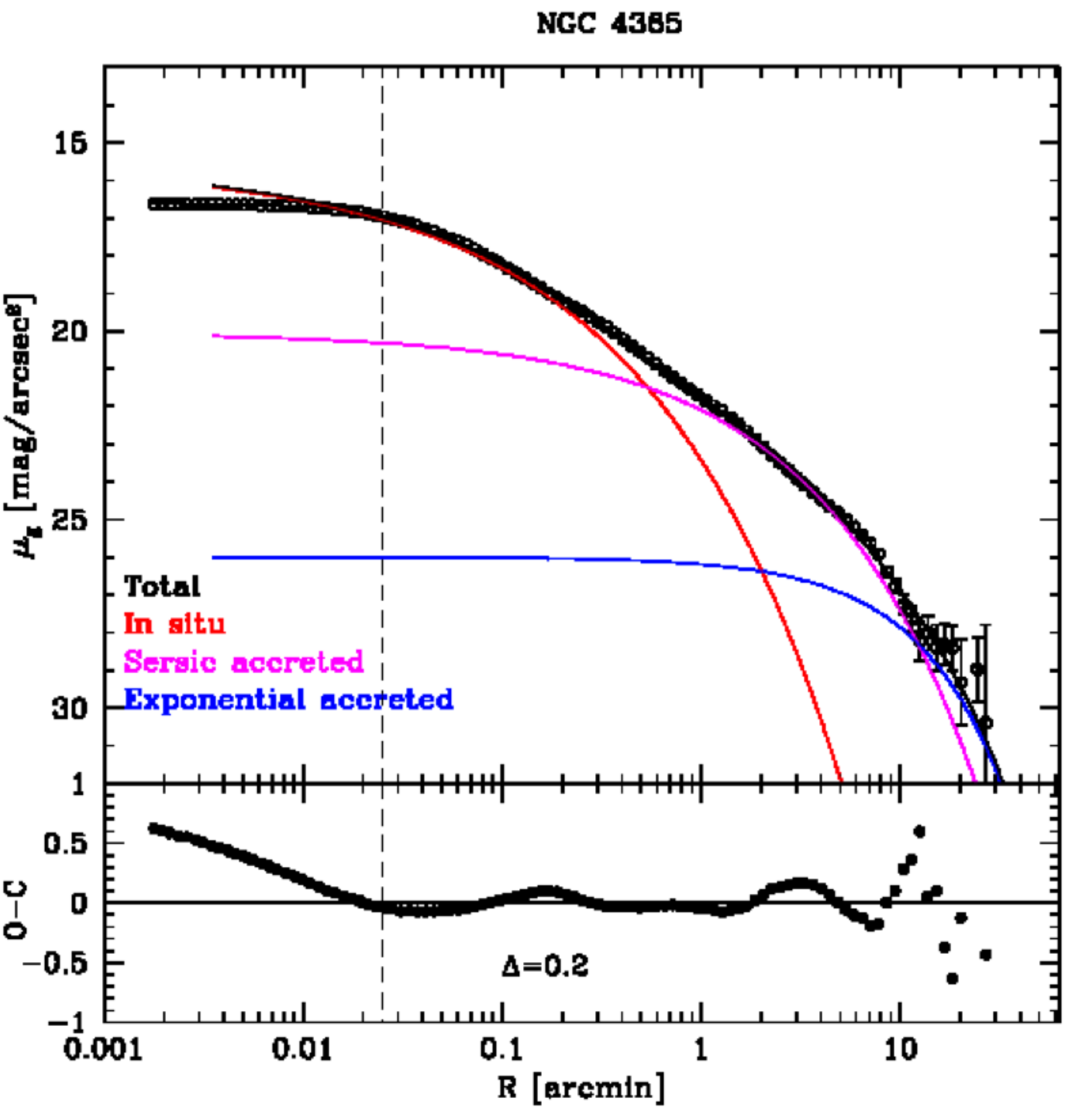}
\includegraphics[width=7.5cm,]{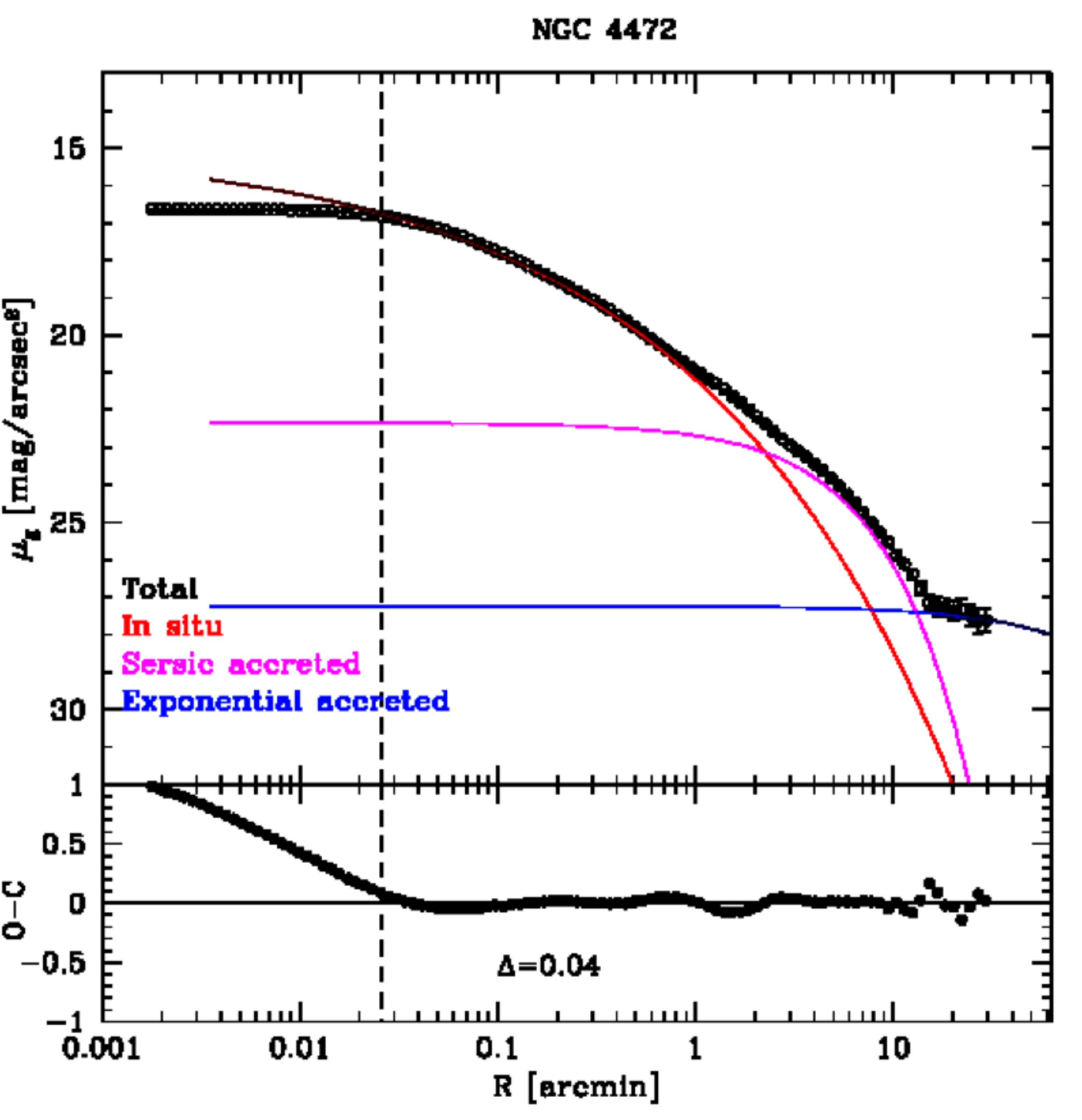}
\includegraphics[width=7.5cm,]{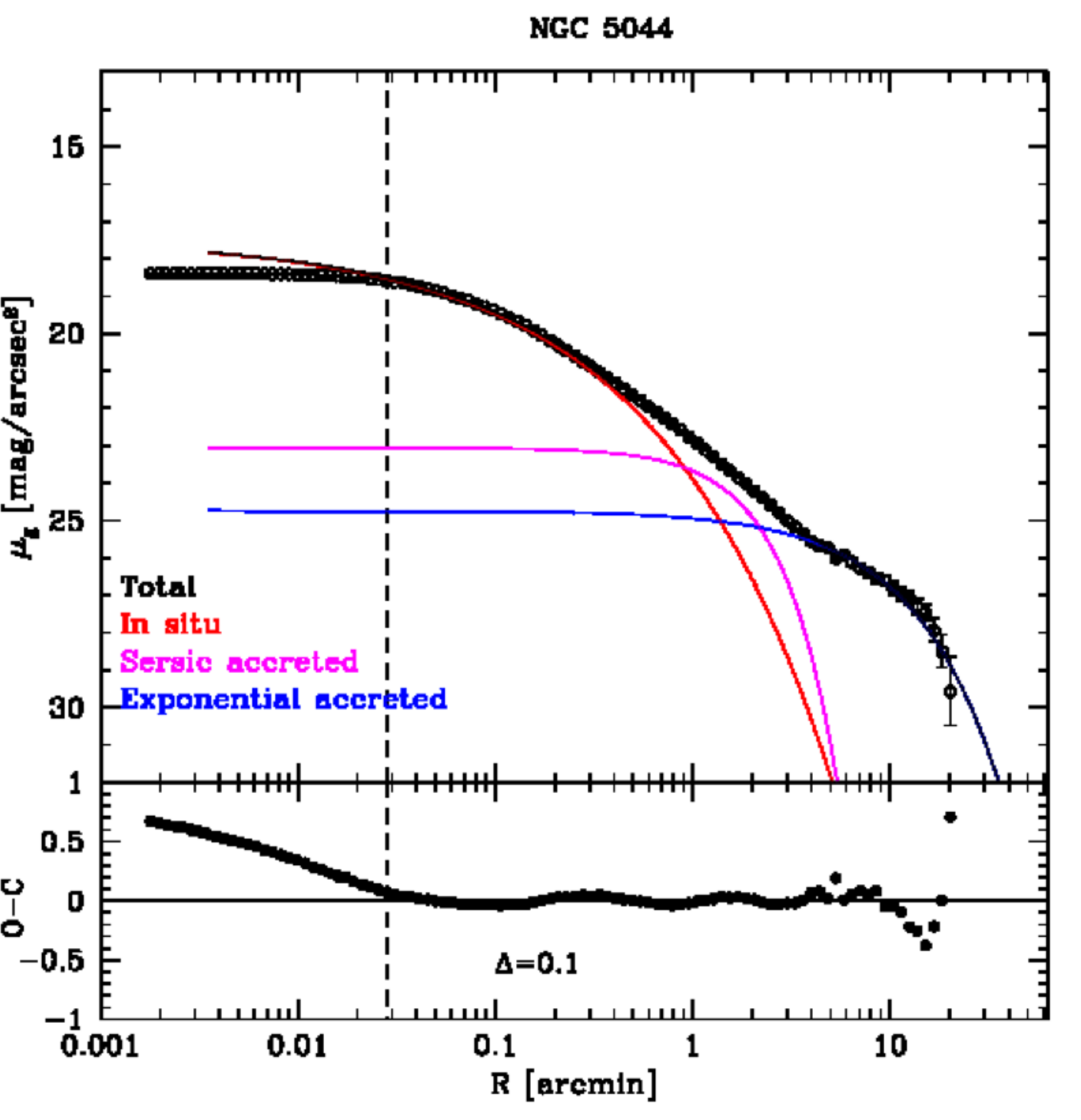}
\includegraphics[width=7.5cm,]{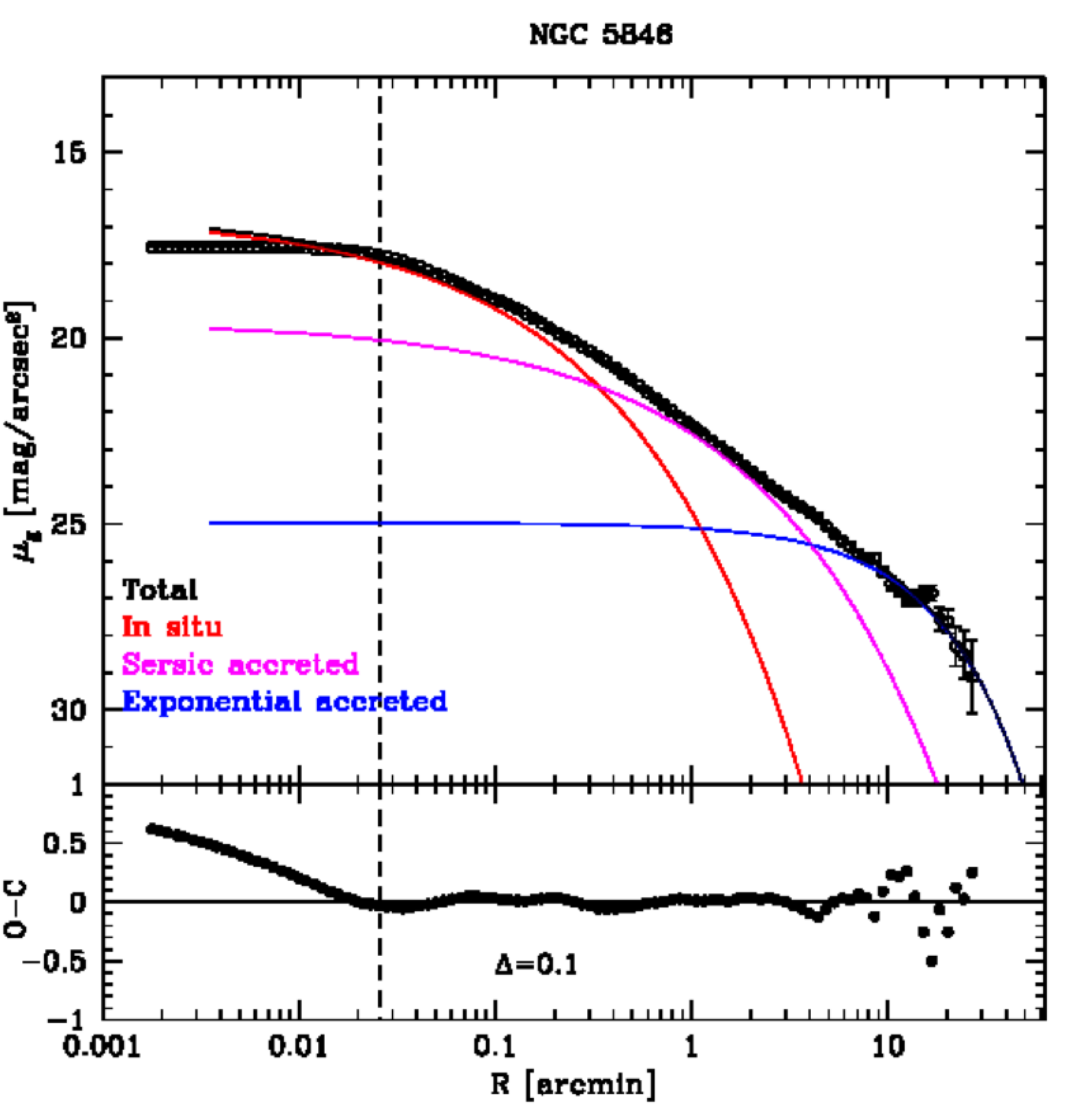}
\caption{VST {\it g} band profiles of NGC 1399, NGC 3923, NGC 4365,
  NGC 4472, NGC 5044, and NGC 5846, fitted with a three component
  model motivated by the predictions of theoretical simulations.}\label{fit3comp}
\end{figure*}

The results of these fits are shown in Fig. \ref{fit3comp}, and the best
fitting parameters are reported in Tab. \ref{tabfit3comp}. {  As
  already explained in Sec. \ref{2compfit}, even if we extract
  azimuthally averaged light profiles, the components fit to the
  radial profile are
  not necessarily elliptical structures.} From this plot it
appears that, as argued by \citet{Cooper15}, the radius $R_{tr}$ identified in
Fig. \ref{fit} marks the transition between different accreted components in
different states of dynamical relaxation, rather than that between in situ and
accreted stars. We found that the S{\'e}rsic indices for the accreted
components in this case range between 1 and 2. These values appear somewhat
lower than those reported by \citep{Cooper13}, but this is most likely because
we have included a third component that accounts for a significant fraction of
the outermost regions of the profile.  In support of this, in appendix
\ref{app}, we report parameters obtained with this theoretically motivated
approach \textit{without} the outer exponential component, i.e. from a
\textit{two}-component S{\'e}rsic fit in which the inner component is fixed to
$n\sim 2$. In that case the index of the outer component ranges $1.45<n_{2}<6.10$.
We have derived the total accreted mass fraction $f_{h,T}$ in Tab.
\ref{tabfit3comp} using the sum of both the `accreted' components in our
three-component fit. As discussed further in the next section, this makes it
easier to compare our results directly with theoretical predictions for this
quantity.

\citet{Huang13} also report multicomponent fits of a sample of 94 bright ETGs,
but without imposing constraints based on simulations. From their analysis they
identified three subcomponents: a compact inner component, and intermediate
component, and an outer and extended envelope. All the subcomponents in their
work had average S{\'e}rsic indices $n \simeq\ 1-2$. In our work we have
imposed a restricted range of models using numerical simulations as `prior' or
guide for the decomposition, simply to mitigate the degeneracy in
multicomponent fits to ETG profiles. Nevertheless, our approach gives results
fully consistent with \citet{Huang13}. 

\begin{table*}
\setlength{\tabcolsep}{1.0pt}
\begin{center}
\small
\caption{Best fitting structural parameters for a three component fit in which the Sersic index of the inner component is fixed.} \label{tabfit3comp}
\vspace{10pt}
\begin{tabular}{lcccccccccccccccc}
\hline\hline
Object & $\mu_{e1}$ &$r_{e1}$&$n_{1}$& $\mu_{e2}$ &$r_{e2}$&$n_{2}$&$\mu_{0}$
&$r_{h}$&$m_{T,1}$&$m_{T,2}$&$m_{T,3}$& $f_{h,T}$&$R_{tr1}$&$R_{tr2}$\\
    & [mag/arcsec$^{2}$] &[arcsec]& & [mag/arcsec$^{2}$] & [arcsec]&&
    [mag/arcsec$^{2}$] & [arcsec]&[mag]&[mag]&[mag]& &[arcmin]&[arcmin]\\
\hline \vspace{-7pt}\\
NGC 1399  & 19.44$\pm$0.08 & 14$\pm$1  &   2&22.7
$\pm$0.2&83$\pm$7&1.1$\pm$0.3& 23.33$\pm$0.06 &291$\pm$4
&10.33&9.74&9.02&84\%&0.5$\pm$0.1 &2.3$\pm$0.2\\
 NGC 3923    &    20.15$\pm$0.10   &   17$\pm$4
 &2&22.8$\pm$0.3&94$\pm$11&1.5$\pm$0.6& 24.1$\pm$0.3 &269$\pm$19
 &10.61&9.51&9.95&82\%&0.5$\pm$0.2 &3.6$\pm$0.8\\
 NGC 4365     &20.00$\pm$0.09  &17$\pm$2 &2.2&23.6$\pm$0.2
 &162$\pm$12&1.8$\pm$0.4& 26.0$\pm$1.1 &350$\pm$10
 &10.51&9.16&11.28&80\%&0.5$\pm$0.2 &12$\pm$2\\
 NGC 4472    & 21.04$\pm$0.03  &56$\pm$2 &2.5&24.05$\pm$0.06
 &280$\pm$6&0.9$\pm$0.1&27.2$\pm$0.2 &5446$\pm$3200
 &8.90&8.42&6.56&91\%&2.3$\pm$0.1 &13.1$\pm$1.3\\
NGC 5044   &21.44$\pm$0.08  &23$\pm$2 &2&24.07 $\pm$ 0.27&82 $\pm$ 7&
0.6 $\pm$ 0.2&24.74$\pm$0.08 &320$\pm$10& 11.24&11.11&10.22&79\%&0.9$\pm$0.2 &2.2$\pm$0.2\\
 NGC 5846   &20.67$\pm$0.09  &15$\pm$3 &2&23.6 $\pm$ 0.2&110 $\pm$ 13&
 2.0 $\pm$ 0.9&25.0$\pm$0.2 &452$\pm$29&11.41&10.02&9.7&89\%&0.4$\pm$0.2 &4.2$\pm$0.9\\
\hline
\end{tabular}

\tablefoot{Columns 2, 3 and 4 report effective magnitude and effective
radius for the inner component of each fit. The S{\'e}rsic index for
the in-situ component has
been fixed to $n\sim 2$, by using the models as a prior \citep{Cooper13}.  We allowed small variations of $\pm 0.5$ around the mean
  value of $n=2$. This would bracket the range of $n$ in the
  simulations, and allows us to obtain a better fit. Columns
5, 6 and 7 list the same parameters for the second component, whereas columns 8 and 9 list the
central surface brightness and scale length for the outer
exponential component. Columns 10, 11 and 12 report the
total magnitude of the inner S{\'e}rsic ($m_{T ,1}$ ) and outer
components ($m_{T ,2}$ and $m_{T ,3}$). Column 13 gives the
total accreted mass fraction derived from our three-component fit,{ 
while columns 14 and 15 report the transition radii between two fit components.}}
\end{center}
\end{table*}

Moreover, looking at the rms scatter $\Delta$, of each fit, we can clearly see that by
  adding the third component we achieve an improvement of
  at least 10\% for each galaxy. Since the expected value of $\Delta$
  scales as $\sqrt{(m-k)/m}$ (see \citealt{Seigar07}), where $m$ is
  the number of measured points ($\sim$ 70 in our case) and $k$ is the
  number of free parameters, we would need 18 free parameters to
  obtain an improvement of 10\%. This means that the improvement we
  obtain in our fits is not only due to the introduction of additional
  free parameters, as already shown by \citet{Seigar07}.

\section{Comparison with theoretical predictions for accreted mass fractions}\label{teor}

In the previous section, we identified inflections
in the surface brightness profiles of galaxies in our sample that may
correspond to transitions between regions dominated by debris from different
accreted progenitors (or ensembles of progenitors) in different dynamical
states. From our fitting procedure, we estimated the contributions of
outer exponential `envelopes' to the total galaxy mass, which
range from 28\% to 60\% for the galaxies in our sample, and the fraction of
total accreted mass, ranging from 83\% to 95\%.

We can convert these results to estimates of the stellar mass fractions in
the different components by assuming appropriate stellar mass-to-light
ratios. {  To obtain typical colours for the central galaxies
  regions and for the outer envelopes, we measured the mean {\it g-i}
  color for each galaxy in regions where each of those components
  dominates. The mean colors are reported in Tab. \ref{mass}, and have
been estimated for $R_{e}\leq R \leq 2R_{e}$, where $R_{e}$ is the
effective radius derived by the 2-component fit (see Tab. \ref{fit2comp}) for the inner and the
outer component.}


We used stellar population synthesis models \citep{Vazdekis12,
Ricciardelli12}, with a Kroupa IMF, to derive the mass-to-light ratios
corresponding to the average colors, and hence the stellar mass of the
whole galaxy, of the outer envelope, and of the accreted component.
These results are summarized in Table \ref{mass}.

\begin{table*}
\setlength{\tabcolsep}{2.5pt}
\begin{center}
  \caption{Total and accreted stellar masses of galaxies in our sample.} \label{mass}
\vspace{10pt}
\begin{tabular}{lcccccccccccccccc}
\hline\hline
Object & $({\it g-i})_{in}$ &$({\it g-i})_{out}$&$(M/L)_{in}$& $(M/L)_{out}$
&$M^{*}_{tot}$&$M^{*}_{envelope}$&$M^{*}_{total\ accreted}$\\
    & [mag] &[mag]&[$M_{\odot}/L_{\odot}$] &[$M_{\odot}/L_{\odot}$] &[$M_{\odot}$]&[$M_{\odot}$]&[$M_{\odot}$]\\
\hline \vspace{-7pt}\\
 NGC 1399    &    1.2   &   0.8  &   5.2&  0.6 & $5.8\times10^{11}$&$3.7\times10^{11}$&$4.9\times10^{11}$\\
NGC 3923    &    1.3   &   1.4  &   6.2&   9.1 & $1.5\times10^{12}$ &$4.9\times10^{11}$&$1.2\times10^{12}$\\
 NGC 4365     &   1.1  &    0.9  &  3.4&   0.9 &$4.3\times10^{11}$ &$1.5\times10^{11}$&$3.4\times10^{11}$\\
 NGC 4472    &    1.2   &  1.1  &    5.2&  3.4  &$8.6\times10^{11}$ &$2.3\times10^{11}$&$7.8\times10^{11}$\\
NGC 5044   &     1.2  &   0.7   &   5.2&  0.5  &$5.7\times10^{11}$ &$2.3\times10^{11}$&$4.5\times10^{11}$\\
 NGC 5846   &    1.4  &    0.7 & 9.1&   0.5  &$9.6\times10^{11}$ &$5.8\times10^{11}$&$8.5\times10^{11}$\\
\hline
\end{tabular}
\tablefoot{Columns 2-5 show the mean, {  extinction corrected,} {\it g-i} color of the inner and
  outer regions of each galaxy and the relative mass-to-light
  ratios in the {\it g} band. Columns 6 is the galaxy stellar mass,
  while column 7 and 8 are the outer envelope and total accreted
  stellar masses, derived from the second component of the
  two-component fit and from both the accreted ones in the
  three-component fit, respectively.} 
\end{center}
\end{table*}

\begin{figure}
\centering
\hspace{-0.cm}
 \includegraphics[width=9cm, angle=-0]{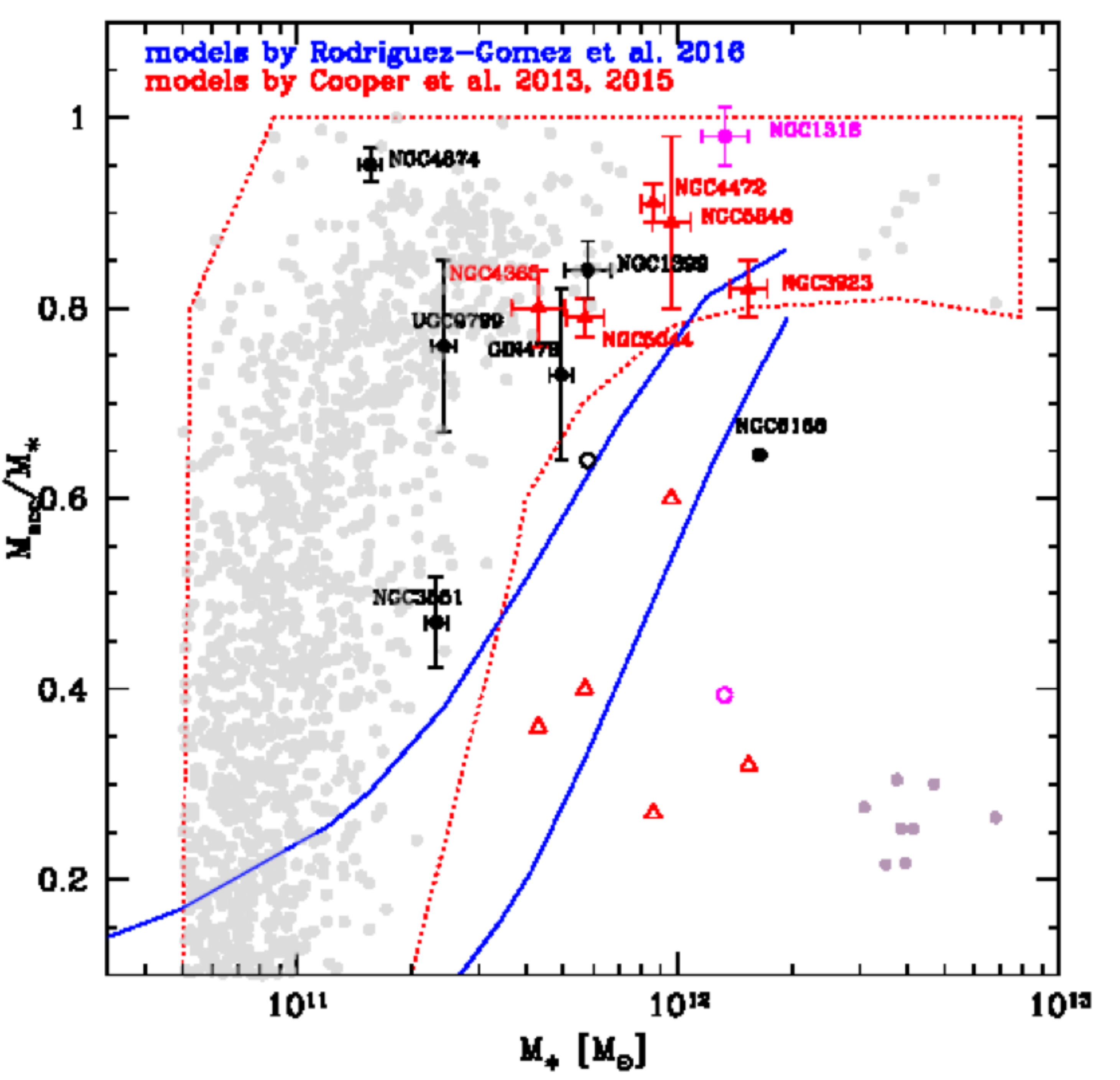}

 \caption{Accreted mass fraction versus total stellar mass for ETGs. Our
 VEGAS measurements are given as red filled and open triangles (see text
 for details). Black circles correspond to other BCGs from literature
 \citep{Seigar07, Bender15, Iodice16} and the magenta dot is to NGC 1316
 (Iodice et al., in preparation). Red and blue regions indicate the
 predictions of cosmological galaxy formation simulations by
 \citet{Cooper13,Cooper15} and \citet{Rodriguez15}, respectively.
 Purple-grey points show the mass fraction associated with the
 streams from Tab. 1 in \citet{Cooper15}, for comparison to the
 observations shown by open symbols.}\label{ratio} 
 
\end{figure}

In Fig. \ref{ratio} we compare the accreted mass ratios we infer from
our observations (filled red triangles) with other observational
estimates for BCGs by \citet{Seigar07, Bender15} and \citet{Iodice16}, and
with theoretical predictions from semi-analytic particle-tagging
simulations by \citet{Cooper13,Cooper15}, and the Illustris cosmological
hydrodynamical simulations \citep{Rodriguez15}. We find that the stellar
mass fraction of the accreted component derived for galaxies in our sample is
fully consistent both with published data for other BCGs (despite
considerable differences in the techniques and assumptions involved) and with
the theoretical models by \citet{Cooper13,Cooper15}. 

In the same figure we also include the halo mass fraction
obtained for NGC 1316, the VEGAS photometry for which will be
published in a forthcoming paper (Iodice et al. in preparation). NGC 1316 is a
peculiar galaxy in the Fornax cluster, with a very perturbed morphology.
The presence of shells, tidal tails and dust indicates that the galaxy has
undergone a recent interaction event.  The accreted mass fraction and the total
stellar mass for this object are consistent with those of NGC 4472 and NGC
3923, which also show peculiar features, such as shells and dust lanes, and
evidences of ongoing interactions.

In Fig. \ref{ratio} we also compare the stellar mass fractions obtained
for the outermost exponential component of our multicomponent fit (open red
triangles) with the mass fraction associated with unbound debris
streams from surviving cluster galaxies in the simulations of
\citet{Cooper15}. We found that the mass fraction in this component of
our fits is consistent with these values from the simulations, suggesting
that such components may give a crude estimate of the mass distribution
associated with dynamically unrelaxed components originating from disrupting
or recently disrupted galaxies, as argued by \citet{Cooper15}.

\section{Discussion and conclusions}\label{disc}

We have presented new deep photometry in the {\it g} and {\it i} bands
for four giant ETGs in the VST Early-type Galaxy Survey (VEGAS):
NGC 3923, NGC 4365, NGC 5044 and NGC 5846. In particular, we have studied the
shapes of their surface brightness profiles to obtain evidence of
structural variations that may constrain their assembly history.  We have also
combined the results of this photometric study with those presented in
Paper I, for NGC 4472, and in \citet{Iodice16}, for NGC 1399.

The data used were collected with the VST/OmegaCAM in March and April of 2015
within the Italian Guaranteed Time Observation (GTO). The large field of view
(FOV) of OmegaCAM mounted on VST (one square degree), together with its
high efficiency and spatial resolution, allows us to map the surface brightness
of galaxies out to many effective radii and to very faint levels.
We can therefore probe regions associated with the faint stellar envelopes
of these galaxies, which, due to their long dynamical timescales, may
retain signatures of their assembly history.

Our analysis suggests that the surface brightness profiles of the
galaxies in our study are best reproduced by multi-component models.
As already pointed out by \citet{Mendez2016}, this kind of
  photometric decompositions need a ``human-supervised'' approach, and
  the choice of the number of components included in the fit is a
  relevant source of uncertainty. Since the surface brightness profiles of
  many galaxies could be fit equally well by a whole range of
  alternative models, the results of these photometric decompositions,
especially for featureless early type galaxies, strongly depends on
the model adopted.
We took two approaches to constructing such models. Adopting the an
empirically-motivated two-component approach most common in the literature we
find that four of our six galaxies (NGC 3923, NGC 4365, NGC 4472, NGC 1399) can
be modelled by an inner S{\'e}rsic component and an outer exponential, while
the other two objects (NGC 5044 and NGC 5846) are best fitted with a double
S{\'e}rsic model.  The inner S{\'e}rsic components for each galaxy have an
average effective radius $r_{e} \sim 12$ kpc and S{\'e}rsic index $n\sim 4.3$,
quite similar to the values found (with a similar fitting
philosophy) by \citet{Gonzalez03} and \citet{Donzelli11},  $r_{e}\sim 5 - 15$
kpc and $n\sim 4.4$.

We also used an alternative approach, motivated by the predictions of
numerical simulations, in which we fitted the surface brightness profiles of
our galaxies with three components: two dominant S{\'e}rsic components
in and an outer exponential component. To mitigate some of the degeneracy in
this approach, we fixed the S{\'e}rsic index of the inner component to a
representative value from simulations. In those simulations, this component
corresponds to a mixture of stars formed in situ and the remnants of mergers
subject to violent relaxation.  Compared to the traditional empirical fit, this
approach allows us to make a more meaningful estimate of the total contribution
of accreted stars.
The accreted stellar mass fractions we infer for galaxies in our sample
with this approach are consistent with the theoretical predictions of
\citet{Cooper13}, in which the fraction of accreted stars, and hence the
S{\'e}rsic index of the overall surface brightness profile, increase strongly
with stellar mass. 

The most massive galaxies in our study, NGC 3923 and NGC 4472, are in
the mass range typically identified with BCGs and ICL phenomena, as is NGC 1316
(Iodice et al., in preparation).  These three galaxies show clear
signs of ongoing interaction and recent accretion events, indicating that their
outer regions are still being assembled, consistent with
theoretical expectations for such galaxies \citep[e.g.][]{Cooper15}. In
this regard it is interesting that the mass fractions in the exponential
components of our two-components profile decompositions are in
good agreement with the mass fractions associated with streams from
surviving galaxies in the simulations of \citet{Cooper15}. This
suggests these outer exponential components may give a crude estimate of the
stellar mass fraction associated with recently disrupted galaxies. We find
values for this fraction ranging from 28\% to 64\%.

Previous works, for example by \citet{Gonzalez05} and \citet{Zibetti05},
have identified outer components similar to those in our fits with so-called
diffuse intracluster light (ICL).  Typical claims for the mass fractions in ICL
range from $10 - 30\%$. From a theoretical perspective, it is not clear that it
is physically meaningful to attempt to distinguish such a component from the
outer envelopes of BCGs \citep[see for example the discussion in][]{Cooper15}.
A wide variety of definitions of ICL have been adopted in both the
observational and theoretical literature, according to many of which such
components cannot be unambiguously identified from photometry alone. We
therefore leave open the question of how the outer-component light fractions we
derive relate in detail to previous claims regarding the nature of ICL.

For all the galaxies in our study we can identify at least one
inflection in the surface brightness profile. These inflections occur at very
faint surface brightness levels ($24.0 \leq \mu_{g} \leq 27.8$
mag/arcsec$^{2}$). They appear to correlate with changes in the trend of
ellipticity, position angle and color with radius, the isophotes becoming
flatter and misaligned and the colors bluer beyond the inflections.  This
suggests that these inflections mark transitions between physically distinct
components (or ensembles of similar components) in different states of
dynamical relaxation. This sense of the surface brightness inflection is
upwards (i.e. a break to a shallower slope) in NGC 3923, NGC 4365 and NGC 4472,
and downwards (i.e. a break to a steeper slope) in the other galaxies.

A variety of possible interpretations for such features have been
suggested based on theoretical models.  Upward inflections (shallower outer
slopes) might reflect transitions between either an inner in situ-dominated
region and an outer accretion-dominated region, or else between two accreted
components \citep{Abadi06,Font11}. Downward inflections can occur within the
profiles of debris from a single progenitors alone, corresponding to the
characteristic apocentric radius of its stars \citep{Cooper10, Deason13,
Amorisco15} and therefore need not represent a transition between two separate
components; however they need not always correspond to a single progenitor,
because they can also be created by a drop in the density of only one of a
small number of contributions having roughly equal density interior to the
break. It has been claimed that the strength and radius of downward inflections
in the accreted component encode information on the number of progenitors as
well as the orbital properties of the dominant progenitor and its disruption
timescale.  For example, \citet{Deason13} examine aspects of these
relationships in the context of satellite accretion onto Milky Way analogues,
and claim that the properties of the break can be used to infer the accretion
time of the progenitor. (In massive ETGs  the picture is further complicated by
violent relaxation of the central potential \citet[e.g.][]{Hilz12}.)

Since there are many ways to produce these inflections, any given
profile might in principle be expected to have several, in both directions. In
practice (as seen in the profile decompositions in \citealt{Cooper10} and
\citealt{Cooper15}) the fact that the accreted component is the superposition
of many contributions tends to blur out most of the individual dynamical
downward inflections but nevertheless to preserve a (more gradual) average
steepening of the profile to larger radius. 
{  In support of such predictions, \citet{Dsouza14} found that the transition radius between the different accreted components,
$R_{tr}$, varies systematically with the stellar mass, being on average
smaller for more massive galaxies, in agreement with the trends predicted by
\citet{Cooper13,Rodriguez15}.}
We also find a possible
correlation between the slope of the surface brightness profiles beyond the
inflection point and the density of the field around each galaxy, indicating
that objects in more dense fields (hence perhaps more massive dark matter
halos) have steeper outer profiles.

For our galaxies with upward profile inflections, the inflection point is well
beyond $2 R_{e}$, suggesting that these features are attributable to an excess
in the number of weakly-bound stars with large orbital apocenters, perhaps
associated with more recent accretion events.  We do not observe significant
changes in surface brightness or other properties at the transition point
between the inner two components of our theoretically-motivated three-component
fits. This is perhaps not surprising, because simulations predict that the in
situ — accreted transition is hardly noticeable in very massive galaxies, for
the most part because the in situ stars account for a relatively small fraction
of the total mass even in the bright body of the galaxy.  In the most massive
galaxies, the accreted mass is expected to be made up from $\sim10$ significant
contributions with roughly similar mass, and of a similar mass to the in situ
component; this is in contrast to Milky Way-mass galaxies which are expected to
have stellar halos dominated by only one or two progenitors much less massive
than the in situ contribution \citep{Cooper13}.  Moreover the same models
predict that the ages and metallicities of in situ stars  in massive ETGs are
quite similar to those of the most significant ``relaxed'' accreted components
\citep{DeLucia07, Cooper13}.

That we see a variety of profile inflections in our photometric investigation
of this small subset of the VEGAS sample, and that these are broadly consistent
with the expectations of state-of-the-art theoretical models, is encouraging.
Our results suggests that, with the complete sample of extremely deep surface
brightness profiles from the full survey, we will be able to investigate the
late stages of massive galaxy assembly statistically, distinguishing
dynamically evolved systems from those still reaching dynamical equilibrium and
probing the balance between in situ star formation and accretion across a wide
range of stellar mass.  This is a promising route to constraining cosmological
models of galaxy formation such as those we have compared with here, which
predict fundamental, relatively tight correlations between the present-day
structure of massive galaxies and the growth histories of their host dark
matter halos.

\begin{acknowledgements}
    This work is based on visitor mode observations taken at the ESO
    La Silla Paranal Observatory within the VST Guaranteed Time
    Observations, Program IDs 090.B-0414(D), 091.B-0614(A), 094.B-
    0496(A), 094.B-0496(B), 094.B-0496(D) and
    095.B-0779(A). The authors wish to thank the
anonymous referee for his or her comments and suggestions
that allowed us to greatly improve the paper. The authors wish to tank ESO for the financial
    contribution given for the visitor mode runs at the ESO La Silla
    Paranal Observatory. M. Spavone wishes to thank the ESO staff of the
    Paranal Observatory for their support during the observations at
    VST. APC is supported by a COFUND/Durham Junior Research Fellowship under
EU grant [267209] and acknowledges support from STFC (ST/L00075X/1).
The data reduction for
this work was carried out with the computational infrastructure of
  the INAF-VST Center at Naples (VSTceN). This research made
  use of the NASA/IPAC Extragalactic Database (NED), which is operated
  by the Jet Propulsion Laboratory, California Institute of
  Technology, under contract with the National Aeronautics and Space
  Administration, and has been partly supported by the PRIN-INAF
  ``Galaxy evolution with the VLT Survey Telescope (VST)'' (PI
  A. Grado). NRN, EI and MP have been supported by the PRIN-INAF 2014
  ``Fornax Cluster Imaging and Spectroscopic Deep Survey''
  (PI. N.R. Napolitano). MS, EI and M. Cantiello acknowledge finacial
  support from the VST project (P.I. M. Capaccioli). 
\end{acknowledgements}


\begin{appendix} 

\begin{table*}
\setlength{\tabcolsep}{2.0pt}
\begin{center}
\small
\caption{Best fitting structural parameters for a two component fit
  in which $n$ for the inner (in situ) component is fixed.} \label{fit2compfix}
\vspace{10pt}
\begin{tabular}{lcccccccccccccccc}
\hline\hline
Object & $\mu_{e1}$ &$r_{e1}$&$n_{1}$& $\mu_{e2}$ &$r_{e2}$&$n_{2}$&$m_{T,1}$&$m_{T,2}$&$f_{h}$\\
    & [mag/arcsec$^{2}$] &[arcsec]& & [mag/arcsec$^{2}$] & [arcsec]&& [mag] & [mag]&\\
\hline \vspace{-7pt}\\
NGC 1399 & 19.76$\pm$0.06   &19.48$\pm$0.89  &  2&24.60$\pm$0.06 &382$\pm$11&1.45$\pm$0.08& 9.92 &8.30 &82\%\\
 NGC 3923    &20.41$\pm$0.14   &17.34$\pm$1.90  &2&24.11$\pm$0.15 &215$\pm$16&3.36$\pm$0.45&10.83 &9.06 &84\%\\
 NGC 4365     &20.06$\pm$0.24  &10.72$\pm$1.60 &2&23.56$\pm$0.15 &159$\pm$11&3.96$\pm$0.36&11.52 &9.17&90\%\\
 NGC 4472    &21.08$\pm$0.33  &39.95$\pm$7.50 &2&25.32$\pm$0.49 &547$\pm$118&6.10$\pm$0.69&9.68 &8.24 &79\%\\
NGC 5044   &22.16$\pm$0.05  &39.3$\pm$1.8 &2.5&27.03 $\pm$ 0.17&678 $\pm$ 55& 1.54 $\pm$ 0.27&10.80 &9.49 &77\%\\
 NGC 5846   &21.11$\pm$0.16  &18.9$\pm$1.1 &2&26.87 $\pm$ 0.28&757 $\pm$ 101& 5.47 $\pm$ 0.68&11.34 &9.09 &89\%\\
\hline
\end{tabular}

\tablefoot{Columns 2, 3 and 4 report effective magnitude, \APC{effective
radius} and S{\'e}rsic index for the inner component of each fit. The S{\'e}rsic index for
the in-situ component has
been fixed to $n\sim 2$, by using the models as a prior
\citep{Cooper13}. We allowed small variations of $\pm 0.5$ around the mean
  value of $n=2$. This would bracket the range of $n$ in the
  simulations, and allows us to obtain a better fit.\APC{Columns
5, 6 and 7 list} the same parameters for the
outer S{\'e}rsic component. \APC{Columns} 8 and 9 \APC{report} the
total magnitude of the inner ($m_{T ,1}$) and outer S{\'e}rsic component ($m_{T ,2}$). \APC{Column} 12 gives the relative \APC{fraction} of the outer
component with respect to the \APC{total luminosity} of the galaxy.}
\end{center}
\end{table*}

\section{Masking of bright sources}
  \label{mask}
\begin{figure*}
\centering
\hspace{-0.cm}
\includegraphics[width=8.cm]{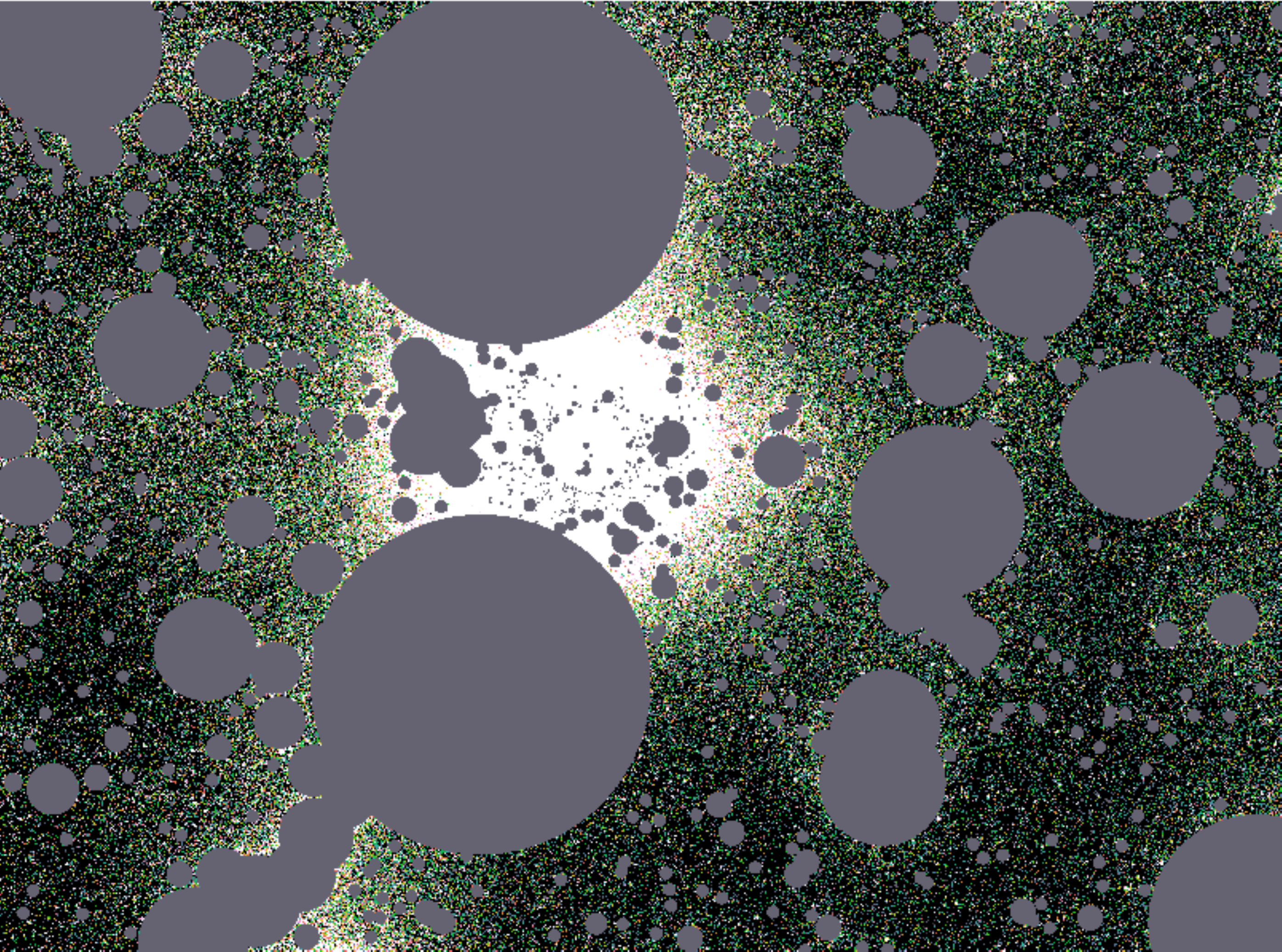}
\includegraphics[width=8.cm, height=6.cm]{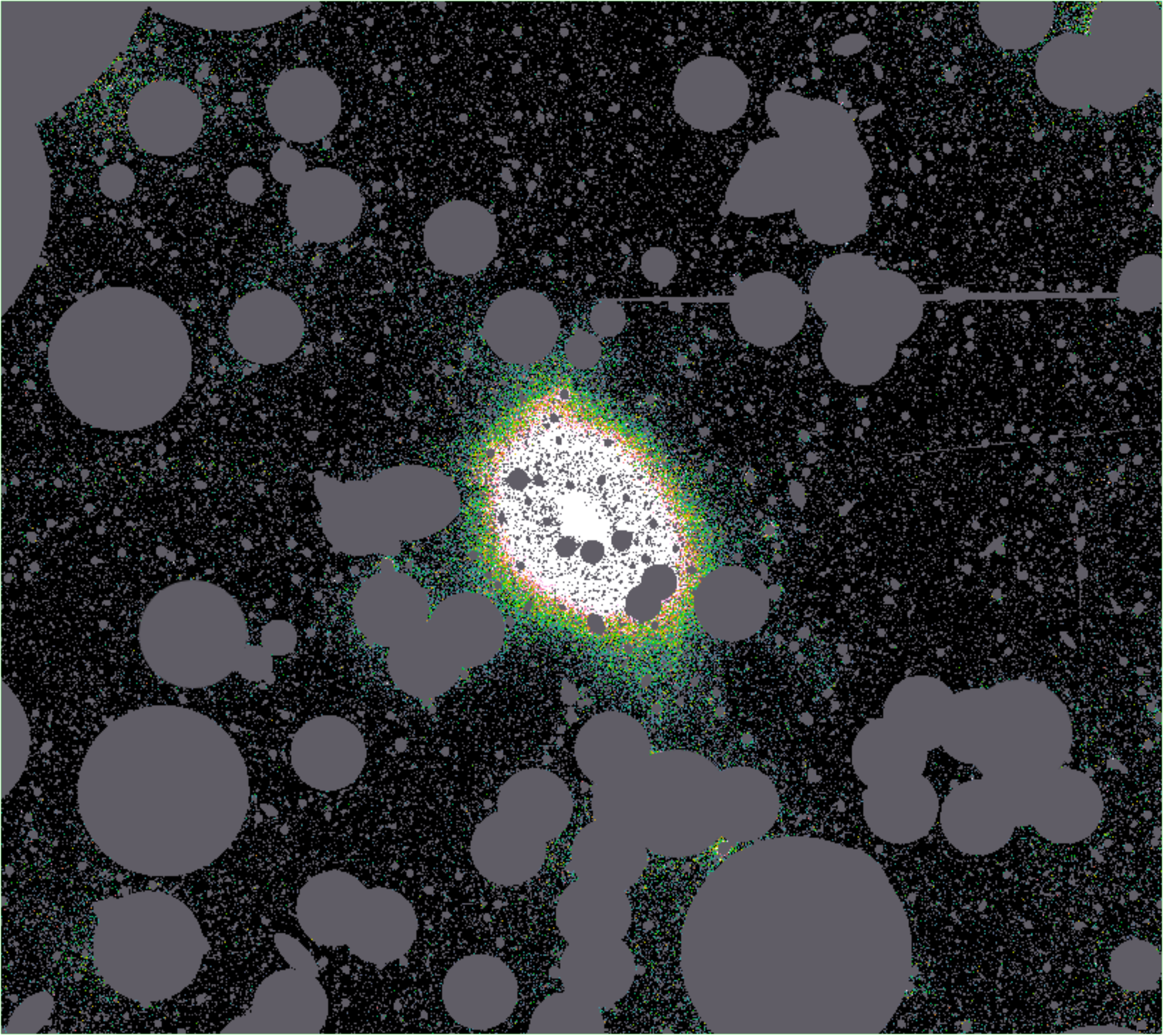}
\includegraphics[width=8.cm, height=8.32 cm]{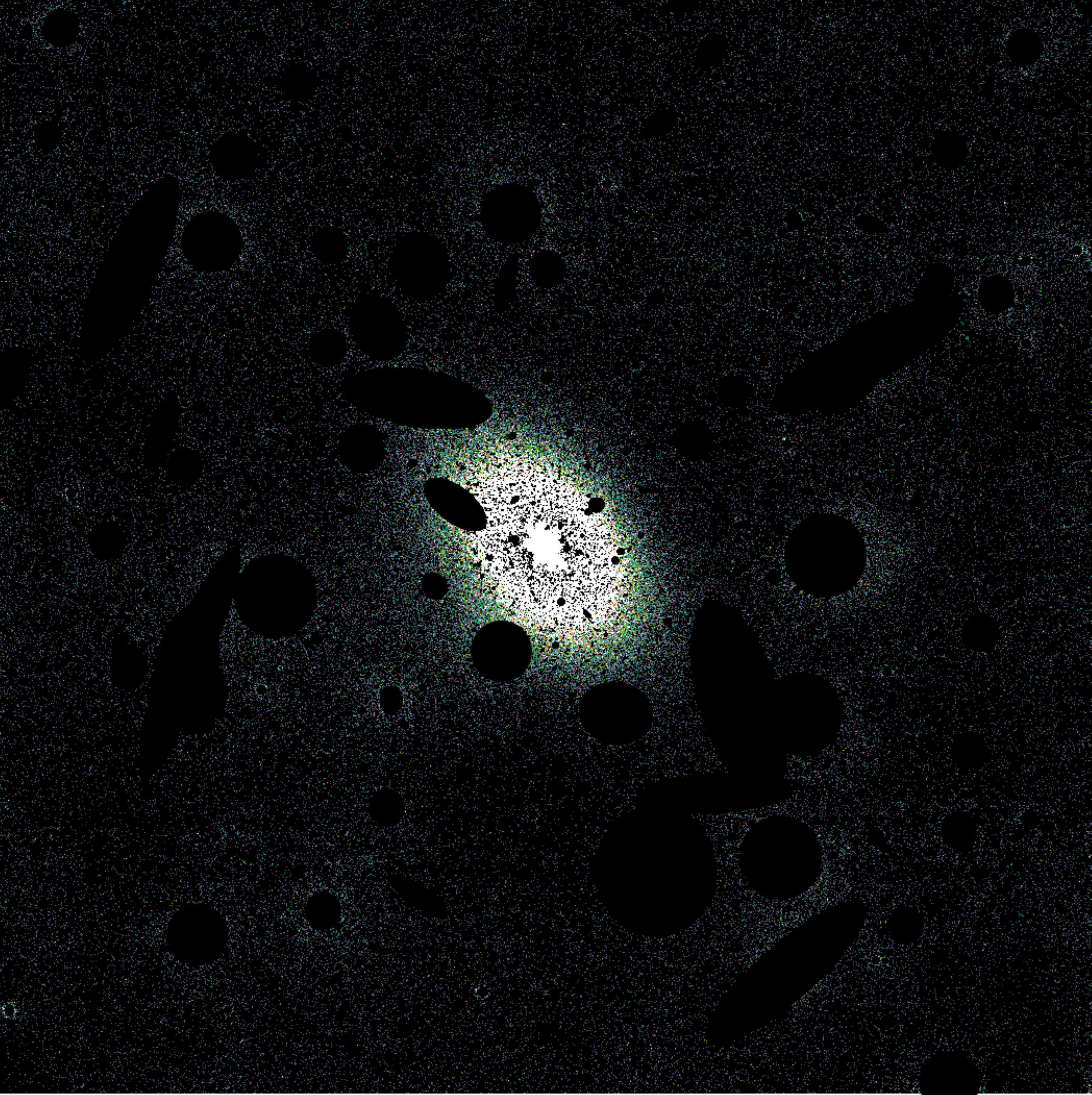}
\includegraphics[width=8.cm]{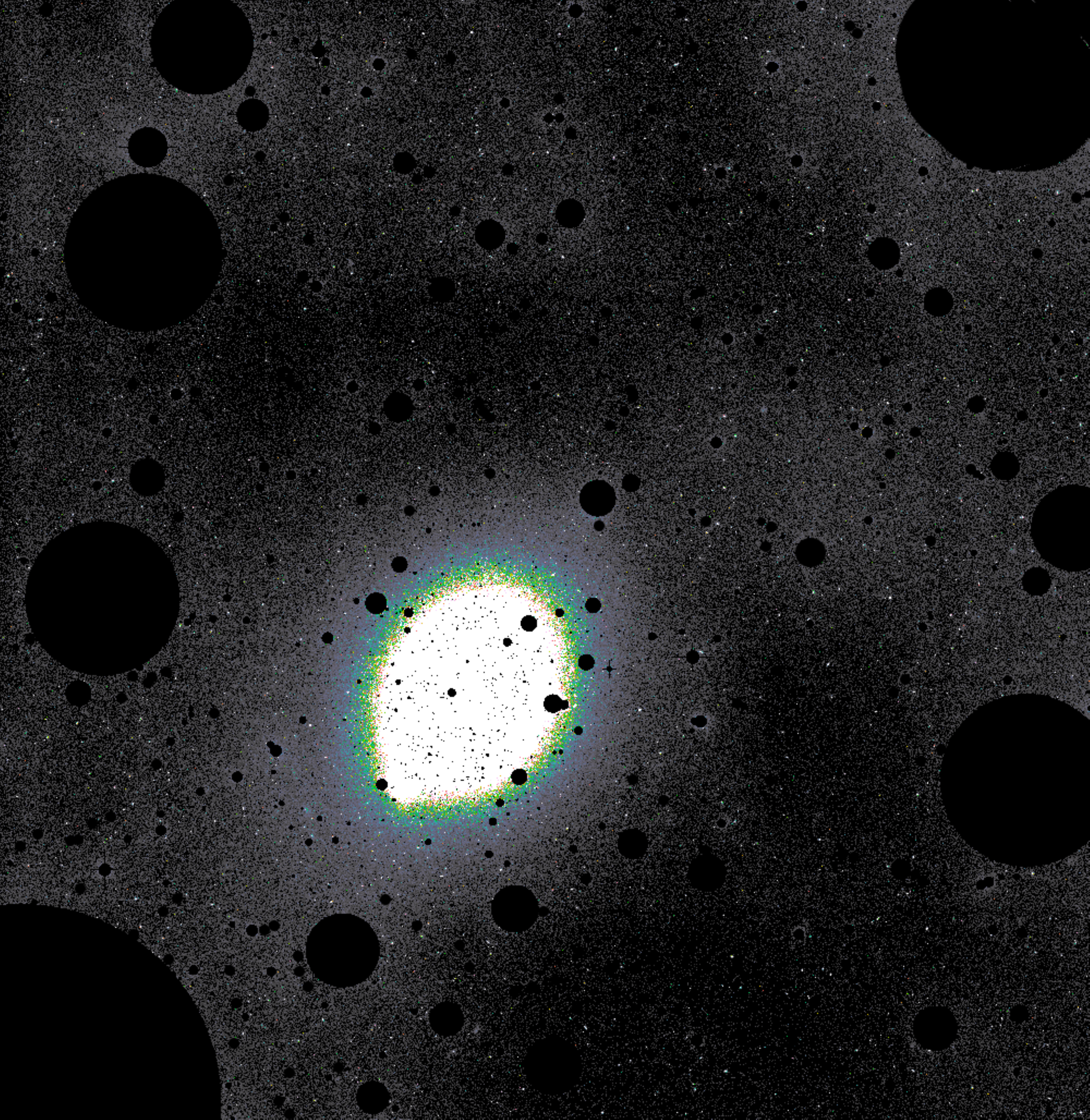}
\includegraphics[width=8.cm]{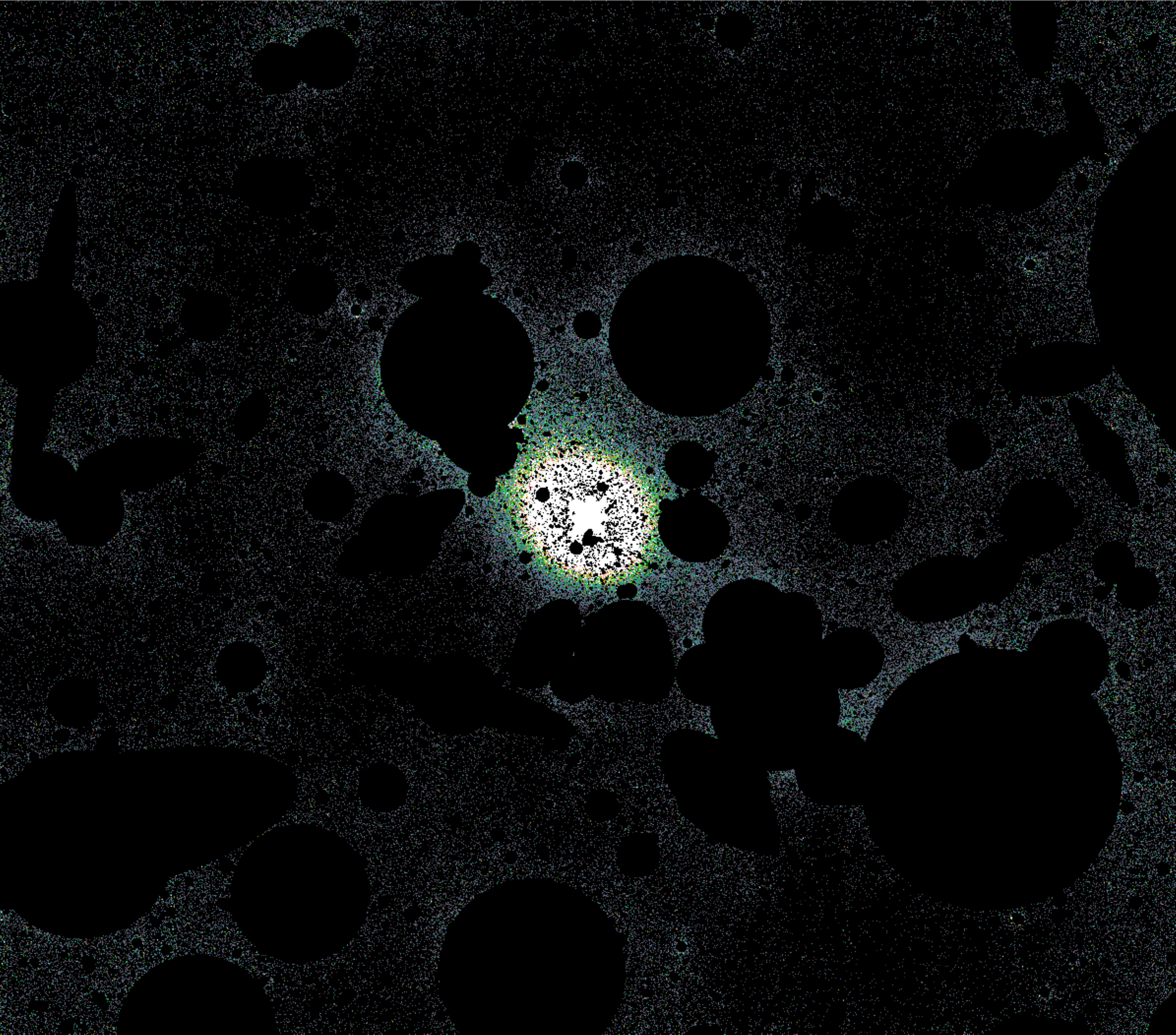}
\includegraphics[width=8.cm]{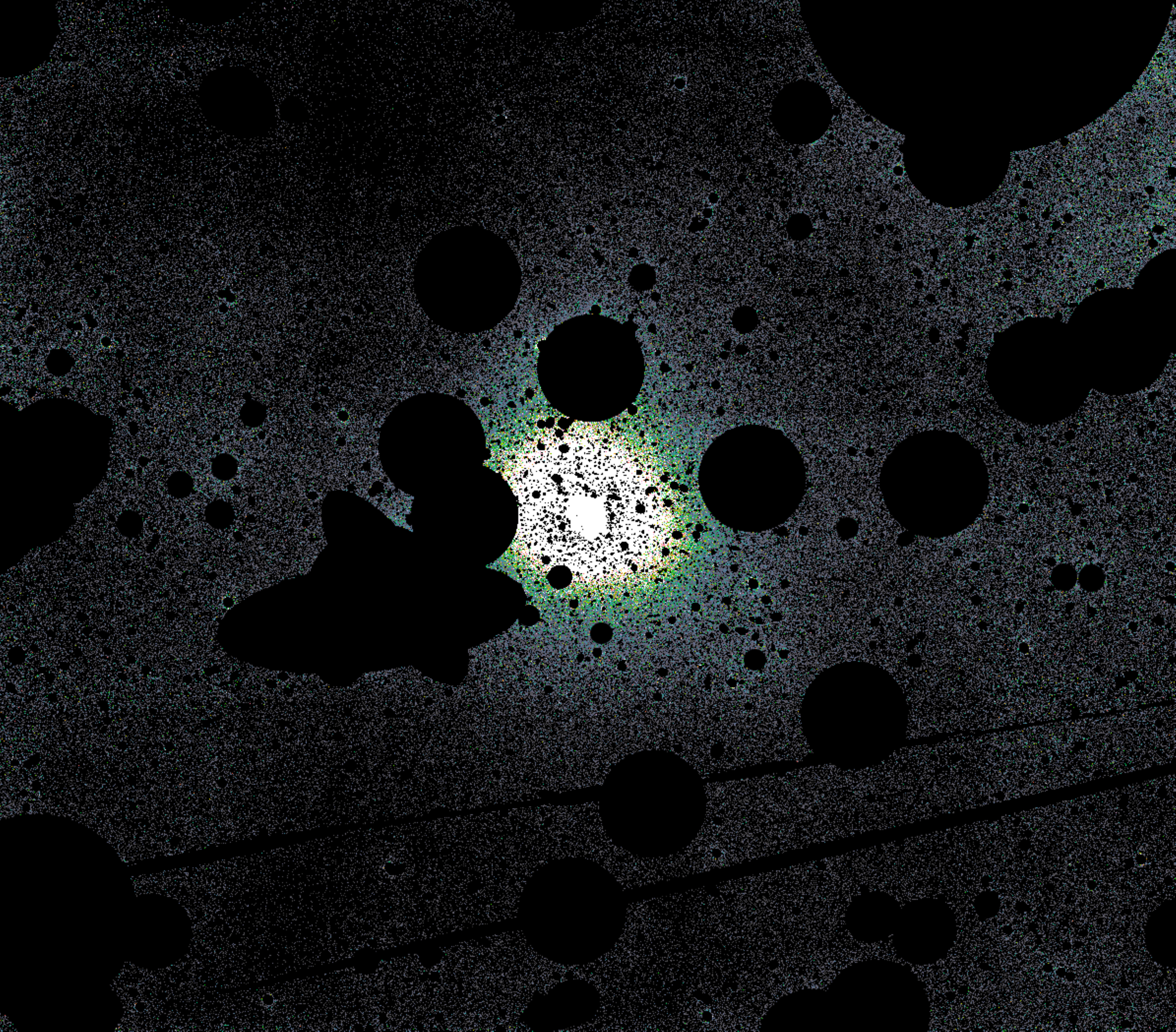}
\caption{VST {\it g} band mosaic of NGC 1399, NGC 3923, NGC 4365, NGC
  4472, NGC 5044 and NGC 5846, showing the masking of bright sources in the field.}
\label{masks}
\end{figure*}

In Fig. \ref{masks}, we show the masks used in this work for the
photometric analysis of NGC 1399, NGC 3923, NGC 4365, NGC 4472, NGC 5044, and
NGC 5846, respectively. All sources have been masked out by using ExAM\footnote{ExAM is a code developed by Z. Huang during his PhD. A detailed description of the code can be found in his PhD thesis, available at the following link: \url{http://www.fedoa.unina.it/id/eprint/8368}} \citep{Huang11}, a program based on
SExtractor \citep{Bertin96}, which was developed to accurately mask background and 
foreground sources, reflection haloes, and spikes from saturated
stars. 
 
\section{Surface brightness profiles in azimuthal sectors}\label{sect}

We measured the surface brightness profiles for each galaxy in four
azimuthal sectors, in order to test the reliability of the profile
measurements at large galaxy radii, and to quantify the combined
systematic uncertainties due to masking, assumption of elliptical
isophotes, stray-light, etc. The same has been already done for NGC
1399 by \citet{Iodice16}.

We found that the surface brightness profiles in the outskirts of almost all
galaxies are on average the same as the azimuthally averaged ones. The
only exception is NGC 5044 in the SW sector, where it shows an excess of light. Such excess is
observed at bright magnitudes ($\sim$ 25 mag/arcsec$^{2}$) and it is in the direction where most of the galaxies belonging
to the group are located. Since NGC 5044 belongs to a fossil group of
galaxies \citep{Falten05}, the excess of light observed in the SW
direction could be the signature of the presence of intracluster light.

 \begin{figure*}
\centering
\hspace{-0.cm}
\includegraphics[width=8.cm]{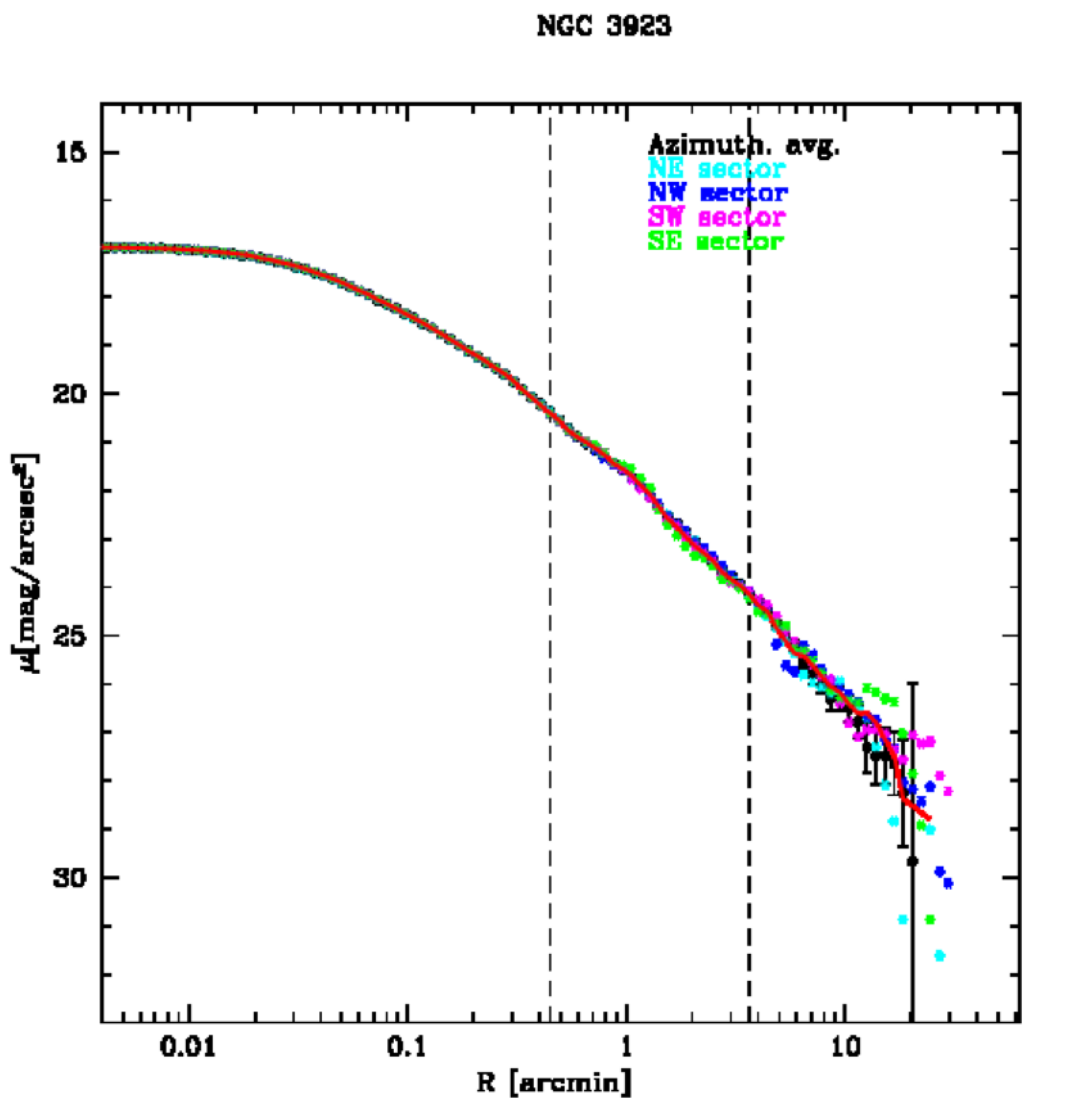}
\includegraphics[width=8.cm]{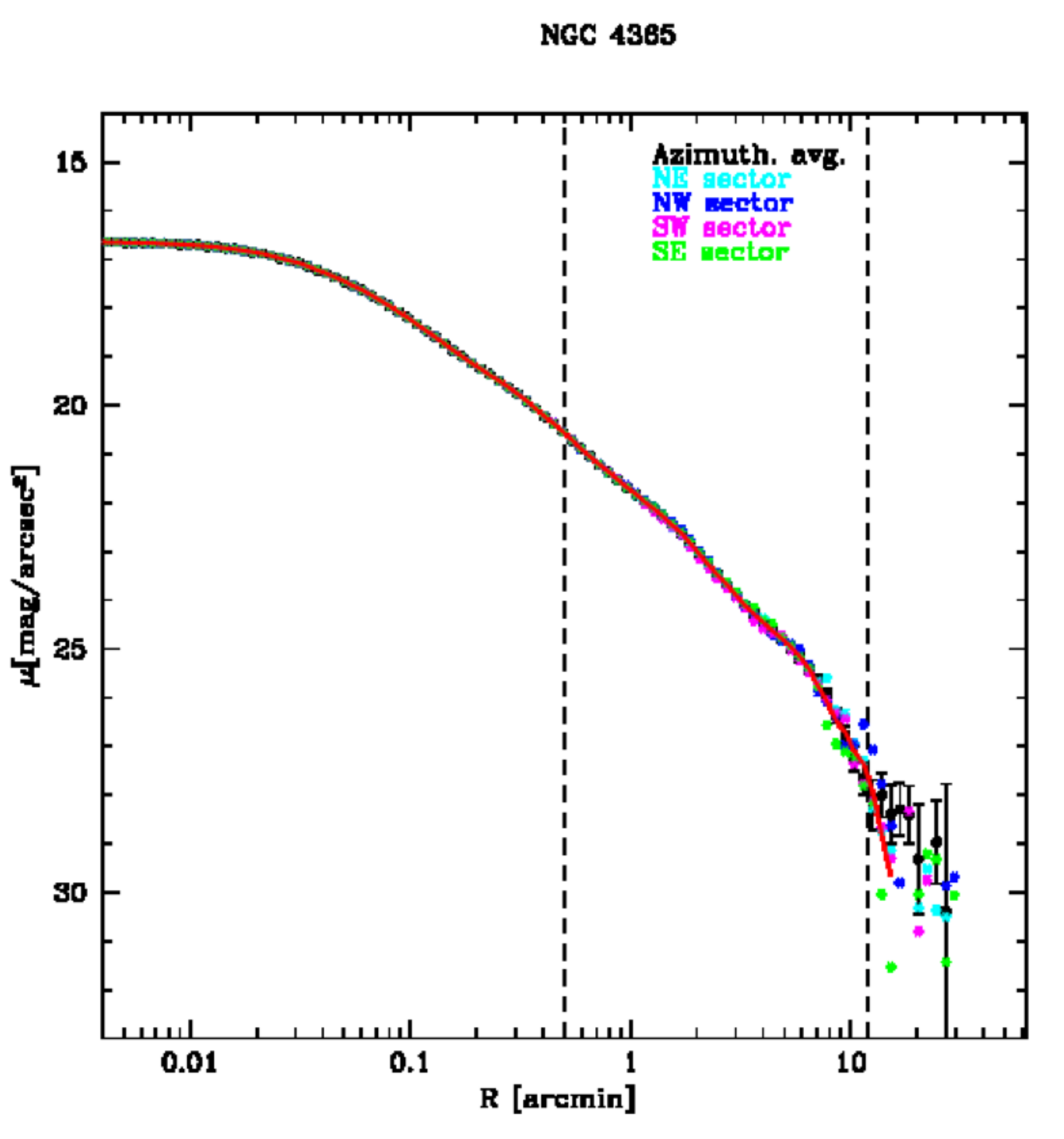}
\includegraphics[width=8.cm]{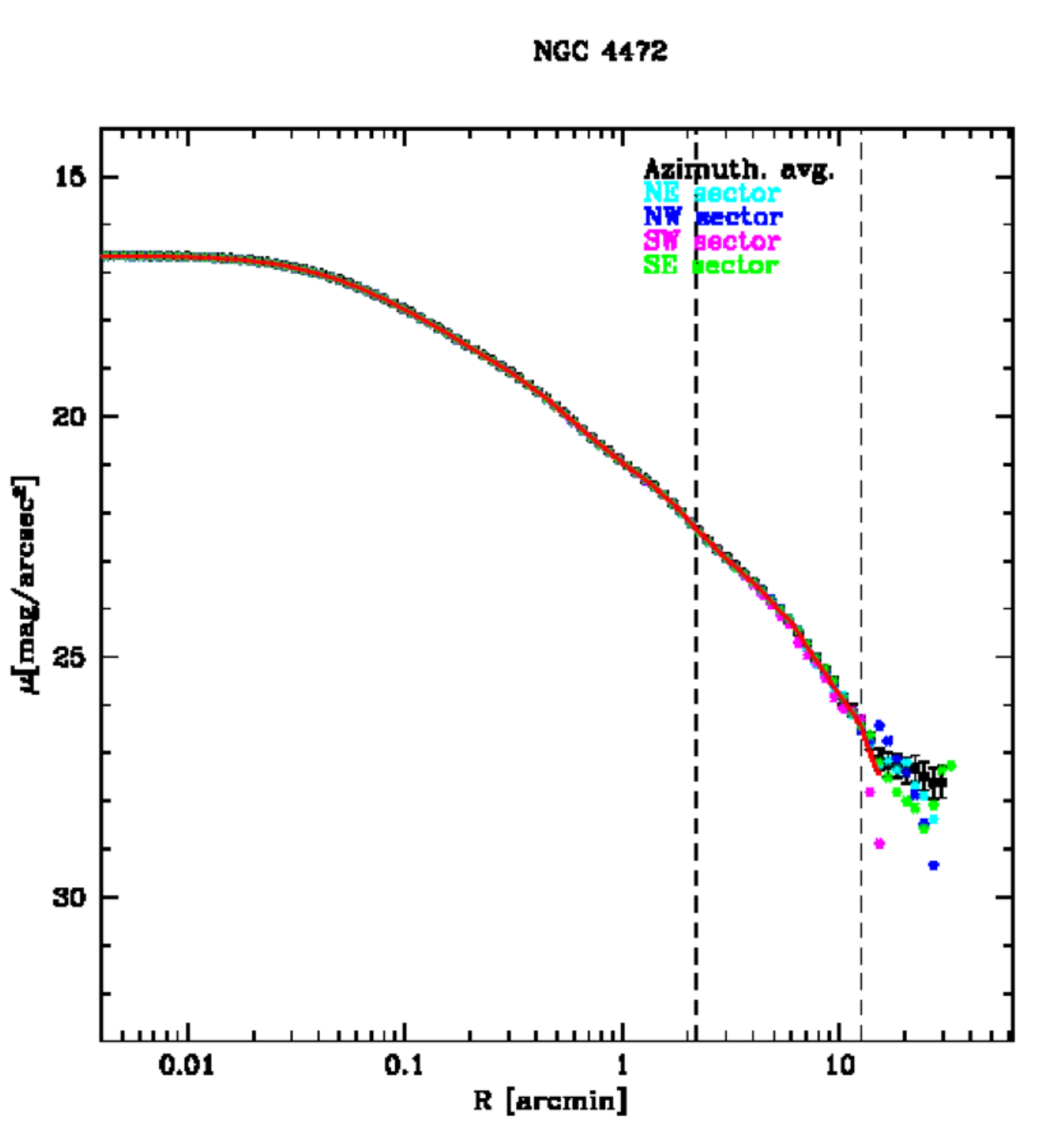}
\includegraphics[width=8.cm]{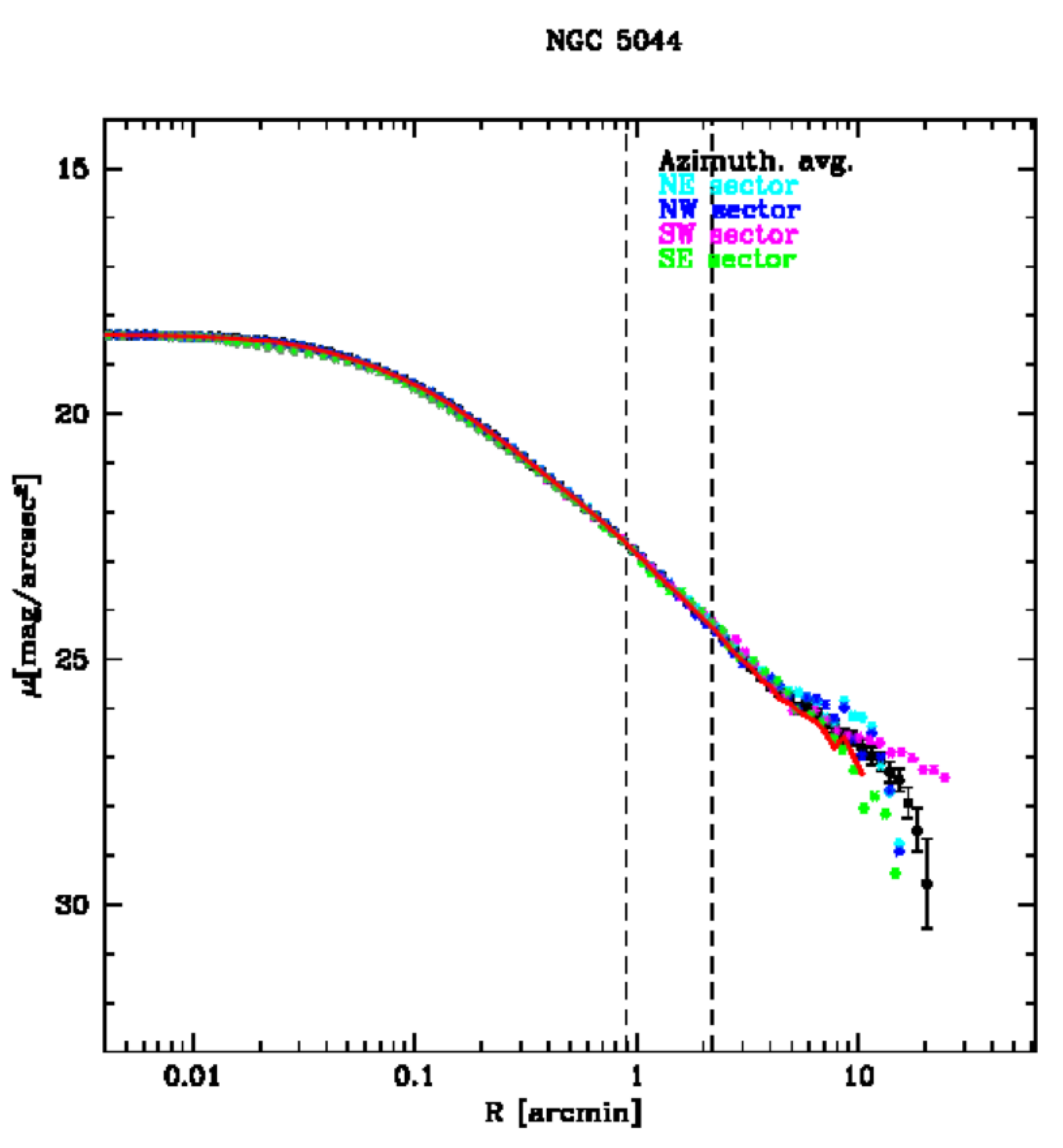}
\includegraphics[width=8.cm]{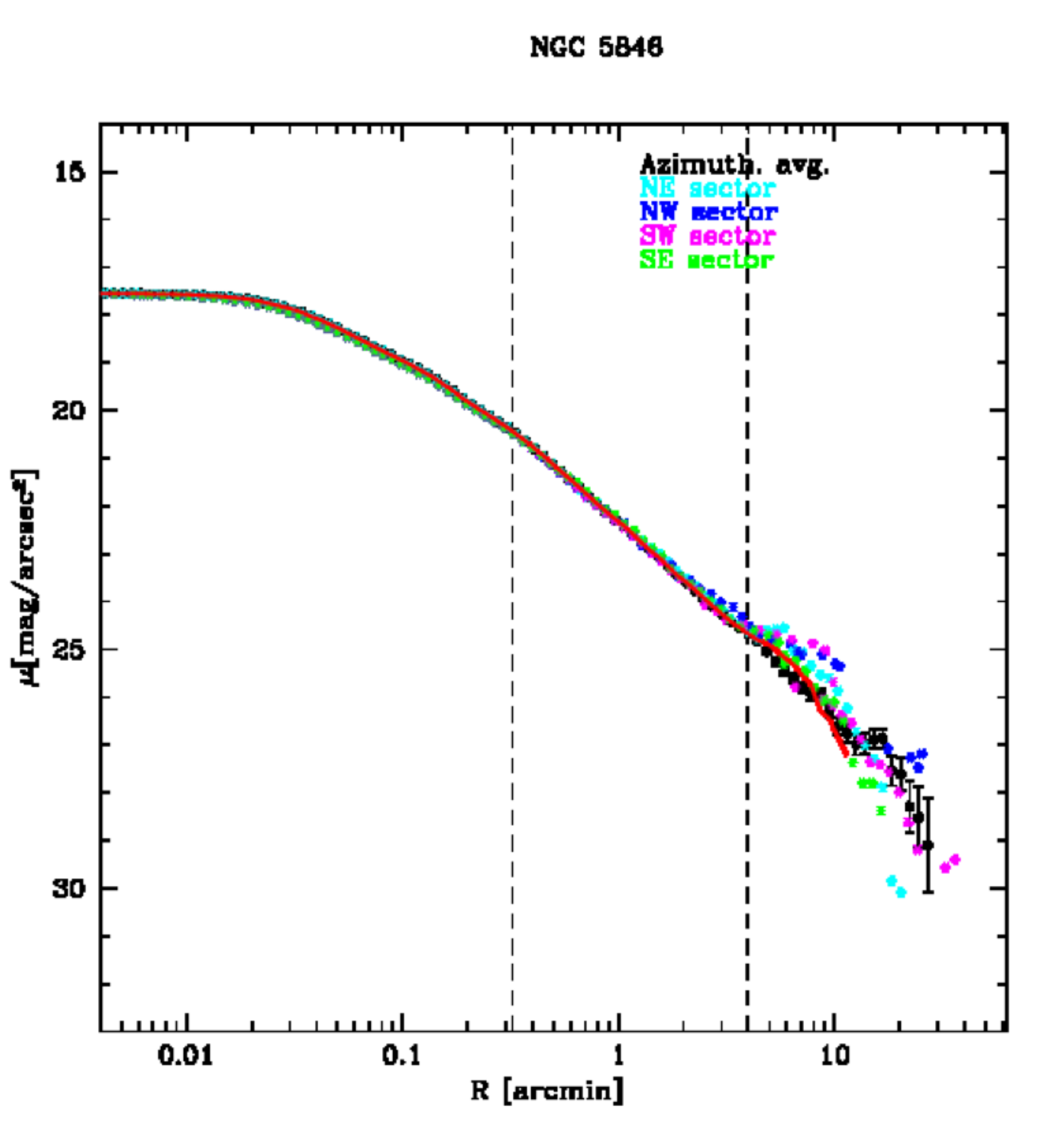}
\caption{VST {\it g} band surface brightness profiles of NGC 3923, NGC 4365, NGC
  4472, NGC 5044 and NGC 5846, extracted in azimuthal sectors and
  compared with the azimuthally averaged profile. {  The red line is the
  average profile, while the dashed lines indicate the location of the
transition radii.}}
\label{sectors}
\end{figure*}

\section{Two-component models fixing the innermost (in situ) component}
  \label{app}

\begin{figure*}
\centering
\hspace{-0.cm}
\includegraphics[width=8.cm,  height=7.5cm]{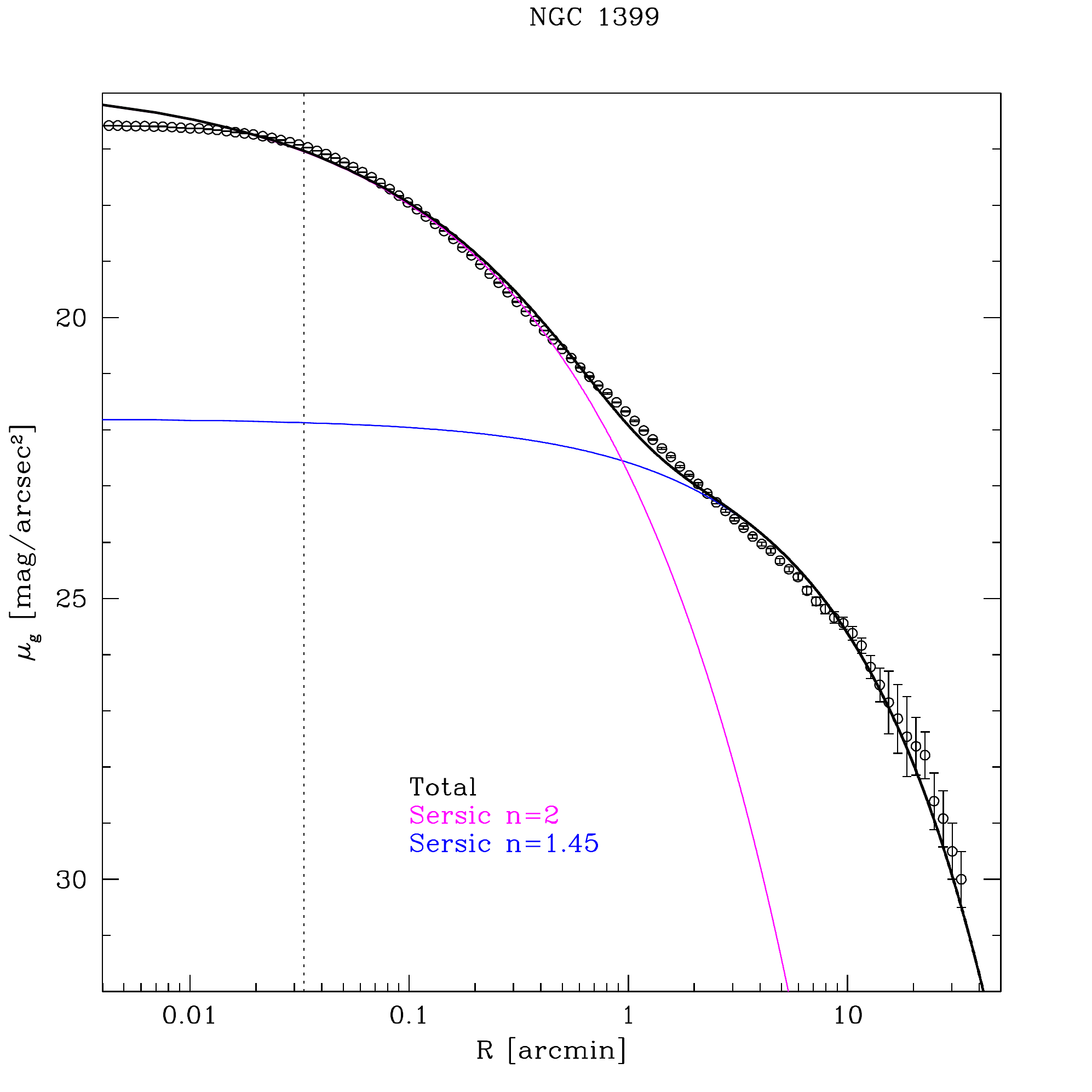}
 \includegraphics[width=8cm,  height=7.5cm]{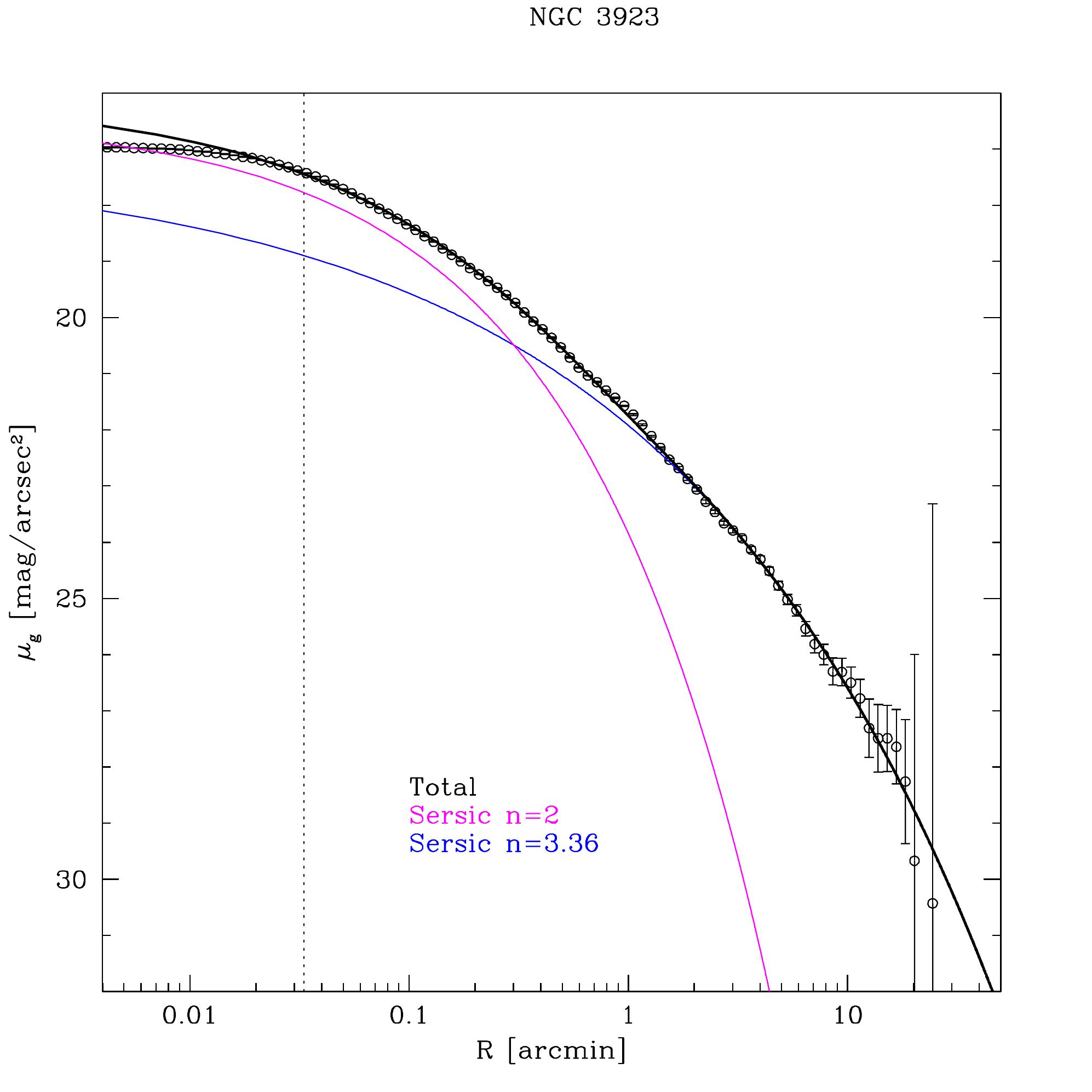}
 \includegraphics[width=8.cm,  height=7.5cm]{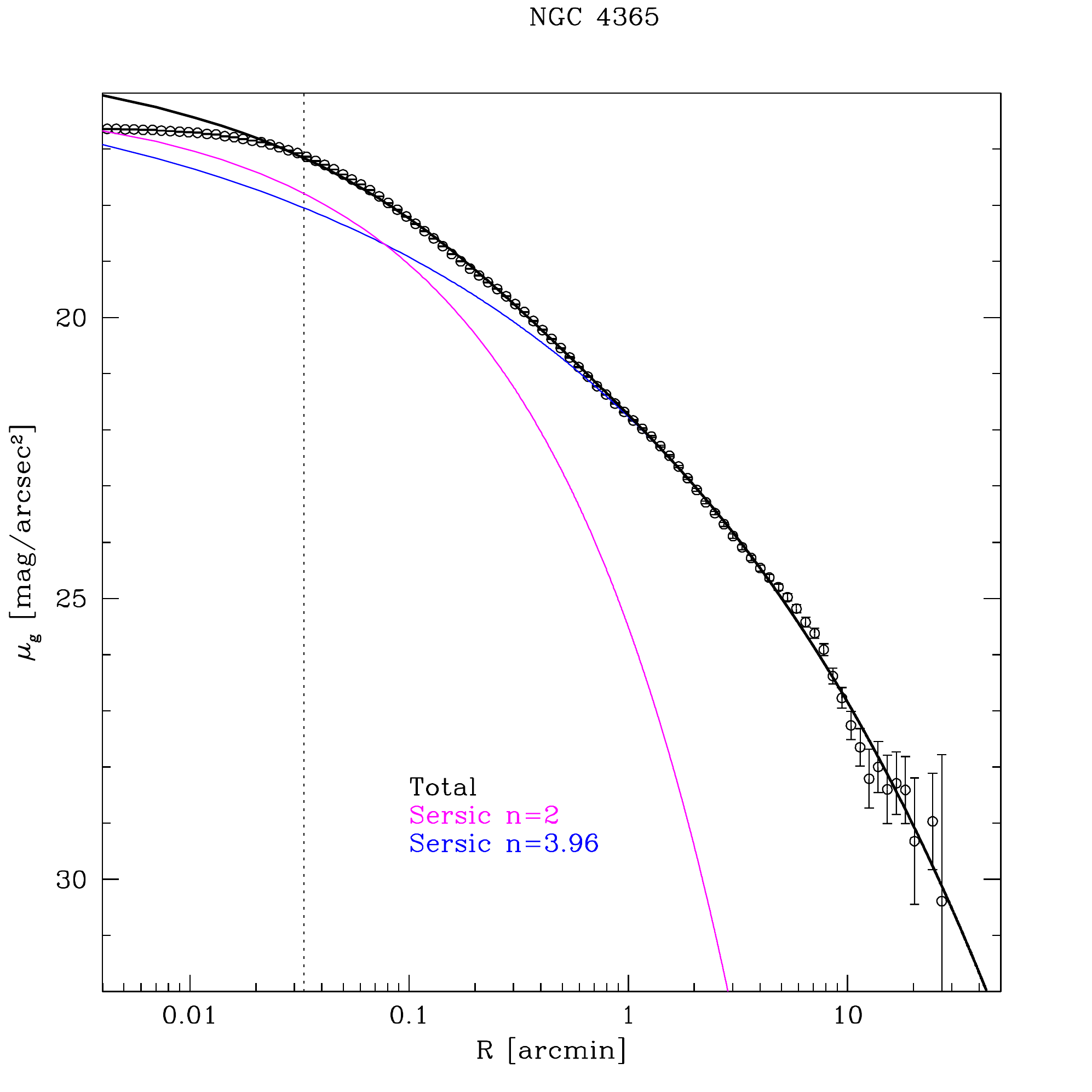}
\includegraphics[width=8.cm,  height=7.5cm]{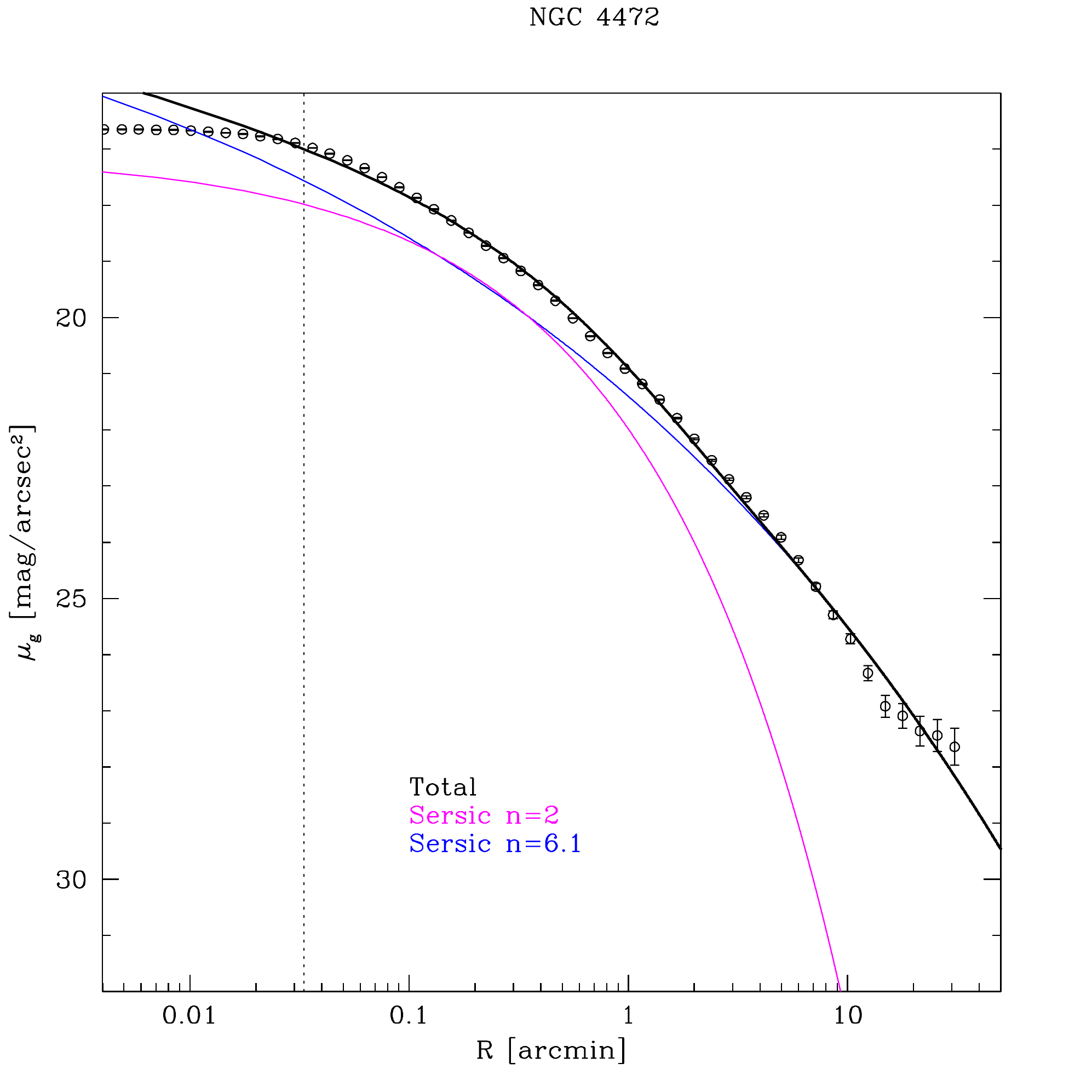}
\includegraphics[width=8.cm,  height=7.5cm]{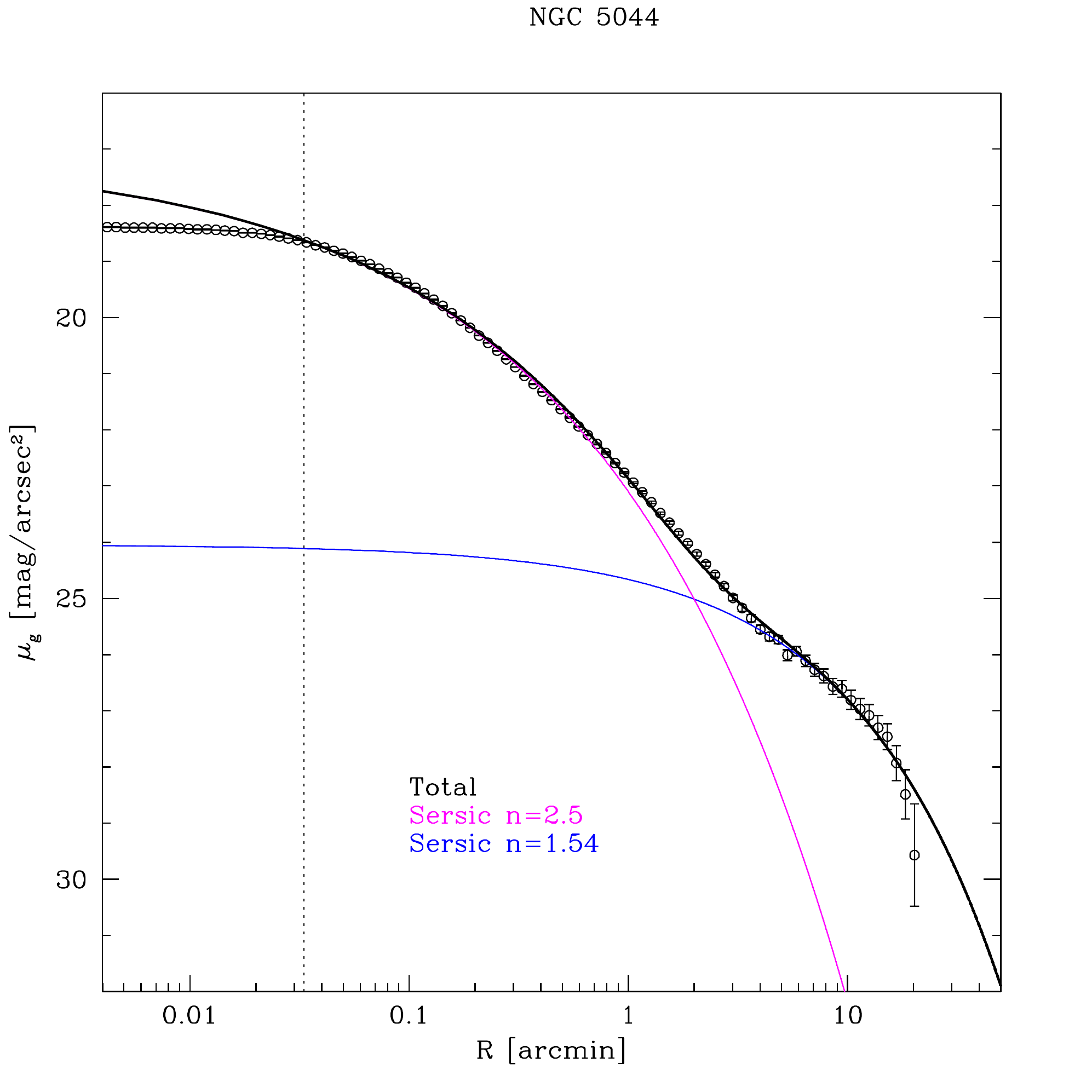}
\includegraphics[width=8.cm,  height=7.5cm]{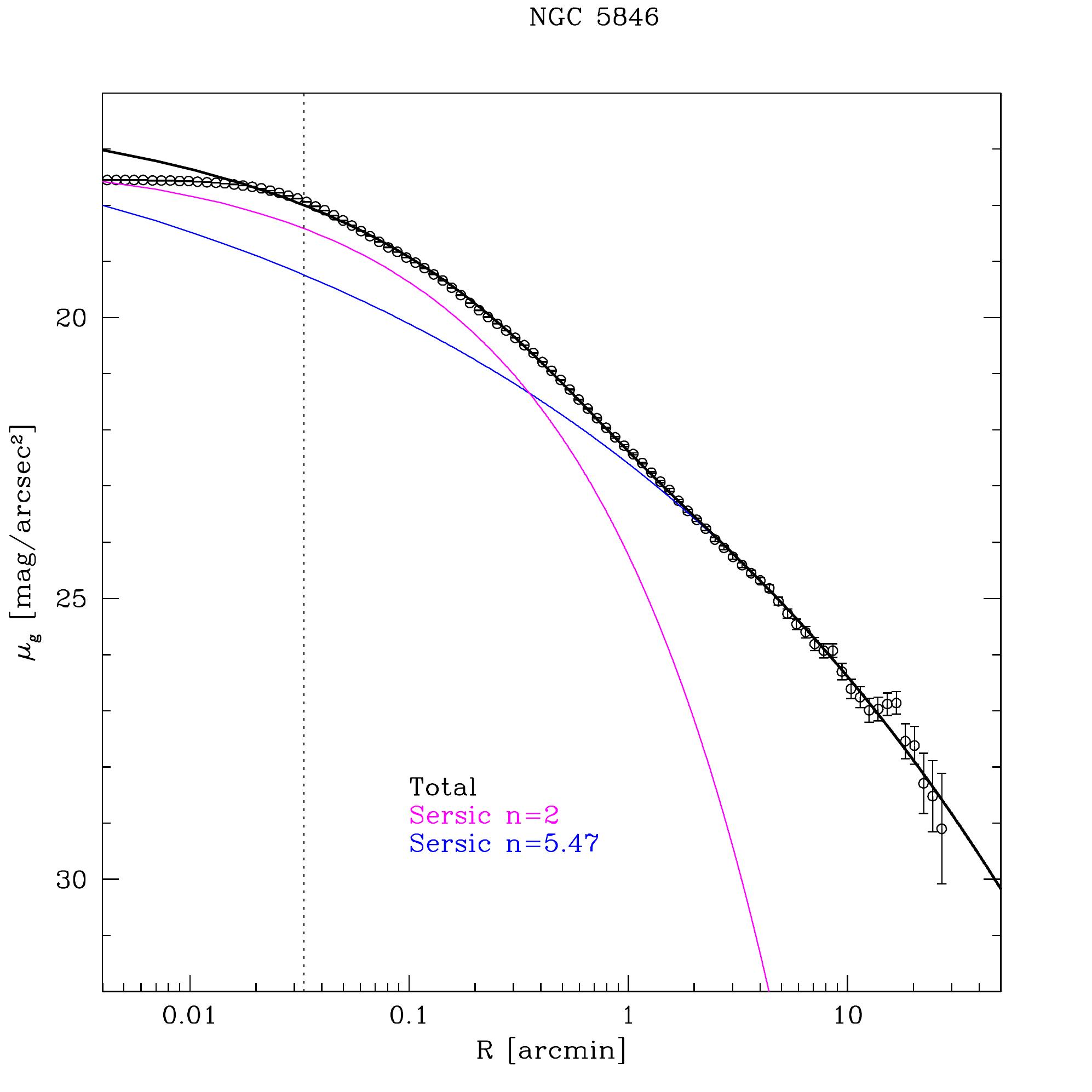}
\caption{VST {\it g} band profiles of NGC 1399, NGC 3923, NGC 4365, NGC 4472,
NGC 5044, and NGC 5846 plotted \APC{on a} logarithmic scale. The blue line is
\APC{a} fit \APC{to} the outer regions. The magenta line is \APC{a} fit \APC{to} the inner
regions with a S{\'e}rsic profile with $n$ fixed from theory \citep{Cooper13}, and the black line is the \APC{sum of the
components in each} fit. The dotted line marks the core of the galaxy
($R\sim 2$ arcsec), which
has been excluded \APC{in the} fit.}\
\label{appendix_fit}

\end{figure*}

  In Sec.~\ref{sec:threecomp} we presented three-component fits to our sample
  of \etg{} light profiles using a theoretically motivated constraint on the
  Sersic index of the innermost component (representing stars formed in situ)
  and the outermost component (representing unrelaxed debris).  In
  Fig.~\ref{appendix_fit} and Tab.~\ref{fit2compfix} we present analogous
  results for the dominant component (representing relaxed accreted stars) when
  we \textit{do not} include an exponential outer component. The parameters
  obtained in this way are more directly comparable to those of other work
  using two-component fitting (note in particular the higher values of $n_{2}$)
  but less representative of the behaviour of the outermost part of the
  profiles in our deep observations. 

\end{appendix}

\end{document}